\documentclass[preprint]{aastex}

\usepackage{graphicx}
\usepackage{epsfig}
\usepackage{multirow}
\usepackage{longtable}

\def\ha{\relax \ifmmode {\rm H}\alpha\else H$\alpha$\fi}
\def\pa{\relax \ifmmode {\rm Pa}\alpha\else Pa$\alpha$\fi}
\def\arcsec{\hbox{$^{\prime\prime}$}}
\def\nii{\relax \ifmmode {\rm N\,{\sc ii}}\else N\,{\sc ii}\fi}
\def\hii{\relax \ifmmode {\rm H\,{\sc ii}}\else H\,{\sc ii}\fi}
\def\hi{\relax \ifmmode {\rm H\,{\sc i}}\else H\,{\sc i}\fi}
\def\deg{\hbox{$^{\circ}$}}

\shorttitle{Breaks in thin and thick discs}
\shortauthors{Comer\'on et al.}

\begin{document}


\title{Breaks in thin and thick discs of edge-on galaxies imaged in the {\it Spitzer} Survey of Stellar Structure in Galaxies (S$^4$G)}


\author{
S\'ebastien Comer\'on,\altaffilmark{1,2}
Bruce G.~Elmegreen,\altaffilmark{3}
Heikki Salo,\altaffilmark{1}
Eija Laurikainen,\altaffilmark{1,4}
E.~Athanassoula,\altaffilmark{5}
Albert Bosma,\altaffilmark{5}
Johan H.~Knapen,\altaffilmark{6,7}
Dimitri A.~Gadotti,\altaffilmark{8}
Kartik Sheth,\altaffilmark{9}
Joannah L.~Hinz,\altaffilmark{10}
Michael W.~Regan,\altaffilmark{11}
Armando Gil de Paz,\altaffilmark{12}
Juan-Carlos Mu\~noz-Mateos,\altaffilmark{9}
Kar\'in Men\'endez-Delmestre,\altaffilmark{13}
Mark Seibert,\altaffilmark{14}
Taehyun Kim,\altaffilmark{9}
Trisha Mizusawa,\altaffilmark{15,16}
Jarkko Laine,\altaffilmark{1}
Luis C.~Ho,\altaffilmark{14}
and Benne Holwerda\altaffilmark{17}}

\altaffiltext{1}{Astronomy Division, Department of Physics, P.O.~Box 3000, FIN-90014 University of Oulu, Finland}
\altaffiltext{2}{Korea Astronomy and Space Science Institute, 776 Daedeokdae-ro, Yuseong-gu, Daejeon 305-348, Republic of Korea}
\altaffiltext{3}{IBM T.~J.~Watson Research Center, 1101 Kitchawan Road, Yorktown Heights, NY 10598, USA}
\altaffiltext{4}{Finnish Centre of Astronomy with ESO~(FINCA), University of Turku, V\"ais\"al\"antie 20, FI-21500, Piikki\"o, Finland}
\altaffiltext{5}{Laboratoire d'Astrophysique de Marseille - LAM, Universit\'e d'Aix-Marseille \& CNRS, UMR7326, 38 rue F. Joliot-Curie, 13388 Marseille Cedex 13, France}
\altaffiltext{6}{Instituto de Astrof\'isica de Canarias, E-38200 La Laguna, Spain}
\altaffiltext{7}{Departamento de Astrof\'isica, Universidad de La Laguna, E-38205 La Laguna, Tenerife, Spain}
\altaffiltext{8}{European Southern Observatory, Casilla 19001, Santiago 19, Chile}
\altaffiltext{9}{National Radio Astronomy Observatory, Charlottesville, VA, USA}
\altaffiltext{10}{Department of Astronomy, University of Arizona, Tucson, AZ, USA}
\altaffiltext{11}{Space Telescope Science Institute, Baltimore, MD, USA}
\altaffiltext{12}{Departamento de Astrof\'isica, Universidad Complutense de Madrid, Madrid, Spain}
\altaffiltext{13}{Observatorio do Valongo, Universidade Federal de Rio de Janeiro, Ladeira Pedro Ant\^onio, 43, Sa\'ude CEP 20080-090, Rio de Janeiro - RJ - Brasil}
\altaffiltext{14}{The Observatories of the Carnegie Institution for Science, Pasadena, CA, USA}
\altaffiltext{15}{Spitzer Science Center, Pasadena, CA, USA}
\altaffiltext{16}{California Institute of Technology, Pasadena, CA, USA}
\altaffiltext{17}{European Space Agency, ESTEC, Keplerlaan 1, 2200 AG, Noordwijk, the Netherlands}


\begin{abstract}

Breaks in the radial luminosity profiles of galaxies have been until now mostly studied averaged over discs. Here we study separately breaks in thin and thick discs in 70 edge-on galaxies using imaging from the Spitzer Survey of Stellar Structure in Galaxies. We built luminosity profiles of the thin and the thick discs parallel to midplanes and we found that thin discs often truncate (77\%). Thick discs truncate less often (31\%), but when they do, their break radius is comparable with that in the thin disc. This suggests either two different truncation mechanisms -- one of dynamical origin affecting both discs simultaneously and another one only affecting the thin disc -- or a single mechanism that creates a truncation in one disc or in both depending on some galaxy property. Thin discs apparently antitruncate in around 40\% of galaxies. However, in many cases, these antitruncations are an artifact caused by the superposition of a thin disc and a thick disc with the latter having a longer scale length. We estimate the real thin disc antitruncation fraction to be less than 15\%. We found that the ratio of the thick and thin stellar disc mass is roughly constant ($0.2<M_{\rm T}/M_{\rm t}<0.7$) for circular velocities $v_{\rm c}>120\,{\rm km\,s^{-1}}$, but becomes much larger at smaller velocities. We hypothesize that this is due to a combination of a high efficiency of supernova feedback and a slower dynamical evolution in lower-mass galaxies causing stellar thin discs to be younger and less massive than in higher-mass galaxies.

\end{abstract}




\section{Introduction}

Thick discs are defined to be disc-like components with a scale height larger than that of the ``canonical'' or ``classical'' discs. They are most easily detected in close to edge-on galaxies in which they appear as a roughly exponential excess of light a few thin disc scale heights above the midplane. They were first detected and described by Tsikoudi (1979) and Burstein (1979). The Milky Way was soon found to also host a thick disc (Gilmore \& Reid 1983) made of old (Bensby et al.~2005), metal-poor, and $\alpha$-enhanced stars (Fuhrmann 1998; Prochaska et al.~2000). Observations now suggest that thick discs in galaxies in the Local Universe are ubiquitous (Yoachim \& Dalcanton 2006; Comer\'on et al.~2011a).

We recently showed evidence of thick discs being significantly more massive than previously considered with stellar masses ranging from 0.3 to 3.0 times that of the thin disc depending on the galaxy mass and the assumed mass-to-light ratios (Comer\'on et al.~2011b; hereafter CO11b). This study confirmed that galaxies dominated by the thick disc are usually those with lower masses (Yoachim \& Dalcanton 2006). Hosting such a large fraction of the stellar mass makes thick discs likely reservoirs of some of the ``local missing baryons''. A few galaxies, such as NGC~4013, host an additional extended component which could be interpreted as a second thick disc or a flattened halo (Comer\'on et al.~2011c).

The properties of those more massive than expected thick discs give some constraints on their origin. Higher mass galaxies could have their thick discs with a low relative mass explained by internal secular heating (see, e.g., Bovy et al.~2012), by the heating caused by the interaction with a satellite galaxy (see, e.g., Qu et al.~2011), by the accretion of stars from satellite galaxies (see, e.g., Abadi et al.~2003), and/or because of the radial migration of stars (see, e.g., Sch\"onrich \& Binney 2009; Loebman et al.~2011), but the very large thick disc mass fraction found in low-mass galaxies indicates an in situ origin during or shortly after the buildup of the galaxy for a large fraction of their thick disc stars.

Two in situ and high-redshift thick disc formation models have been put forward. Elmegreen \& Elmegreen (2006) and Bournaud et al.~(2009) proposed a scenario where the thick disc forms from the primordial galaxy discs, which contained large clumps of star formation whose scale height was comparable to that of present-day thick discs. Those large clumps would have dissolved into what we know as thick discs. Brook et al.~(2007) proposed thick discs to be made of stars in gas-rich proto-galactic fragments which merged and of stars created during that merger. In both scenarios, the thin disc would have formed afterwards from gas remaining from the initial galaxy formation processes and from new gas accreted through cold flows. These two models predict different properties for the thick discs. If they have been created from the dissolution of giant star-forming clusters, one would expect them to be genuine discs, with a high rotation speed and a luminosity profile similar to that of thin discs. If thick discs 
result from the merger of proto-galactic fragments they would, at least in some cases, host a significant fraction of lagging and/or counterrotating stars.

Our comprehensive detection and modeling of thick+thin discs in edge-on galaxies started in CO11b allow us to study breaks, both truncations and antitruncations, in the radial profile of the thick discs, hitherto not possible. A radical truncation in the radial light profile was first discussed in edge-on galaxies by van der Kruit (1979) and van der Kruit and Searle (1981a, b; 1982). The break phenomenon is widely observed in local galaxies (van der Kruit and Searle 1981b; Barteldrees and Dettmar 1994; Pohlen et al.~2000; de Grijs et al.~2001; Florido et al.~2001; Kregel et al.~2002; Kregel and van der Kruit 2004; Pohlen and Trujillo 2006; Florido et al.~2006a,b) and at high redshift (P\'erez 2004; Trujillo and Pohlen 2005). See for reviews of these optical studies van der Kruit (2001) and Pohlen et al.~(2004b). 

Galaxies have been divided into three types depending on their disc properties; Type I, with an unbroken exponential (Freeman 1970), Type II (truncation), hosting a downbending break, and Type III (antitruncation) hosting a transition to a shallower exponential profile (Pohlen \& Trujillo 2006; Erwin et al.~2008), based on averaged radial light profiles. This opens the possibility that some of the observed characteristics may be due to the superposition of the thin and thick discs in these galaxies. Due to the low-surface-brightness nature of the thick disc, this was left mostly unexplored. Pohlen et al.~(2004a, 2007) do suggest a weakening of the truncation with height moving from the thin to the thick disc-dominated regime. This has been confirmed by Hubble Space Telescope observations over two truncations for all stellar populations (de Jong et al.~2007, Radburn-Smith et al.~2012).

The origin of truncations and antitruncations is under debate. The former are hypothesized to be formed due to star-formation thresholds (Kennicutt 1989), the point of maximum angular momentum of the original primordial cloud from which the galaxy has formed (van der Kruit 1987), and/or by bar angular momentum redistribution (Debattista et al.~2006). Laurikainen \& Salo (2001) proved that many M~51-like galaxies have an antitruncation caused by the stripping of stars and gas from the disc during the interaction. They numerically showed that these antitruncations could last for several Gyr. Also, a few antitruncations ($~15\%$) seem to be caused by the contribution of an extended bulge to the luminosity profile (Maltby et al.~2012).

The main purpose of this paper is to obtain the luminosity profiles of thin and thick discs of nearly edge-on galaxies parallel to their midplane to further constrain their origin. We also study truncations and antitruncations in thin and thick discs as separate features by fitting their luminosity profiles with a generalization of the function proposed by Erwin et al.~(2008) for the description of discs with breaks.

We used images from the Spitzer Survey of Stellar Structure in Galaxies (S$^4$G, Sheth et al.~2010) which has imaged more than 2000 galaxies representative of the nearby Universe (radial velocity $V_{\rm radio}<3000\,{\rm km\,s^{-1}}$) in $3.6\mu{\rm m}$ and $4.5\mu{\rm m}$ using the Infrared Array Camera (IRAC; Fazio et al.~2004). The S$^4$G reaches a typical surface brightness of $\mu_{3.6\mu{\rm m}}({\rm AB})(1\sigma)\sim27\,{\rm mag\,arcsec^{-2}}$ with a pixel size of $0.75\arcsec$ and traces mostly the light of old stars with little dust absorption, which makes it ideal for the study of edge-on galaxies.

This paper is structured as follows. We present the selected sample in Section~\ref{secsample} and we describe the fitting procedure in Section~\ref{secfitting}. We continue in Section~\ref{seccaveat} by addressing what would happen if some assumptions made when producing the fits were not applicable to our sample of galaxies. We show and discuss our results in Section~\ref{secresults} and Section~\ref{secdiscussion} and we present our conclusions in Section~\ref{secconclusions}. 

\section{Galaxy sample}

\label{secsample}

We looked at each of the 2132 galaxies available in the S$^4$G archive by February 15, 2012 and we manually selected disc galaxies appearing to be edge-on with morphological types $-3\leq T<8$ ($T$ from HyperLEDA; Paturel et al.~2003). Later type galaxies were rejected because of their generally ill-defined structure. We have assumed that $T$ can be ``reasonably reliably'' determined for edge-on galaxy as claimed by Buta (2012).

We also rejected galaxies with distorted morphologies, with indications of being not quite edge-on (with visible spiral arms or resonance rings), or discs that are too dim to certainly assess their orientation. When possible, Sloan Digital Sky Survey Data Release 8 (Aihara et al.~2011) imaging was used in order to ensure close to edge-on galaxy orientation by looking at the position of the midplane dust.

The resulting sample has 169 galaxies which includes 29 out of 30 galaxies studied in detail in CO11b. The remaining galaxy, IC~1970, was excluded due to some hint of spiral structure. All those galaxies underwent the image processing and fitting procedure described in Sections~\ref{secmakever} and \ref{secverfit}. In Section~\ref{sectrim1} we describe how, after fitting thin+thick+gas disc profiles to galaxy luminosity profiles perpendicular to their midplanes, we applied additional selection criteria aiming to remove faint galaxies, galaxies with extended envelopes, and/or badly fitted galaxies, in order to ensure the data quality and we built up a final sample with 70 galaxies.

\section{Luminosity profile fitting}

\label{secfitting}

In order to obtain luminosity profiles of thin and thick discs parallel to galaxy midplanes and to study them we followed a procedure that is described in detail in the next subsections:
\begin{itemize}
 \item We prepared a grid of synthetic luminosity profiles perpendicular to the galaxy midplanes, as done in CO11b. The selected profile was that of a combined thin, thick, and gas disc gravitationally coupled in local isothermal equilibrium for galaxies with $T\geq1$. For earlier-type galaxies we only considered a thin and a thick disc. (Section~\ref{sectsynt}).
 \item We extracted the observed luminosity profiles perpendicular to galaxy midplanes using an average of 3.6 and 4.5 $\mu{\rm m}$-band S$^4$G images (vertical luminosity profiles; Section~\ref{secmakever}).
 \item We fitted the observed luminosity profiles by comparing them with the grid of synthetic models as done in CO11b (Section~\ref{secverfit}).
 \item Using the vertical luminosity profile fits, we determined the height, $z_{\rm s}$, above which the thick disc dominates the light emission (Section~\ref{secdefrs}).
 \item By looking at the properties of vertical luminosity profile fits and the resulting $z_{\rm s}$ we trimmed the 169 galaxy sample down to 70 in order to ensure the quality of the data. The galaxy rejection criteria were aimed to remove galaxies with bad fits, having extended envelopes, or being too faint and thus having noisy luminosity profiles (Section~\ref{sectrim1}).
 \item Knowing $z_{\rm s}$ has allowed us to determine the region dominated by the thin and thick disc light; thus, we have been able to prepare luminosity profiles of the thin and thick discs parallel to the galaxy midplanes. We also prepared a profile including light from both discs (horizontal luminosity profiles; Section~\ref{secmakehor}).
 \item The horizontal luminosity profiles of both the thin and the thick discs, as well as those including light from both, were fitted with a generalization of the function used by Erwin et al.~(2008) to describe discs with breaks (Section~\ref{secfithor}).
\end{itemize}

\subsection{Synthetic luminosity profiles perpendicular to galaxy midplanes}

\label{sectsynt}

In order to prepare thin and thick disc luminosity profiles parallel to the midplane of a galaxy, it is fundamental to know in which range in heights above the midplane each disc dominates. To do so, we fitted  luminosity profiles perpendicular to galaxy midplanes with thin+thick+gas disc functions for galaxies with $T\geq1$. For earlier-type galaxies we set the gas disc to have a zero density, which is equivalent to assuming a thin+thick disc function. Assuming little or no gas in these galaxies solves one of the fitting biases discussed in Section~\ref{secnogas}, namely that of gas-depleted galaxies with a thick disc mass overestimated because being fitted with a function accounting for gas.

As in CO11b, we assumed that galaxy discs are relaxed structures whose particles behave like those of a fluid in equilibrium. We also assumed that discs are made of three gravitationally coupled baryonic discs -- gas disc, thin disc and thick disc -- which feel the effect of something acting like a dark matter halo. Then we can write following the equation from Narayan \& Jog (2002):
\begin{equation}
\label{equilibrium}
 \frac{{\rm d}^2\rho_{\rm i}}{{\rm d}z^2}=\frac{\rho_{\rm i}}{\sigma^2_{\rm i}}\left(-4\pi{\rm G}\left(\rho_{\rm t}+\rho_{\rm T}+\rho_{\rm g}\right)+\frac{{\rm d}K_{\rm DM}}{{\rm d}z}\right)+\frac{1}{\rho_{\rm i}}\left(\frac{~{\rm d}\rho_{\rm i}}{{\rm d}z}\right)^2
\end{equation}
where $t$ refers to the thin disc, $T$ to the thick disc and $g$ to the gas disc; the subindex $i$ can be either $t$, $T$, or $g$, $\rho$ stands for the mass density, and $\sigma$ for the vertical velocity dispersion of the component. ${\rm d}K_{\rm DM}/{\rm d}z$ is the term describing dark matter effects. This is a set of three coupled second order differential equations which we solved by using the Newmark-$\beta$ method with $\beta=0.25$ and $\gamma=0.5$. $\beta$ and $\gamma$ are internal parameters of the algorithm which have been set to make it unconditionally stable (Newmark 1959).

When integrating Equation~(\ref{equilibrium}) we assumed that the isothermal hypothesis is true. This hypothesis implies that for a given galactic radius, $r$, the vertical velocity dispersion is constant with height, $z$ ($\sigma(r,z)=\sigma(r)$). We also assumed that the effect of dark matter is negligible (${\rm d}K_{\rm DM}/{\rm d}z=0$; see justification in CO11b and Section~\ref{seccaveat}). Line-of-sight effects were avoided by assuming that all of the discs have similar scale lengths and that the scale heights remain roughly constant with varying $r$. This last assumption is based on the fact that S$^4$G discs do not flare significantly within the optical radius and has been widely used in the literature (van der Kruit \& Searle 1981a; Yoachim \& Dalcanton 2006; CO11b). We also assumed that the face-on gas surface mass density  for galaxies with $T\geq1$ is 0.2 times that of the thin disc at all radii. A $20\%$ gas fraction is slightly higher than that found in the Milky Way at solar radius (Banerjee \& Jog 2007) and we have shown in CO11b that it yields slightly better fits than when not accounting for gas. It is also a fraction of gas representative of what is found in our galaxy sample according to our gas fraction calculations presented in Section~\ref{secv120}. We set $\rho_{\rm g}=0$ for galaxies with $T<1$. Finally, we assumed $\sigma_{\rm g}=1/3\sigma_{\rm t}$ in rough accordance with Milky Way velocity dispersion measurements (Spitzer 1978; Stark 1984; Clemens 1985; Lewis \& Freeman 1989).

Equation~(\ref{equilibrium}) provides mass density profiles. For obtaining luminosity profiles some assumptions about $\Upsilon_{\rm t}$ and $\Upsilon_{\rm T}$, the mass-to-light ratios of the thin and the thick discs, were needed. Different $\Upsilon_{\rm T}/\Upsilon_{\rm t}$ values can be calculated from different star formation histories for the thin and the thick discs. Reasonable star formation histories yield $1.2<\Upsilon_{\rm T}/\Upsilon_{\rm t}<2.4$ (see in discussion in Section~3.3 in CO11b). In this paper we use the most conservative value -- that with lower relative thick disc masses -- $\Upsilon_{\rm T}/\Upsilon_{\rm t}=1.2$.

When integrating Equation~(\ref{equilibrium}), six boundary conditions are needed. Since the maximum density is found in the midplane, we set ${\rm d}\rho_{\rm i}/{\rm d}z|_{z=0}=0$. The other boundary conditions we used are the mass densities of each disc in the midplane $\rho_{\rm i}(z=0)$ (hereafter $\rho_{\rm i0}$).

We solved Equation~(\ref{equilibrium}) for a grid of models with different central density ratios ($\rho_{\rm T0}/\rho_{\rm t0}$) and different vertical velocity dispersion ratios, $\sigma_{\rm T}/\sigma_{\rm t}$. We made integrations for 150 values of $\rho_{\rm T0}/\rho_{\rm t0}$, equally spaced from 0.015 to 2.25, and for 150 values of $(\sigma_{\rm T}/\sigma_{\rm t})^2$, equally spaced from 1.1 to 16.0. The solutions in this grid were then compared to observed luminosity profiles as described in the following subsections.

\subsection{Observed luminosity profiles perpendicular to galaxy midplanes}
\label{secmakever}

We selected the 3.6 and 4.5 $\mu$m-band images for the 169 galaxies in our original sample and we subtracted the background from each image. The background was measured by selecting several tens of $5\times5$ pixel boxes far away from the midplane of the galaxy and bright stars and then by obtaining the median value of the pixels in the box. The selected background value was that of the median of the individual medians found for each box. The angular distance between the boxes in which we measured the background and the midplane of the galaxy for galaxies fitting in the IRAC chip ($D_{25}<3.3'$; Sheth et al.~2010) is limited by IRAC's field of view ($5.2'\times5.2'$) and is typically around 100\arcsec. For the few galaxies which required a mosaic, the sky has been measured in boxes as close to the border of the mosaic as possible in the direction perpendicular to the midplane.

The median distance of the galaxies for which we studied the breaks is 29.8\,Mpc (see Section~\ref{sectrim1} for details on the original sample trimming). At that distance, 100\arcsec\ correspond to 14.4\,kpc, which may cause our estimated sky value to be actually the stellar halo level. Bakos \& Trujillo (2012) found that haloes start affecting the luminosity profiles of galaxies at $r'\sim28\,{\rm mag\,arcsec^{-2}}$. Because we have not found color indexes including IRAC filters for stellar haloes in the literature, we have used globular cluster colors as a proxi for halo colors in order to test whether this error in the sky subtraction could affect our analysis. The reddest globular cluster in the sFample of Spitler et al.~(2008) has $R(\rm AB)-\mu_{3.6\mu{\rm m}}(\rm AB)\sim0.8$. This globular cluster has $V-R\sim0.55$ (Spitler et al.~2008) and using the color transformations for Population~II stars in Jordi et al.~(2006), we can afirm that $r'({\rm AB})\sim R({\rm AB})$ within $0.1$\,mag. That implies that the halo would start affecting our luminosity profiles at $\mu_{3.6\mu{\rm m}}(\rm AB)\sim27\,{\rm mag\,arcsec^{-2}}$. Since our break detection limit is within the range $\mu_{3.6\mu{\rm m}}=25-26\,{\rm mag\,arcsec^{-2}}$ (see Section~\ref{sectrunc}) stellar haloes are not likely to affect our analysis of the background for subtraction.

Typical one-$\sigma$ errors of the background determination of the addition of the 3.6 and 4.5$\mu$m images are around 0.0015\,counts, corresponding to a surface brightness level of $\mu=27.5\,{\rm mag\,arcsec^{-2}}$. This error bar was measured by finding the standard deviation of different background values obtained using bootstrapping over the local background values in the individual $5\times5$ pixel boxes. The value of the measured background does not vary significantly when increasing the box size from $5\times5$ to $9\times9$ pixels. However, the background error rarely is the limiting factor when fitting the luminosity profiles; the actual main limiting factor is the quality of the mask as explained at the end of Section~\ref{secverfit}.

Using the average of the sky-subtracted 3.6 and 4.5 $\mu$m-band images we produced four luminosity profiles for each galaxy: at each side of the bulge, central cluster or presumed center along the disc long axis for projected galactocentric distances $0.2r_{25}<|R|<0.5r_{25}$ and $0.5r_{25}<|R|<0.8r_{25}$ (top panel in Figure~\ref{example}). We used $r_{25}$ values from HyperLEDA except for NGC~4111, for which we preferred the Third Reference Catalog of Bright Galaxies value (RC3; de Vaucouleurs et al.~1991) due to the HyperLEDA value being clearly underestimated. The profiles were prepared by adding the counts above and below the midplane in order to get a unique profile for each bin and averaging for each $z$ over non-masked pixels. We used manually refined masks from the S$^4$G Pipeline~2 (Sheth et al.~2010). The flux was transformed into magnitudes using a zero point $z_{\rm p}=20.472$.

We increased the signal-to-noise at low surface brightness levels by smoothing the luminosity profiles. To do so, we found the height of the first profile data point for which the statistical error of the photometry was larger than 10\% of the signal. We named that height $z_{\rm c}$. Then, for every height, the photometry was obtained by averaging over the range of heights $z-p\,z/z_{\rm c}\leq z\leq z+p\,z/z_{\rm c}$, where $p=0.75\arcsec$ is the pixel size. Thus, smoothing was small close to the midplane ($z<z_{\rm c}$), but significant for low signal-to-noise regions, where $z\geq z_{\rm c}$.

\subsection{Comparison between observed and synthetic luminosity profiles}
\label{secverfit}

\begin{figure*}
\begin{center}
\begin{tabular}{c c}
\multicolumn{2}{c}{\includegraphics[width=0.9\textwidth]{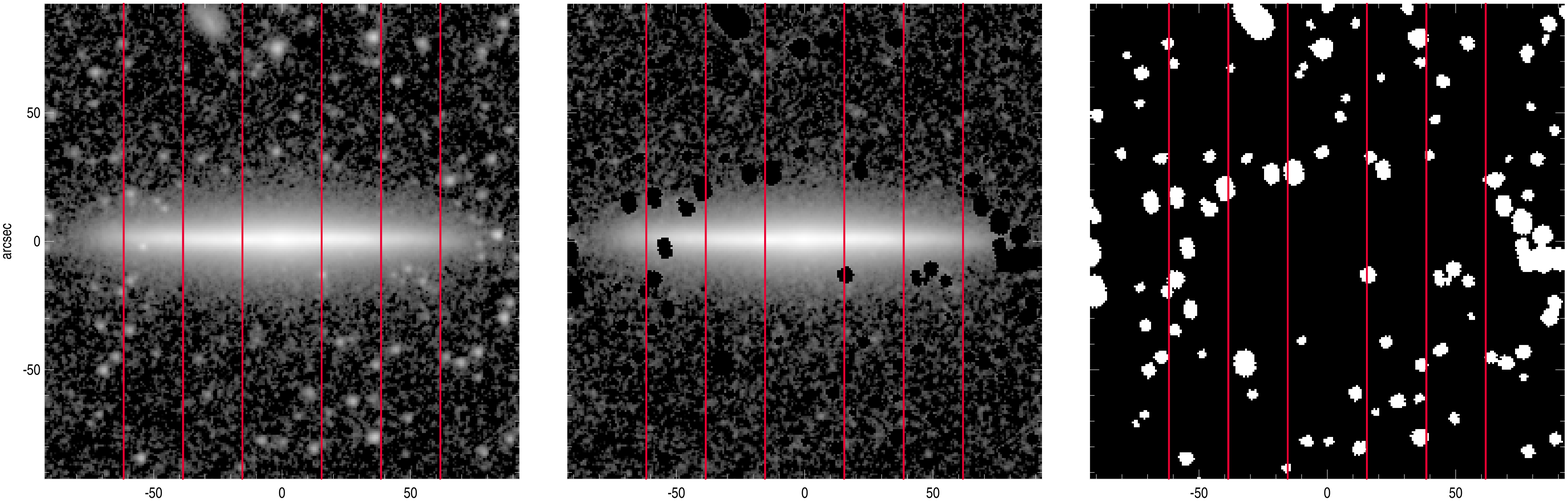}}\\
\includegraphics[width=0.45\textwidth]{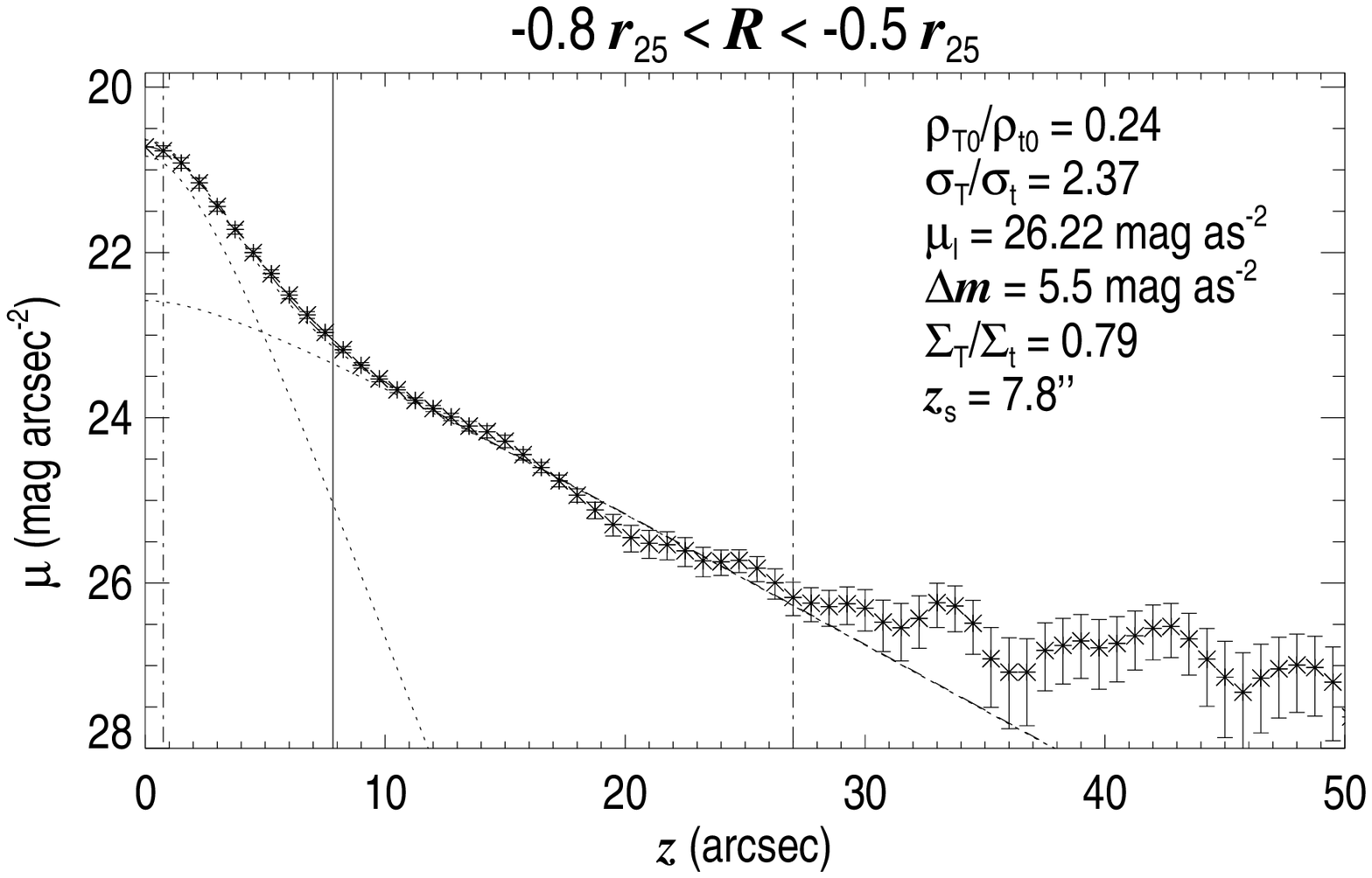}&
\includegraphics[width=0.45\textwidth]{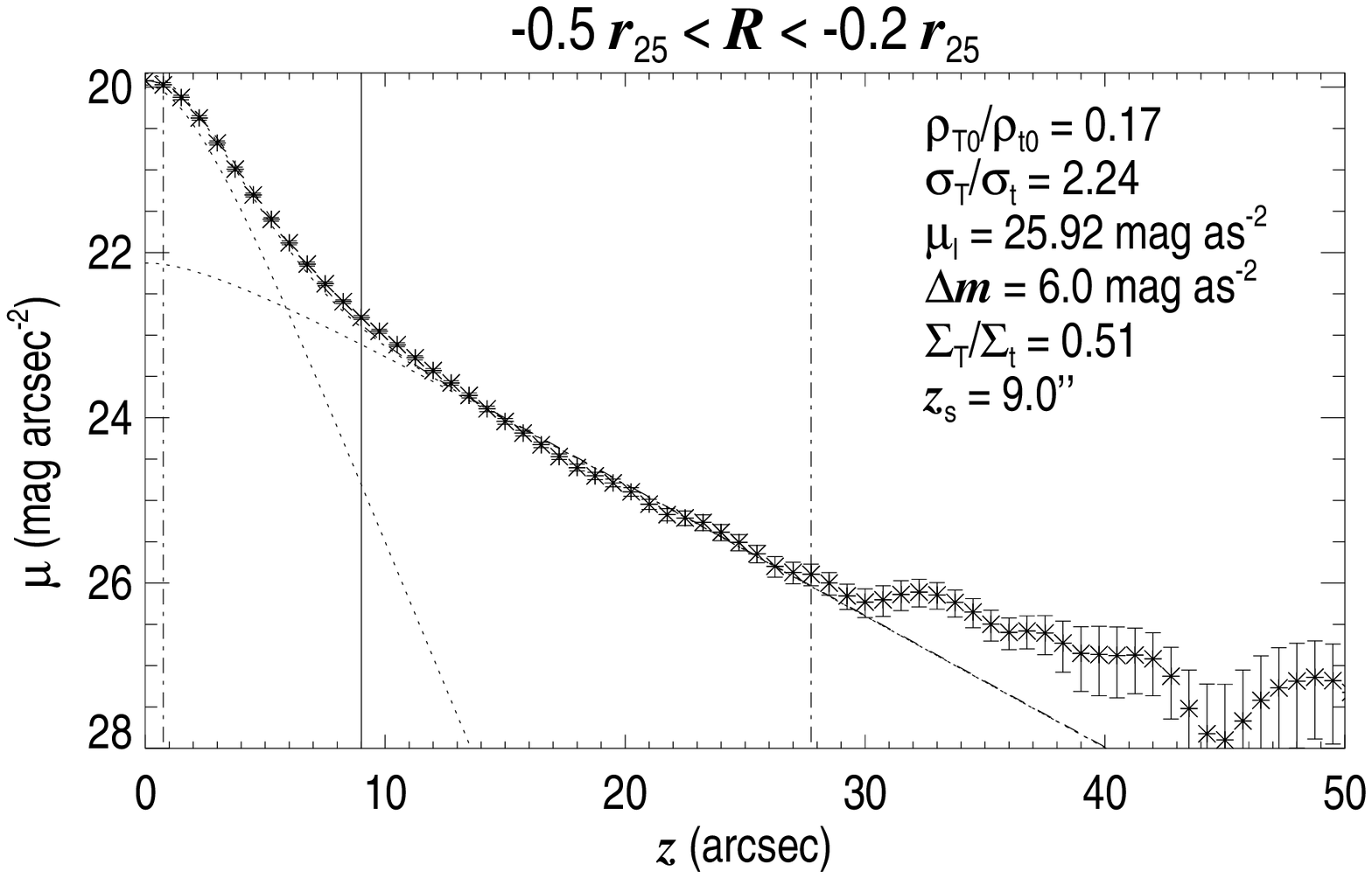}\\
\includegraphics[width=0.45\textwidth]{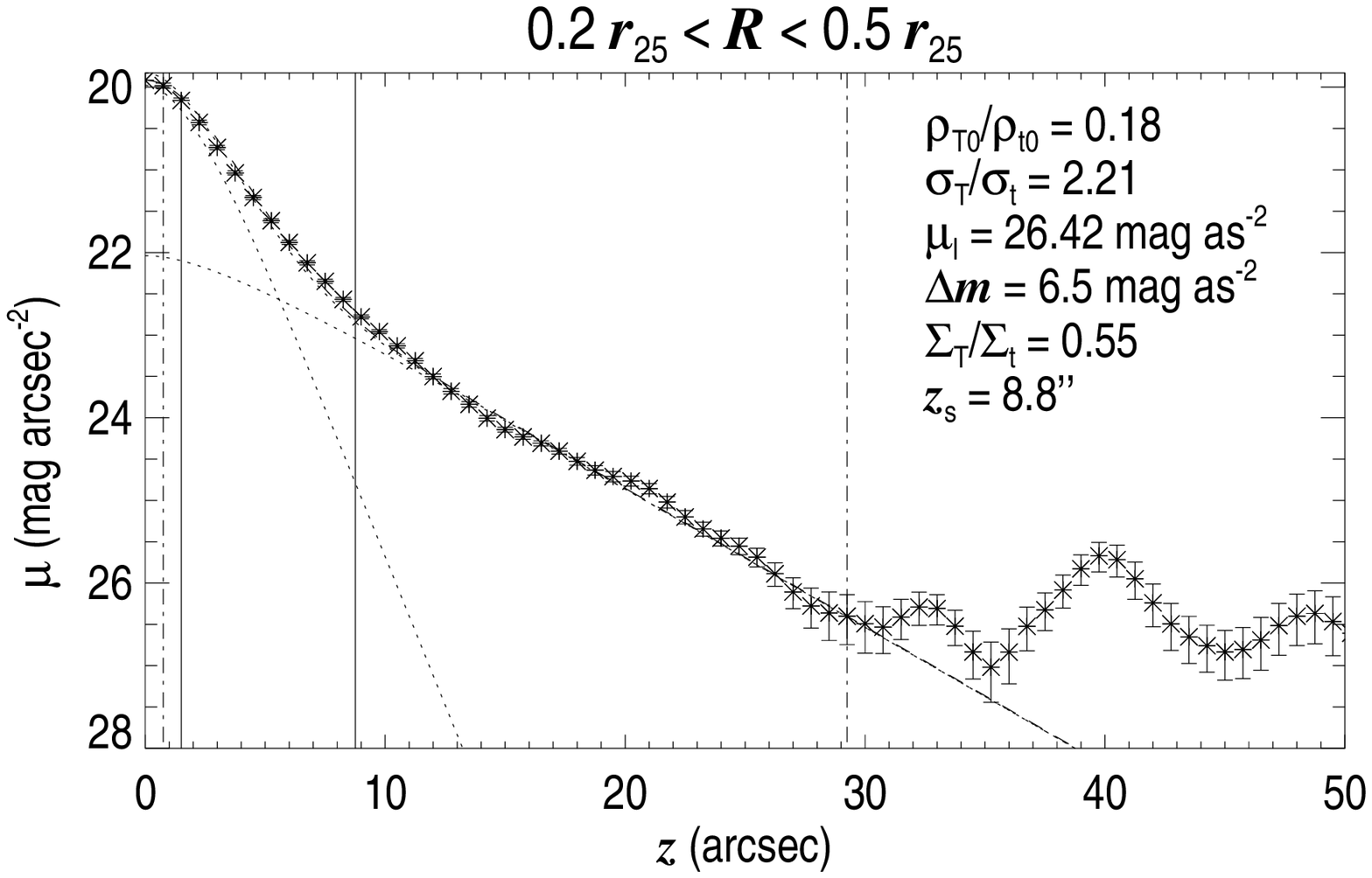}&
\includegraphics[width=0.45\textwidth]{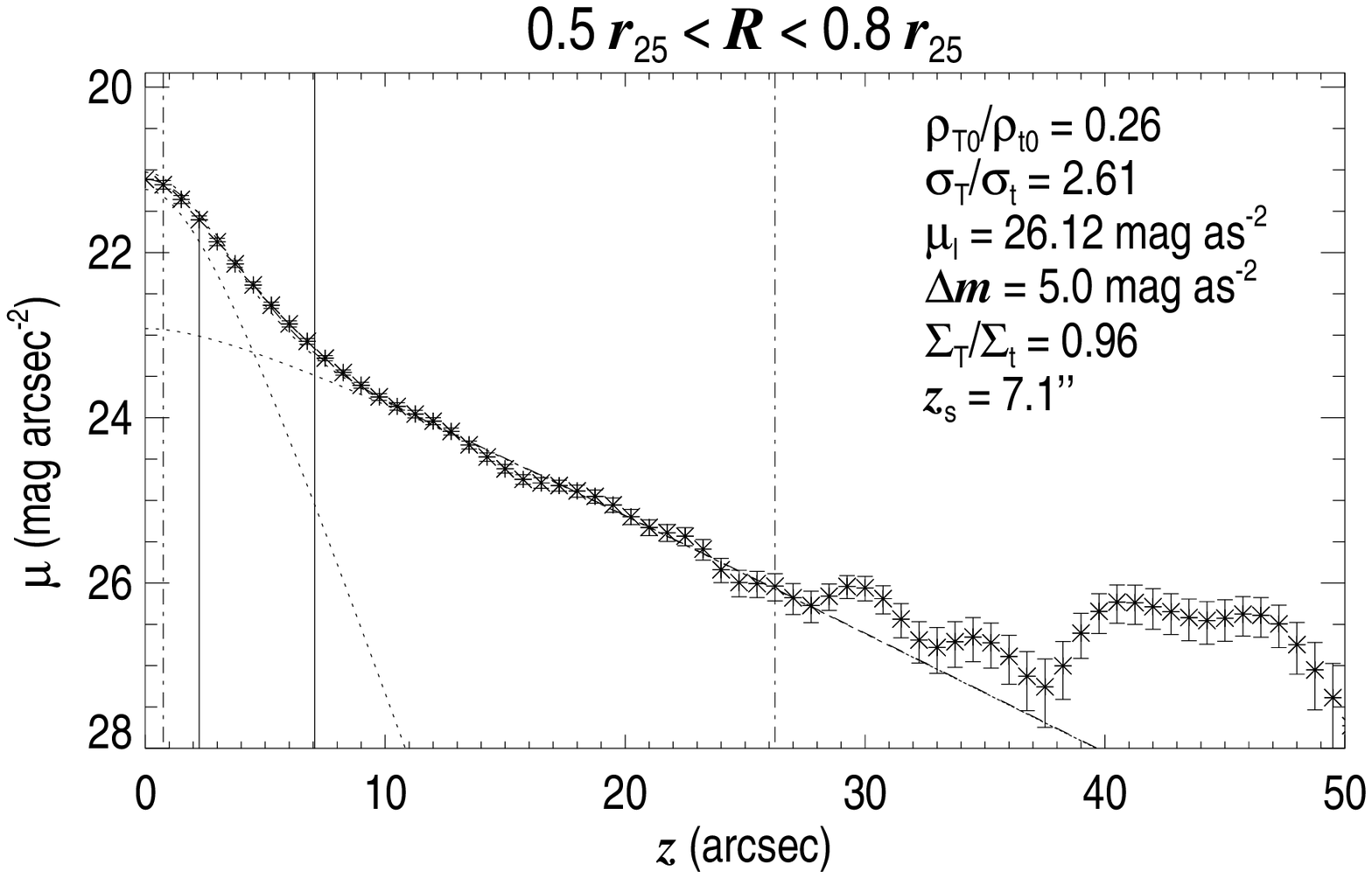}
\end{tabular}
\caption{\label{example} Vertical luminosity profile fit for NGC~5470. The three images in the top row show the average of the 3.6 and 4.5$\mu{\rm m}$-band background-subtracted S$^4$G frames (left), the same image after masking (center), and the used mask (right). The vertical red lines indicate the limits of the fitted vertical bins, the central one being ignored due to the possible presence of a bulge. The other panels show the fits to the luminosity profiles in these bins. The data points have $2\sigma$ statistical error bars, the dashed curve represents the best fit, the dotted curves indicate the contributions of the thin and the thin discs. The dash-dotted vertical lines indicate the limits of the range in vertical distance above the mid-plane used for the fit. The vertical solid line indicates $z_{\rm s}$ for each bin.}
\end{center}
\end{figure*}

Both synthetic and observed luminosity profiles were scaled in the same way before the actual fitting. In both cases the midplane surface brightness was set to be one. In CO11b we had  stretched the $z$ axis in such a way that the surface brightness at $z=200$ was equal to one tenth of the surface brightness at $z=0$, $I(z=200)=0.1I(z=0)$. Thus, the synthetic and the observed luminosity were forced to intersect at the point for which the luminosity is one tenth of the midplane one. In the present paper, we made four different vertical scalings, $I(z=200)=0.1I(z=0)$, $I(z=200)=0.2I(z=0)$, $I(z=200)=0.3I(z=0)$, and $I(z=200)=0.4I(z=0)$. For practical purposes, what this stretching does is to set the luminosity at which the synthetic and the observed luminosity profiles intersect. The synthetic fits were convolved with a Gaussian kernel with a FWHM of $2\arcsec.2$, which is the approximate FWHM of both 3.6 and 4.5$\mu{\rm m}$-band S$^4$G point-spread functions made by stacking several hundreds non-saturated 
stars (S$^4$G ``super-PSF''; Sheth et al.~2010).

The fitting of the observed luminosity profiles was done by following the procedure described in Section~3.5 and Figure~4 in CO11b. Basically, after being scaled, the observed luminosity profiles for each radial bin in each galaxy were compared to the grid of synthetic models. That was done by minimizing the differences of the brightest section of the profiles starting at a dynamic range of $\Delta m=4.5\,{\rm mag\,arcsec^{-2}}$ and then going to fainter levels in steps of $0.5\,{\rm mag\,arcsec^{-2}}$ down to $28\,{\rm mag\,arcsec^{-2}}$. We selected as the ``correct'' fit the one with the largest $\Delta m$ having a $\chi^2<0.01\,\left({\rm mag\,arcsec^{-2}}\right)^2$, where $\chi^2$ is the mean squared difference in magnitudes between both profiles. This procedure was only done down to $26\,{\rm mag\,arcsec^{-2}}$ in CO11b. The lower fitted surface brightness for a given fit is $\mu_{\rm l}$ and the midplane luminosity is $\mu_0$. Thus, for a given fit, the dynamical range $\Delta m$ over which a fit has been produced is $\Delta m=\mu_{\rm l}-\mu_0$. $\mu_{\rm l}$ is reached at a height $z_{\rm l}$ which is typically on the order of the smoothing scale height, $z_{\rm c}$.

In some cases, no $\chi^2<0.01\,\left({\rm mag\,arcsec^{-2}}\right)^2$ fit was found for a given bin and thus this bin was flagged to be non-fittable.

The code also includes a module for computing the effect of some dust absorption in the midplane as described in CO11b.

An example of a vertical luminosity profile fit is presented in Figure~\ref{example} and all the vertical luminosity profiles for our final sample galaxies appear in Figure~\ref{verticalfits} of Appendix~\ref{appendixver}. The errors in the photometric profile depict the statistical error of the averaging used at obtaining each photometric point. In spite of the care which has been taken at determining the background level, several profiles show a significant flux at large height (at a level of $\sim27-26\,{\rm mag\,arcsec^{-2}}$). There are several possible explanations for this: in some cases, the galaxies are surrounded by a crowd of globular clusters and/or foreground stars whose extended wings are hard to distinguish from thick disc light and may have not been properly masked. In the case of galaxies with a larger angular size, large-scale background variations as those described in Comer\'on et al.~(2011a) may be present and may cause artifacts in some profiles. Also, large saturated stars affect some of the profiles. Finally, some galaxies may have faint extended haloes at a brightness level comparable to our detection threshold. The restrictions applied at choosing right fits (mainly selecting as good fits those with $\chi^2<0.01\,\left({\rm mag\,arcsec^{-2}}\right)^2$), precludes this residual light from affecting our results.

\subsection{Determination of the height above which the thick disc dominates the luminosity profile}
\label{secdefrs}

A key parameter needed for producing the luminosity profiles of the thin and the thick discs parallel to midplanes is the range of heights for which each disc dominates the light emission. We defined $z_{\rm s}$ to be the height above which a fraction $f_{\rm T}=90\%$ of the light is emitted by the thick disc according to the fits to the vertical luminosity profiles. This value was calculated for each of the correctly fitted bins.

In the example fit presented in Figure~\ref{example}, the $z_{\rm s}$ for each bin is indicated by a solid vertical line in each of the middle and lower row panels.

\subsection{Selection of reliable vertical luminosity profile fits}
\label{sectrim1}

In order for a fit made for a given bin in a given galaxy to be further considered for use in this paper, several conditions were implemented:
\begin{itemize}
 \item A few galaxies, such as NGC~4013, have relatively bright components in addition to the thin and thick discs (CO11b; Comer\'on et al.~2011c). These components are likely to be a second thick disc or a bright squashed halo. In order to avoid including galaxies with wrong fits due to these components, we excluded fits for which $\mu_{\rm l}<24.5\,{\rm mag\,arcsec^{-2}}$. This threshold ensures that, for most cases, possible additional components are dim enough not to significantly affect the fit.
 \item The statistical uncertainties for fits made in very faint galaxies are large. To minimize this problem, we have only flagged as valid those fits for which the midplane surface brightness is $\mu_0<22\,{\rm mag\,arcsec^{-2}}$.
 \item In a few cases, as described in CO11b, the fit is compatible with a single disc. This is most likely to happen for the fits done over a small $\Delta m$, for which the fitted dynamical range is not large enough to allow distinguishing between both discs. We rejected fits compatible with a single disc. The criterion to detect those fits is slightly more restrictive than in CO11b, namely that the mean squared difference between the fitted solution and a single-disc profile is $\chi^2<0.5\,\left({\rm mag\,arcsec^{-2}}\right)^2$.
\end{itemize}

One of our assumptions when producing the fits is that the scale lengths of both thin and thick discs are similar and that their scale heights do not vary too much with radius. This implies that $z_{\rm s}$ should be roughly constant with the projected radius, $R$. As a final quality check for our fits, we selected those galaxies having more than one radial bin with a ``good'' fit according to the criteria expressed in the list and we looked for the largest and smallest $z_{\rm s}$ among those fits. If the ratio between these two was smaller than 1.5, the galaxy was included in the final sample. If the ratio was larger than 1.5, the galaxy was only included if the two $0.2r_{25}<|R|<0.5r_{25}$ had a ``good'' fit and the ratio between their $z_{\rm s}$ was smaller than 1.5. The fits for the bins with $0.5r_{25}<|R|<0.8r_{25}$ were then rejected and only those with $0.2r_{25}<|R|<0.5r_{25}$ were considered when measuring global properties of the galaxy (such as the average $z_{\rm s}$ and the disc relative masses, $M_{\rm T}/M_{\rm t}$). This was the case for 10 galaxies and was made so as not to exclude those galaxies with flares or truncations potentially happening at $0.5r_{25}<|R|<0.8r_{25}$ or galaxies with noisy profiles at this range of projected radii.

Galaxies with varying $z_{\rm s}$ (and thus excluded from the final sample), have not necessarily intrinsically largely varying scale lengths. For example, in the case of galaxies with a small angular size, they are likely to have a foreground star or a globular cluster covering a large fraction of the fitting bin at some height. In some cases, masking those features may be as harmful for the vertical luminosity profiles as not masking them at all, because it will cause that different heights will have their surface brightness measured in significantly different ranges of projected radii. This would result into unreliable fits which are excluded by the $z_{\rm s}$ stability criterion.

The $z_{\rm s}$ stability selection criterion is also useful at removing galaxies whose thin disc is so thin that it is barely or not resolved.

\begin{figure}
\begin{center}
\begin{tabular}{c}
{\includegraphics[width=0.45\textwidth]{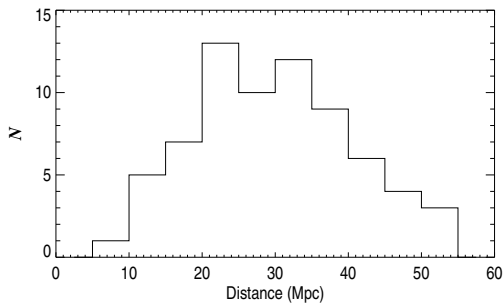}}\\
\end{tabular}
\caption{\label{distances} Histogram of the distances of the 70 galaxies in our final sample.}
\end{center}
\end{figure}

Applying all these criteria trimmed the original sample from 169 to 70 galaxies. The properties of these 70 galaxies are presented in Table~\ref{sampledata}. These restrictive criteria cause eleven out of thirty  galaxies included in the CO11b final sample not to appear in this paper. Fits in galaxies of the final sample which have not been used are labeled as ``Not used'' in Figure~\ref{verticalfits} of Appendix~\ref{appendixver}. The distance distribution of the 70 selected galaxies is presented in Figure~\ref{distances}. The median distance of the sample is 29.8\,Mpc.

For each of the 70 galaxies in our sample, as done in CO11b, the vertical luminosity profile fits were used for estimating the ratio of the stellar mass of the thick and the thin disc for each galaxy, $M_{\rm T}/M_{\rm t}$, using the following expression:
\begin{equation}
\label{eqaverage}
 \frac{M_{\rm T}}{M_{\rm t}}=\frac{\sum_{\rm b}\left(10^{-0.4\mu_{\rm 0b}}\right)\left(\Sigma_{\rm T}/\Sigma_{\rm t}\right)_{\rm b}}{\sum_{\rm b}10^{-0.4\mu_{\rm 0b}}}
\end{equation}
where $\Sigma_{\rm i}$ refers to the edge-on column mass densities, the subindex $b$ refers to the different bins in galactocentric distance for which ``good'' fits have been obtained and $\mu_{0}$ is the midplane surface brightness for a given bin in magnitudes.

\subsection{Observed luminosity profiles parallel to galaxy midplanes}
\label{secmakehor}
\begin{table}
\begin{center}
\caption{\label{zvalues} $z_{\rm s}$ and $z_{\rm u}$ values}
\begin{tabular}{l c c| l c c}
\hline
\hline
ID & $z_{\rm s}$ & $z_{\rm u}$ & ID & $z_{\rm s}$ & $z_{\rm u}$ \\
   & (\arcsec)   & (\arcsec)   &    & (\arcsec)   & (\arcsec)   \\
\hline
ESO~157-049&6.8&23.2&     NGC~4081&8.9&33.5\\
ESO~240-011&17.4&31.9&    NGC~4111&25.4&52.3\\
ESO~292-014&8.3&23.1&     NGC~4330&10.1&34.6\\
ESO~346-001&6.6&14.9&     NGC~4359&12.7&48.4\\
ESO~440-027&10.4&40.6&    NGC~4437&40.8&83.8\\
ESO~443-021&10.7&23.0&    NGC~4565&42.3&109.7\\
ESO~466-014&6.3&16.2&     NGC~4607&7.6&32.3\\
ESO~469-015&6.7&21.1&     NGC~4747&9.8&61.4\\
ESO~533-004&6.9&22.2&     NGC~5084&17.5&144.7\\
ESO~544-027&6.0&18.8&     NGC~5470&8.4&27.8\\
IC~0217&9.0&26.5&         NGC~5529&10.1&26.0\\
IC~0610&7.1&24.1&         NGC~5981&11.4&28.4\\
IC~1197&5.7&23.3&         NGC~6010&16.0&41.7\\
IC~1553&9.1&24.2&         NGC~7347&10.9&20.9\\
IC~1711&10.8&30.3&       PGC~013646&11.6&25.1\\
IC~1913&12.4&20.0&       PGC~028308&8.0&22.3\\
IC~2058&7.4&21.6&        PGC~030591&8.3&16.5\\
IC~2135&6.9&32.3&        PGC~032548&6.9&15.0\\
IC~5176&10.7&41.4&       PGC~052809&8.0&30.9\\
NGC~0489&10.8&22.5&      UGC~00903&7.8&33.3\\
NGC~0522&8.8&26.4&       UGC~01970&10.&23.1\\
NGC~0678&22.2&55.7&      UGC~05347&9.2&14.0\\
NGC~1032&28.7&75.2&      UGC~05689&7.3&21.1\\
NGC~1163&7.7&23.7&       UGC~05958&6.7&16.6\\
NGC~1422&9.7&34.0&       UGC~06526&7.7&21.0\\
NGC~1495&11.2&27.7&      UGC~07086&6.7&31.0\\
NGC~1596&21.7&81.4&      UGC~08737&6.4&27.3\\
NGC~2732&10.9&38.7&      UGC~09448&7.4&18.2\\
NGC~3098&8.6&33.1&       UGC~09665&10.3&23.5\\
NGC~3279&9.3&26.0&       UGC~10043&6.1&17.5\\
NGC~3454&10.2&29.2&      UGC~10288&8.8&32.4\\
NGC~3501&12.0&27.0&      UGC~10297&12.4&16.2\\
NGC~3592&8.8&23.5&       UGC~12518&6.8&21.2\\
NGC~3600&12.8&40.7&      UGC~12692&8.0&21.6\\
NGC~3628&20.0&140.5&     UGC~12857&13.1&26.1\\
\hline
\end{tabular}
\end{center}
\end{table}

For each galaxy in our final 70 galaxy sample, we obtained a single height above which the thick disc dominates the luminosity, the global galaxy $z_{\rm s}$, by averaging the local $z_{\rm s}$ values in all the bins with valid fits. We also calculated $z_{\rm u}$, which is the height at which the $26\,{\rm mag\,arcsec^{-2}}$ level was found averaged over the valid fits. The $z_{\rm s}$ and $z_{\rm u}$ values appear listed in Table~\ref{zvalues}.

The region dominated by the thin disc was defined to be that between $z=0$ and $z=0.5z_{\rm s}$ at all projected radii, $R$ in accordance with the assumption of roughly constant scale heights for all discs. The region dominated by the thick disc was defined to be that between $z=z_{\rm s}$ and $z=z_{\rm u}$ at all projected radii. We averaged the galaxy over its four quarters (top-right, top-left, bottom-left, and bottom-right) taking into account the masking and using for the PA of the disc major axis the value appearing in HyperLEDA, except for a few cases for which that value was obviously off by a few degrees, where we used our own value for the disc PA computed using ellipse fitting.

We used the average of the four quarters of the galaxy for obtaining a luminosity profile parallel to the midplane for both the thin (thin horizontal profile) and the thick disc (thick horizontal profile) by averaging in $z$ over the range of heights they dominate. We produced a third luminosity profile including the light of both the thin and the thick disc by averaging the light from $z=0$ to $z=z_{\rm u}$ (total horizontal profile).

In order to increase the signal-to-noise in the outer parts of the galaxy we used a logarithmic sampling; each data point was measured at a radial distance 1.03 times larger than the previous one. For each of the three horizontal profiles we calculated the projected radius at which the $27\,{\rm mag\,arcsec^{-2}}$ level is found. We defined $R_{\rm f}$ to be the largest of these projected radii and we cut the three profiles down to $R_{\rm f}$.

The horizontal luminosity profiles are displayed in the bottom panels of Figure~\ref{verticalfits} in Appendix~\ref{appendixver}. 

\subsection{Fitting of the horizontal luminosity profiles}
\label{secfithor}

\begin{figure}
\begin{center}
\begin{tabular}{c}
{\includegraphics[width=0.45\textwidth]{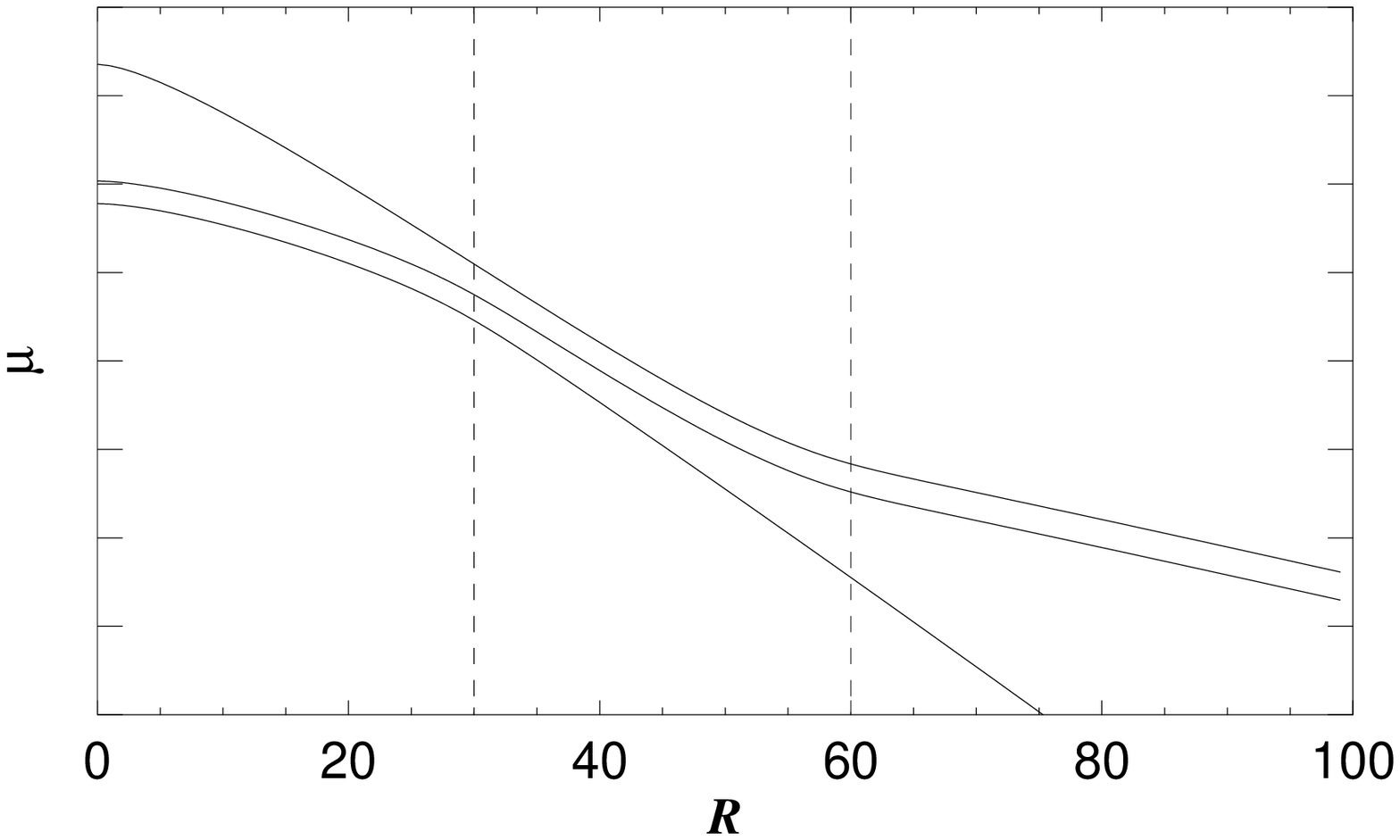}}\\
\end{tabular}
\caption{\label{exampleerwin} Examples of a profile with an antitruncation (top; Type~III profile), with a truncation and antitruncation (middle; Type~II+III profile), and a truncation (bottom; Type~II profile). All profiles have been obtained using Equation~(\ref{integral}). For the top profile the antitruncation radius is $r_{1,2}=60$ and the scale lengths are $h_{1}=10$ and $h_{2}=30$. For the middle profile the break radii are $r_{1,2}=30$ and $r_{2,3}=60$ and the scale lengths are $h_{1}=20$, $h_{1}=10$, and $h_{3}=30$. For the bottom profile the truncation radius is $r_{1,2}=30$ and the scale lengths are $h_{1}=20$ and $h_{2}=10$. The dotted vertical lines indicate the break radii. $I_{0}$ values have been set arbitrarily.}
\end{center}
\end{figure}

Erwin et al.~(2008) described breaks in discs by using what they termed the ``broken-exponential'' function which consists of two exponential pieces joined by a transition of variable ``sharpness'' (their Equation~5). We generalized this function to allow us to describe profiles with more than one break (both truncations and antitruncations):
\begin{equation}
\label{eqerwin}
  I(r)=S\,I_{0}\,{\rm e}^{-\frac{r}{h_1}}\prod_{i=2}^{i=n}\left\{\left[1+{\rm e}^{\alpha_{\rm i-1,i}\left(r-r_{\rm i-1,i}\right)} \right]^{\frac{1}{\alpha_{\rm i-1,i}}\left(\frac{1}{h_{\rm i-1}}-\frac{1}{h_{\rm i}}\right)}\right\}
\end{equation}
where $S$ is a scaling factor defined as
\begin{equation}
 S^{-1}=\prod_{i=2}^{i=n}\left\{\left[1+{\rm e}^{-\alpha_{\rm i-1,i}\,r_{\rm i-1,i}} \right]^{\frac{1}{\alpha_{\rm i-1,i}}\left(\frac{1}{h_{\rm i-1}}-\frac{1}{h_{\rm i}}\right)}\right\}.
\end{equation}
$I_0$ stands for the central brightness of the inner exponential section, $r$ for the radial distance, $h_{\rm i}$ for the scale lengths of the different sections, $n$ is the number of exponential sections in the profile, $r_{\rm i-1,i}$ stands for the break radius between the section with a $h_{\rm i-1}$ slope and that with a $h_{\rm i}$ slope, and $\alpha_{\rm i-1,i}$ is a parameter which controls the sharpness of the breaks.

For profiles with no breaks we used a simple exponential:
\begin{equation}
\label{eqsimple}
  I(r)=I_{0}\,{\rm e}^{-\frac{r}{h_1}}.
\end{equation}

Equations~(\ref{eqerwin}) and (\ref{eqsimple}) cannot be immediately applied for the fitting of horizontal luminosity profiles because they need to be integrated along the line of sight. We did so assuming that the galaxy had a sharp cut-off at $r_{\rm f}=5\,R_{\rm f}$, although the results do not depend on this cut-off radius as long it is large enough ($r_{\rm f}\gtrapprox1.5\,R_{\rm f}$). So, if we define $s$ to be $s\equiv \sqrt{r^2-R^2}$, then the integral along the line of sight is:
\begin{equation}
\label{integral}
 J(R)=2\int_{s=0}^{s=s_{\rm f}}I\left(\sqrt{R^2+s^2}\right){\rm d}s
\end{equation}
where $s_{\rm f}=\sqrt{R_{\rm f}^2-R^2}$.

Since the result of Equation~(\ref{integral}) yields non-analytic results, we fitted our luminosity profiles data in magnitudes with $-2.5{\rm log}\left(J(R)\right)$ using {\sc idl}'s {\sc curvefit} function. For each fit, the number of truncations, $n$, was manually set after observing the luminosity profiles. The initial values for the scale lengths were those obtained from an exponential fit to each of the exponential sections. We found that, especially for profiles with more than two breaks, the number of fitting parameters was too high to yield reliable results, so we set all $\alpha_{\rm i-1,i}=0.5$, which is typical of break sharpnesses found in Erwin et al.~(2008).

The fitting range was set manually by defining the limits that we called $R_{\rm min}$ and $R_{\rm max}$. Special care was taken in order to exclude regions strongly affected by bulges and noisy outskirts regions from the fitting range.

Examples of Equation~(\ref{eqerwin}) integrated over the line of sight are presented in Figure~\ref{exampleerwin}. The top and the bottom profile represent an antitruncated and a truncated profile respectively. The middle profile combines the breaks of the two other profiles. Due to line of sight integration, the profiles do not have a central peak and have a rounded profile for $R\ll h_{1}$.

The fits to the horizontal luminosity profiles are overlayed to the profiles in Figure~\ref{verticalfits} in Appendix~\ref{appendixver} (black lines). The fitted parameters for total, thin and thick disc profiles are presented in Tables~\ref{bigtable1}, \ref{bigtable2}, and \ref{bigtable3} respectively.

\section{What if fits were done in galaxies not fulfilling our assumptions?}
\label{seccaveat}

Throughout the fitting process several assumptions were made. Deviations from these perfect conditions should be tested in order to know whether our results are accurate.

Two parameters from our fits are especially important for our conclusions (and those in CO11b): the ratio of the stellar masses of the thick and the thin disc, $M_{\rm T}/M_{\rm t}$, and $z_{\rm s}$ which needs to be precisely measured so as to obtain a correct separation of the light of the thin and the thick disc. $M_{\rm T}/M_{\rm t}$ was measured by averaging the ratio of the thick and thin disc edge-on column mass densities, $\Sigma_{\rm T}/\Sigma_{\rm t}$, over all the bins with ``good'' vertical luminosity fits (Equation~(\ref{eqaverage})).

We thus have to test the reliability of the fitted values of $z_{\rm s}$ and $\Sigma_{\rm T}/\Sigma_{\rm t}$ if we loosen some of our assumptions:
\begin{itemize}
 \item What happens if the scale lengths of the thin and the thick discs are not similar?
 \item What happens if the scale lengths of the discs are not constant and a disc has breaks?
 \item What happens if the disc is sub-maximal and thus dominated by dark matter within the optical radius?
 \item What happens if the disc of a galaxy is not as edge-on as we think it is?
 \item We computed our results for the case $\Upsilon_{\rm T}/\Upsilon_{\rm t}=1.2$. We know from CO11b that $\Sigma_{\rm T}/\Sigma_{\rm t}$ roughly scales with $\Upsilon_{\rm T}/\Upsilon_{\rm t}$ for reasonable star formation histories, but would $z_{\rm s}$ remain unchanged if we had selected the wrong $\Upsilon_{\rm T}/\Upsilon_{\rm t}$?
 \item We assumed the gas column mass density to scale with that of the thin disc, thus ensuring a constant scale heights for all discs. However, Bigiel et al.~(2008) have showed that in spiral galaxies molecular gas tends to concentrate in the central parts of the galaxy. Their data also show that the atomic gas column mass density within $r_{25}$ varies little with radius and that it dominates the gas column density for $r>0.5r_{25}$. What would happen if indeed the gas distribution was significantly different from the one assumed? 
\end{itemize}

\subsection{A galaxy model in order to test deviations from our assumptions}

A set of model galaxies was created in order to test the result of loosening our assumptions when producing the fits to vertical luminosity profiles. The model galaxies were considered to have $r_{25}=9.2$\,kpc, and an inner thin disc scale length $h_{\rm t1}=0.3r_{25}$, which is the median values in our final 70 galaxy sample. In case of truncated thin discs, the scale length of the outer section has been set to be $h_{\rm t2}=0.12r_{25}$ which again is typical of what is observed in our galaxies. We designed all model galaxies to have a stellar face-on column mass density equal to $S=S_{\rm t}+S_{\rm T}=60M_{\sun}\,{\rm pc^{-2}}$ and a thin disc vertical velocity dispersion $\sigma_{\rm t}=20\,{\rm km\,s^{-1}}$ at $r=0.65r_{25}$, which is on the order of what is found in the solar neighborhood. The absolute value of the column mass density and the velocity dispersion are only relevant when considering the biases due to the gas distribution and the dark matter halo, which as seen later are small compared to those caused by the inclination angle, the thick disc relative scale length, and the thick disc column mass density. The truncations in discs were described using Equation~(\ref{eqerwin}). 

In the literature, for a galaxy with a single disc in equilibrium, discs are ensured to have a roughly constant scale height by setting them to be $z_{\rm t}=\sigma^2_{\rm t}/(\pi {\rm G} S)={\rm constant}$ at all radius (see, e.g., van der Kruit \& Searle 1981a). The ratio of the squared velocity dispersion and the surface mass density does not directly control the disc scale height in a galaxy with two discs, however setting it to be constant ($\sigma^2_{\rm i}/(\pi {\rm G} S)={\rm constant}$) still ensures scale heights not to vary too much with $r$ in the absence of perturbing effects such as those of a dark matter halo. In this context $S$ is the total face-on column mass denisty, also accounting for gas. Thus, we implemented this condition in our models. The models were computed for $r<2.0r_{25}$, but the exact radius at which the computation ends is not important.

Additionally, the computed grid of models has the following properties, which cover the range of possibilities found in our fits to observed galaxies:
\begin{itemize}
 \item Two very different star formation histories for the thick disc: $\Upsilon_{\rm T}/\Upsilon_{\rm t}=1.2$ and $\Upsilon_{\rm T}/\Upsilon_{\rm t}=2.4$.
 \item Three face-on column mass density ratios at $r=0$ yielding $\rho_{\rm T0}/\rho_{\rm t0}$ values compatible with what is observed in our vertical fits: $S_{\rm T}/S_{\rm t}=1/15$, $S_{\rm T}/S_{\rm t}=1/10$, and $S_{\rm T}/S_{\rm t}=1/5$.
 \item Four truncation radii for the thin disc: $r_{\rm t1,2}=0.3r_{25}$, $r_{\rm t1,2}=0.6r_{25}$, $r_{\rm t1,2}=0.9r_{25}$, and $r_{\rm t1,2}=\infty$. We have not considered truncations in thick discs. As shown in later sections, they are usually found to happen at very low surface brightness levels and thus, not likely to affect much the fits.
 \item Two vertical velocity dispersion ratios typical of what we find in our vertical fits: $\sigma_{\rm T}/\sigma_{\rm t}=2.0$ and $\sigma_{\rm T}/\sigma_{\rm t}=2.3$.
 \item Two thick disc scale lengths typical of what we find in our vertical fits: $h_{\rm T}=0.3r_{25}$ and $h_{\rm T}=0.6r_{25}$.
 \item Models with no dark matter halo and models with a pseudo-isothermal dark matter halo (van Albada et al.~1985) with a core radius $r_{\rm c}=3.52$\,kpc and a maximum circular velocity $v_{\rm c}=110{\rm\,km\,s^{-1}}$. This core radius is slightly larger than the inner thin disc scale length $r_{\rm c}=1.28h_{\rm t1}$. The selected circular velocity corresponds to the median velocity in our 70 galaxy sample. The formalism used for describing the dark matter halo (${\rm d}K_{\rm DM}/{\rm d}z$) can be found in Narayan \& Jog (2002) and CO11b.
 \item Two gas distributions: one case with no gas at all and one case in which the face-on gas column mass density is constant with radius and equal to $S_{\rm g}=12M_{\sun}\,{\rm pc^{-2}}$ and with a vertical velocity dispersion $\sigma_{\rm g}=1/3\sigma_{\rm t}$. This second gas distribution is qualitatively similar to that observed in many galaxies in the Bigiel et al.~(2008) sample, but has a higher surface mass density in order to study what extreme gas mass distributions may do to our fits.
 \item Six galaxy inclinations: $i=90\deg$, $i=89\deg$, $i=88\deg$, $i=87\deg$, $i=86\deg$, and $i=85\deg$.
\end{itemize}

Model fits were produced, like in our observed galaxies, for bins $0.2r_{25}<|R|<0.5r_{25}$ and $0.5r_{25}<|R|<0.8r_{25}$ with six different dynamical ranges which cover those found in observed galaxy fits: $\Delta m=4.5$, $\Delta m=5.0$, $\Delta m=5.5$, $\Delta m=6.0$, $\Delta m=6.5$, and $\Delta m=7.0$. In all cases, we fitted the models with the function of two stellar discs and a gas disc, the later having 20\% of the column mass density of the thin disc.

The total number of fits was 27648. We excluded from our analysis 162 wrong fits ($\chi^2>0.01\,\left({\rm mag\,arcsec^{-2}}\right)^2$), 10562 fits compatible with a single-disc distribution according to the criterion presented in Section~\ref{sectrim1}, and 1537 fits for which $(\Sigma_{\rm T}/\Sigma_{\rm t})_{\rm f}>2.1$ because such massive thick discs have not been found in our observed galaxies. The remaining number of fits is 15387.

\begin{figure}
\begin{center}
\begin{tabular}{c}
{\includegraphics[width=0.45\textwidth]{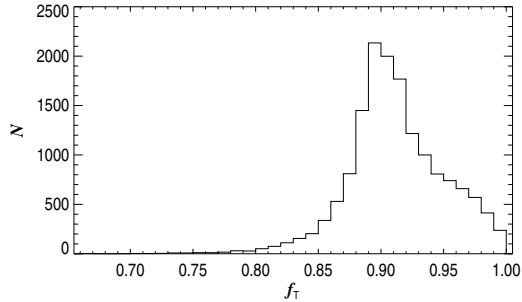}}\\
\end{tabular}
\caption{\label{histoofractions} Histogram representing the fraction of light emitted by the thick disc for heights above the fitted $z_{\rm s}$ ($f_{\rm T}$) in the set of 15387 model galaxies considered for study (see text).}
\end{center}
\end{figure}

\begin{figure*}
\begin{center}
\begin{tabular}{c}
{\includegraphics[width=0.90\textwidth]{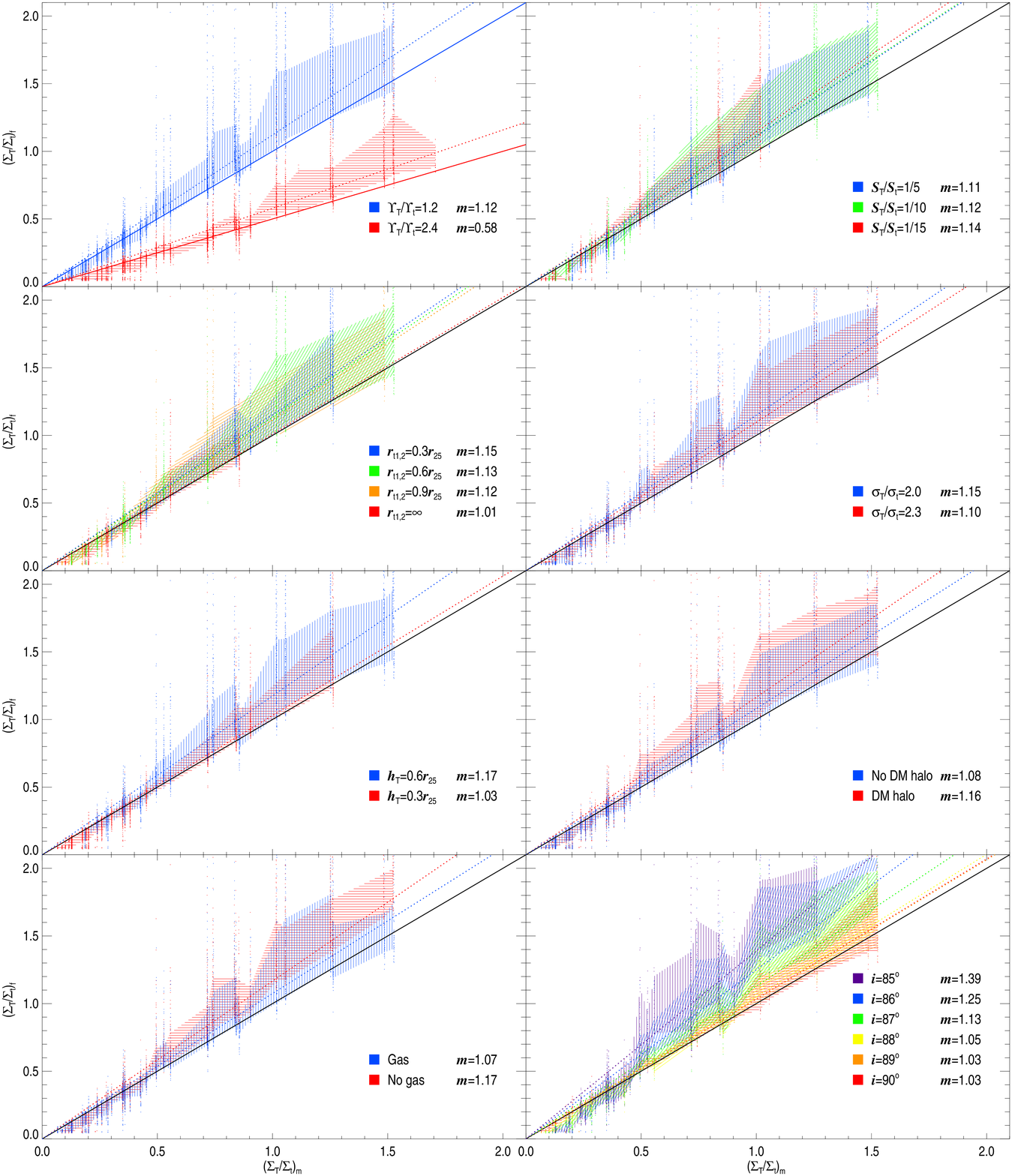}}\\
\end{tabular}
\caption{\label{histoofratios} Comparison of the ratio of the thin and thick disc edge-on column mass densities in 15387 galaxy models with varying properties (see text for the selection criteria of the models). $(\Sigma_{\rm T}/\Sigma{\rm t})_{\rm f}$ is the fitted edge-on column density ratio and $(\Sigma_{\rm T}/\Sigma_{\rm t})_{\rm m}$ the real ratio in the galaxy model. In the first panel, both $\Upsilon_{\rm T}/\Upsilon_{\rm t}=1.2$ and $\Upsilon_{\rm T}/\Upsilon_{\rm t}=2.4$ data are plotted. In the other panels, only the $\Upsilon_{\rm T}/\Upsilon_{\rm t}=1.2$ data points are presented. Solid lines show $(\Sigma_{\rm T}/\Sigma{\rm t})_{\rm f}=(\Sigma_{\rm T}/\Sigma{\rm t})_{\rm m}$, except for the red line in the first panel, which shows $(\Sigma_{\rm T}/\Sigma{\rm t})_{\rm f}=0.5(\Sigma_{\rm T}/\Sigma{\rm t})_{\rm m}$ as it would correspond if the light of the thin and the thick disc were correctly attributed by fits in the case $\Upsilon_{\rm T}/\Upsilon_{\rm t}=2.4$. The dashed lines show linear regressions -- forced to cross the origin -- to models color-coded according to the properties described in the bottom-right corner of each panel. $m$ denotes the slope of those fits. The hatched areas indicate the region which encloses 68.2\% of the data points around the local $(\Sigma_{\rm T}/\Sigma_{\rm t})_{\rm f}$ median value.}
\end{center}
\end{figure*}

\subsection{Effect of loosening our assumptions into the fitted $z_{\rm s}$}

\subsubsection{Reliability of the fitted $z_{\rm s}$}

Because the goal of this paper is to study thin and thick disc horizontal luminosity profiles, it is important to know whether most light above a fitted $z_{\rm s}$ is emitted by the thick disc. This issue is addressed in Figure~\ref{histoofractions}, in which it is shown that the fraction of light above the fitted $z_{\rm s}$ peaks around $f_{\rm T}=0.9$, with $f_{\rm T}>0.8$ for over 99\% of the fits. Only one valid model has $f_{\rm T}<0.7$. It thus seems reasonable to say that the fitted $z_{\rm s}$ is good enough to define the limit of the region dominated by the thick disc.

\subsubsection{$z_{\rm s}$ variations with radius}

When selecting ``valid'' vertical profile fits, we assumed that $z_{\rm s}$ should vary less than a factor 1.5 within the fitted vertical bins. Thanks to our modeling we can test the stability of $z_{\rm s}$ with varying projected radii. For every modeled galaxy, we compared $z_{\rm s}$ in the $0.2r_{25}<|R|<0.5r_{25}$ and $0.5r_{25}<|R|<0.8r_{25}$ bins. We found that if each bin is fitted down to the same $\mu_{\rm l}$, $z_{\rm s}\left(0.2r_{25}<|R|<0.5r_{25}\right)=1.26\pm0.29z_{\rm s}\left(0.5r_{25}<|R|<0.8r_{25}\right)$. This is natural because in most of our models, the thick disc increases its relative mass fraction with increasing radii. In 8\% of our models $z_{\rm s}\left(0.2r_{25}<|R|<0.5r_{25}\right)>1.5z_{\rm s}\left(0.5r_{25}<|R|<0.8r_{25}\right)$, which would lead the outer bins to be ignored when calculating $z_{\rm s}$.

\subsection{Effect of loosening our assumptions into the fitted $\Sigma_{\rm T}/\Sigma_{\rm t}$}

In order to check the effect of deviations from the assumptions in our vertical luminosity profile fits, we have produced the plots shown in Figure~\ref{histoofratios}. In them we compare the ratio of the fitted ratio of thick to thin column mass densities $(\Sigma_{\rm T}/\Sigma_{\rm t})_{\rm f}$ with that in the original model $(\Sigma_{\rm T}/\Sigma_{\rm t})_{\rm m}$. The displayed distributions have a significant scatter, which means that some galaxies may have their thick disc column mass density under or overestimated. The purpose of this subsection is to clarify the origin of those biases. In each panel of Figure~\ref{histoofratios}, we have divided the fits into several color-coded bins, for which we have produced a linear regression which crossed the origin. In the case $\Upsilon_{\rm T}/\Upsilon_{\rm t}=1.2$, the more accurate the fitted $(\Sigma_{\rm T}/\Sigma_{\rm t})_{\rm f}$, the closer the slope of the linear regression, $m$, will be to one. Solid lines denote the position that points associated to a perfect fit should have in the plot. The data points tend to appear above the solid lines, which indicates that our thick disc relative masses tend to be overestimated by a factor of $\sim10\%$ on average as described in the next subsections.

\subsubsection{$\Sigma_{\rm T}/\Sigma_{\rm t}$ roughly scales with $\Upsilon_{\rm T}/\Upsilon_{\rm t}$}

The fitted edge-on thick to thin column mass ratios -- $(\Sigma_{\rm T}/\Sigma_{\rm t})_{\rm f}$ -- are in general very similar but slightly larger than that in the original model galaxy -- $(\Sigma_{\rm T}/\Sigma_{\rm t})_{\rm m}$ when $\Upsilon_{\rm T}/\Upsilon_{\rm t}=1.2$ ($m=1.12$; top-left panel in Figure~\ref{histoofratios}). When $\Upsilon_{\rm T}/\Upsilon_{\rm t}=2.4$, the slope of the linear regression is $m=0.58$, which is not very far from $m=0.5$ and not far from being exactly half of the slope obtained for $\Upsilon_{\rm T}/\Upsilon_{\rm t}=1.2$. This indicates that the fraction of light assigned to the thin and the thick discs does not vary too much for a reasonable range of thick disc star formation histories and that $\Sigma_{\rm T}/\Sigma_{\rm t}$ roughly scales with $\Upsilon_{\rm T}/\Upsilon_{\rm t}$ as already pointed out in CO11b. Because of this close to linear behavior, it is easy to convert results obtained using a given $\Upsilon_{\rm T}/\Upsilon_{\rm t}$ to another $\Upsilon_{\
rm T}/\Upsilon_{\rm t}$. As a consequence we will continue our analysis focusing on the $\Upsilon_{\rm T}/\Upsilon_{\rm t}=1.2$ case.

\subsubsection{Biases of the fitted $\Sigma_{\rm T}/\Sigma_{\rm t}$}
\label{secnogas}

According to the linear fits made in Figure~\ref{histoofratios} the three main reasons for our fitted $\Sigma_{\rm T}/\Sigma_{\rm t}$ to be overestimated are:

\begin{itemize}
 \item{\emph{Inclination angles far from edge-on:} Galaxies with $i<87\deg$ have their $\Sigma_{\rm T}/\Sigma_{\rm t}$ overestimated on average by over a 25\%. However, our selection criteria should prevent many of these galaxies to be included in our sample.}
 \item{\emph{The thick disc scale length being significantly longer than the thin disc scale length:} if both scale lengths are similar, then the the slope of the linear regression between the modeled and the fitted $\Sigma_{\rm T}/\Sigma_{\rm t}$ is $m\sim1$.}
 \item{\emph{The thin disc being truncated:} The smaller the truncation radius, the larger the risk of $\Sigma_{\rm T}/\Sigma_{\rm t}$ being overestimated.}
\end{itemize}

Other parameters which, to a lesser extent, contribute to overestimate the fitted $\Sigma_{\rm T}/\Sigma_{\rm t}$ are:

\begin{itemize}
 \item{\emph{A gas-depleted disc being fitted with a function including a significant gaseous disc:} This justifies fitting the vertical profiles of early-type galaxies with functions which does not include a gas disc.}
 \item{\emph{The disc being submaximal.}}
 \item{\emph{Low $\sigma_{\rm T}/\sigma_{\rm t}$ values.}}
\end{itemize}

Additionally, we noticed that in galaxies with the most dominant fitted thick discs the relative thick to thin disc masses are more easily overestimated. In Figure~\ref{histoofratios}, data points appearing in the left side of a panel have a lower chance to be found   above the $(\Sigma_{\rm T}/\Sigma{\rm t})_{\rm f}=(\Sigma_{\rm T}/\Sigma{\rm t})_{\rm m}$ line than those in the right side of the panel. This is also seen in the data presented in Table~\ref{over}. Yoachim \& Dalcanton (2006) and CO11b describe how the ratio of the stellar mass of the thick and the thin disc, $M_{\rm T}/M_{\rm t}$, decreases with increasing galaxy mass. One could interpret that the slope of this relationship has been overestimated because of the overestimate of thick disc masses in thick-disc dominated galaxies. However, this effect is likely to be compensated by the fact that $\Upsilon_{\rm T}/\Upsilon_{\rm t}$ is probably larger in smaller galaxies, whose thin discs host in the local Universe a larger relative star formation, which lowers $\Upsilon_{\rm t}$. The exact quantification of $\Upsilon_{\rm T}/\Upsilon_{\rm t}$ is under study and will be published in a follow-up paper (Comer\'on et al.~2013, in preparation).

As a conclusion, in most of the cases, the overestimation of the thick disc is relatively small (generally less than a 20\%). This could be compensated by the fact that, among reasonable star formation histories for thin and thick discs, we have selected a mass-to-light ratio that yields the smaller thick disc relative mass ($\Upsilon_{\rm T}/\Upsilon_{\rm t}=1.2$).

\begin{table}
\begin{center}
\caption{\label{over} $\Sigma_{\rm T}/\Sigma_{\rm t}$ overestimate as a function of $\Sigma_{\rm T}/\Sigma_{\rm t}$ values in models with $\Upsilon_{\rm T}/\Upsilon_{\rm t}=1.2$.}
\begin{tabular}{c | c}
\hline
\hline
\multicolumn{2}{c}{As a function of the fitted column mass ratio}\\
\hline
$(\Sigma_{\rm T}/\Sigma_{\rm t})_{\rm f}$  & $(\Sigma_{\rm T}/\Sigma_{\rm t})_{\rm f}/(\Sigma_{\rm T}/\Sigma_{\rm t})_{\rm m}$\\
&(Overestimation factor)\\
\hline
$0.0<(\Sigma_{\rm T}/\Sigma_{\rm t})_{\rm f}\leq0.5$ & $0.90\pm0.20$ \\
$0.5<(\Sigma_{\rm T}/\Sigma_{\rm t})_{\rm f}\leq1.0$ & $1.10\pm0.15$ \\
$1.0<(\Sigma_{\rm T}/\Sigma_{\rm t})_{\rm f}\leq1.5$ & $1.24\pm0.27$ \\
$1.5<(\Sigma_{\rm T}/\Sigma_{\rm t})_{\rm f}\leq2.1$ & $1.44\pm0.36$ \\
\hline
\hline
\multicolumn{2}{c}{As a function of the model column mass ratio}\\
\hline
$(\Sigma_{\rm T}/\Sigma_{\rm t})_{\rm m}$  & $(\Sigma_{\rm T}/\Sigma_{\rm t})_{\rm f}/(\Sigma_{\rm T}/\Sigma_{\rm t})_{\rm m}$\\
&(Overestimation factor)\\
\hline
$0.0<(\Sigma_{\rm T}/\Sigma_{\rm t})_{\rm m}\leq0.5$ & $0.93\pm0.23$ \\
$0.5<(\Sigma_{\rm T}/\Sigma_{\rm t})_{\rm m}\leq1.0$ & $1.19\pm0.30$ \\
$1.0<(\Sigma_{\rm T}/\Sigma_{\rm t})_{\rm m}\leq1.5$ & $1.36\pm0.49$ \\
$1.5<(\Sigma_{\rm T}/\Sigma_{\rm t})_{\rm m}\leq2.1$ & $1.52\pm0.69$ \\
\hline

\end{tabular}
\end{center}
Note.~-- The format of the data is average$\pm1\sigma$.
\end{table}

\section{Results}

\label{secresults}

\subsection{Thick disc relative masses}
\label{secmasses}

\begin{figure}
\begin{center}
\begin{tabular}{c}
{\includegraphics[width=0.45\textwidth]{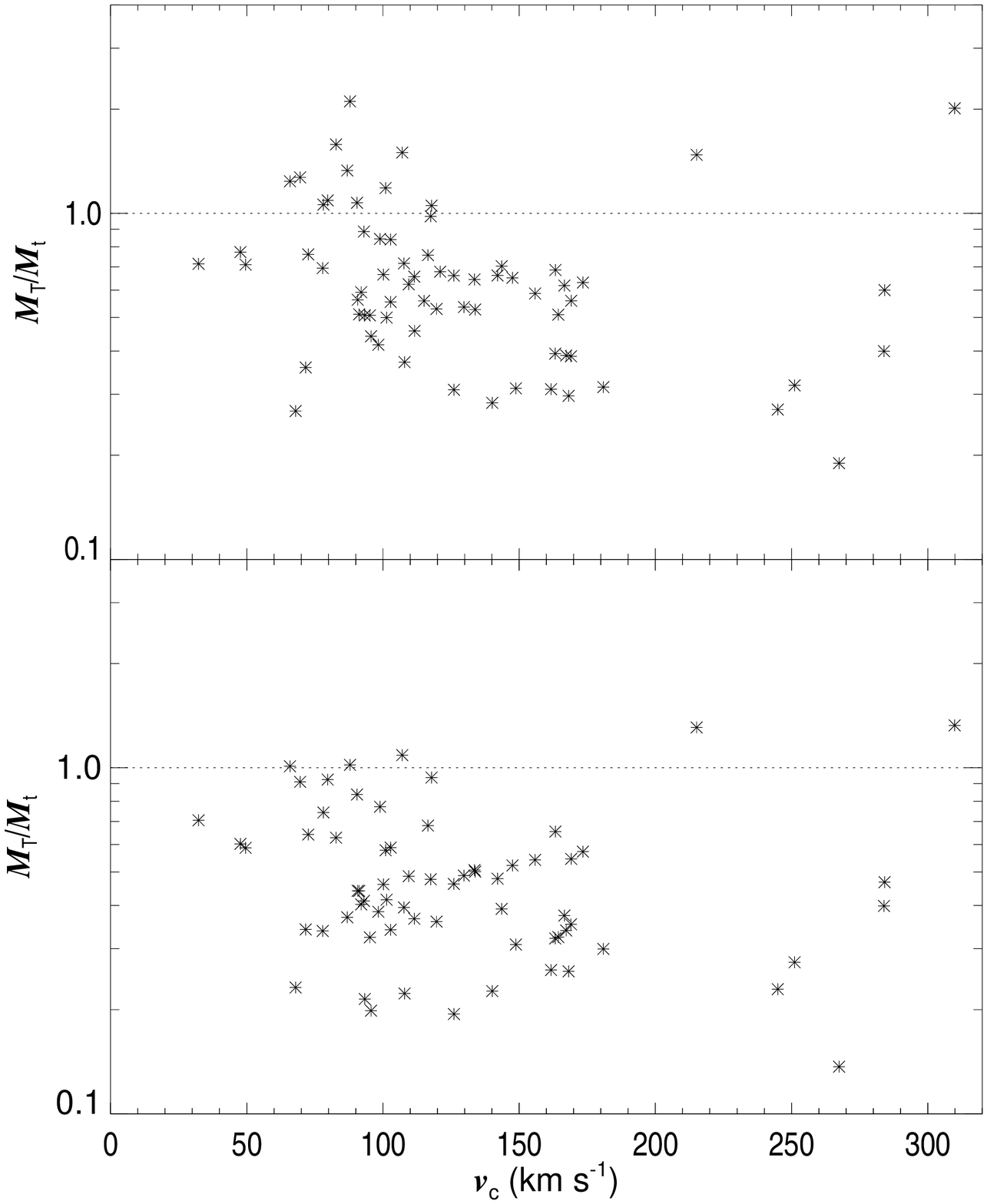}}\\
\end{tabular}
\caption{\label{plotgas} Ratio of the thick and thin disc mass, $M_{\rm T}/M_{\rm t}$, as a function of the circular velocity, $v_{\rm c}$. The top panel shows data} points calculated using only the stellar thin disc and thick discs and the bottom panel shows data points for which the gas disc has been included into the thin disc as explained in Section~\ref{secv120}.
\end{center}
\end{figure}

\begin{figure}
\begin{center}
\begin{tabular}{c}
{\includegraphics[width=0.45\textwidth]{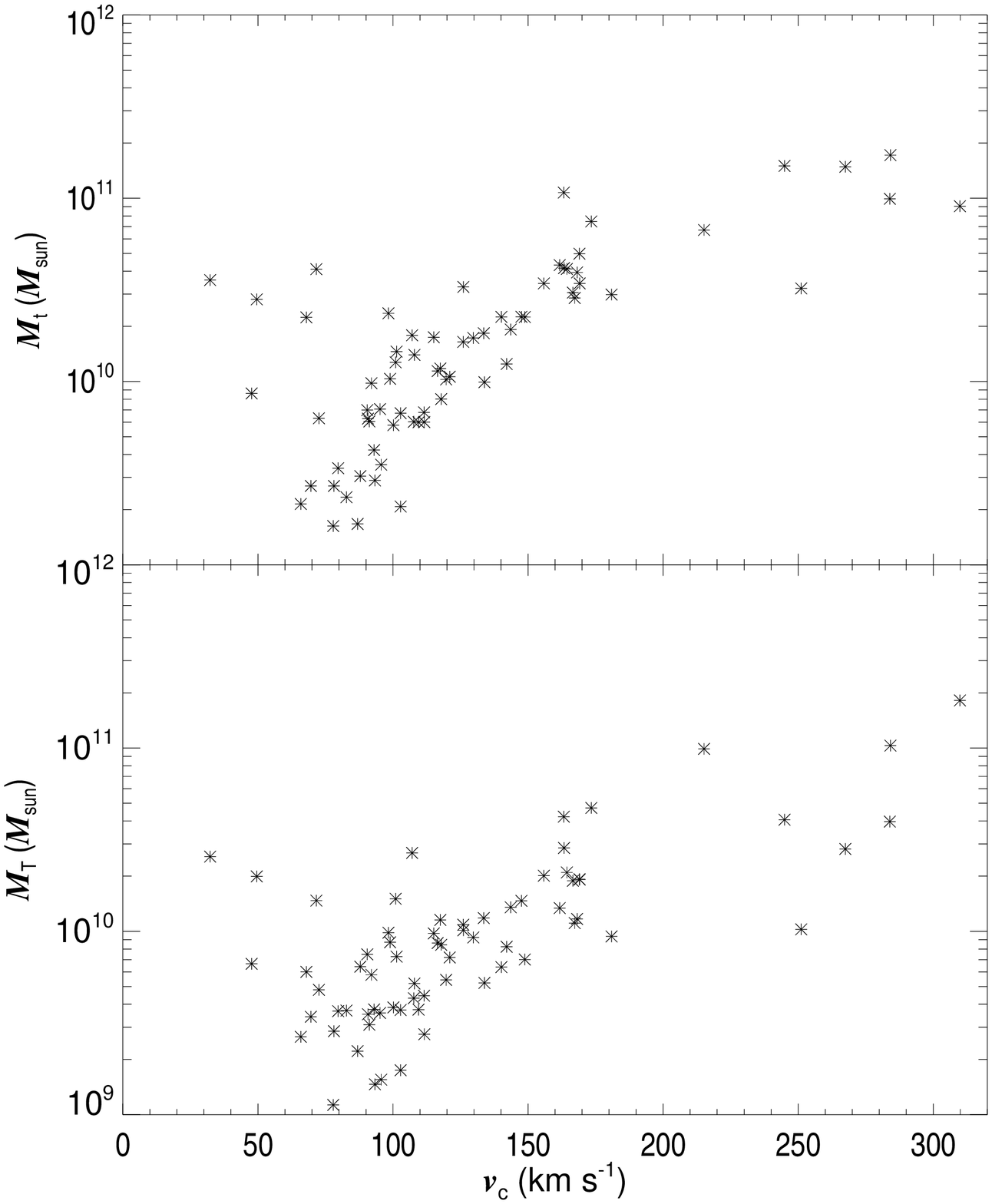}}\\
\end{tabular}
\caption{\label{absmasses} Absolute stellar masses of the thin (top panel) and the thick disc (bottom panel) as a function of circular velocity, $v_{\rm c}$.}
\end{center}
\end{figure}

Our sample is larger than in CO11b and the criteria used to select which vertical luminosity profiles are ``good'' should guarantee our fits to be of better quality than those obtained in CO11b. This is why we replotted the Figure~13 of CO11b --  that relating the relative thick disc stellar mass with the galaxy circular velocity -- in Figure~\ref{plotgas}. We can see how, as seen in Yoachim \& Dalcanton (2006) and CO11b, the relative thick disc stellar mass, $M_{\rm T}/M_{\rm t}$ anticorrelates with the circular velocity, $v_{\rm c}$. The range of relative disc masses, $0.2<M_{\rm T}/M_{\rm t}<2.0$ is also very similar to that found in previous works.

In CO11b, four galaxies were outliers with $v_{\rm c}\sim200\,{\rm km\,s^{-1}}$ and $M_{\rm T}/M_{\rm t}\gtrapprox1$. Only one of these galaxies appears in our new plot. The reason is that two of these galaxies, ESO~079-003 and NGC~4013, have not been included in our final sample because they have extended envelopes and were excluded by the $\mu_{\rm l}<24.5\,{\rm mag\,arcsec^{-2}}$ criterion. A third galaxy, ESO~443-042 has been dropped because of a too noisy luminosity profile due to the vicinity of a nearby bright star. The only one of these outliers remaining is NGC~3628, an interacting galaxy in the Leo Triplet which we visually identified to have an extended component which is faint enough to go through our sample selection criteria. If we were making our selection criteria more restrictive, we would be excluding galaxies which do not have an extended component other than a thin and a thick disc, so we decided to keep NGC~3628 in the sample. The galaxy appearing at $v_{\rm c}\sim300\,{\rm km\,s^{-1}}$ with $M_{\rm T}/M_{\rm t}\sim2$ is NGC~5084, and what we fitted as the thick disc could actually be a very extended bulge. 

Another galaxy with $M_{\rm T}/M_{\rm t}>1$ which is interacting but which does not fall outside the main trend is NGC~4747 which has been perturbed by NGC~4725 (Haynes 1979).

We found that $M_{\rm T}/M_{\rm t}$ is roughly constant for $v_{\rm c}>120\,{\rm km\,s^{-1}}$, with values $0.2<M_{\rm T}/M_{\rm t}<0.7$. For this range of velocities, the mean value of the stellar disc mass ratio is $M_{\rm T}/M_{\rm t}=0.48\pm0.03$ once the two outliers, NGC~3628 and NGC~5084 have been excluded. Below that circular speed, there is a sudden increase of the typical $M_{\rm T}/M_{\rm t}$. This $v_{\rm c}=120\,{\rm km\,s^{-1}}$ limit is the same below which Dalcanton et al.~(2004), suggested that star formation becomes less efficient.

We calculated the thin and thick disc absolute masses by assuming $\Upsilon_{\rm t}=1$, $\Upsilon_{\rm T}/\Upsilon_{\rm t}=1.2$, and that the $3.6\mu{\rm m}$ absolute magnitude of the Sun in the AB system is $M_{\sun}=6.06\,{\rm mag}$ (Oh et al.~2008). We found that both the thin and the thick disc absolute masses increase with increasing $v_{\rm c}$ (Figure~\ref{absmasses}). However, the thin disc mass increases generally faster with  $v_{\rm c}$, which is the reason why $M_{\rm T}/M_{\rm t}$ declines with increasing $v_{\rm c}$. Also, $M_{\rm T}$ values have a larger scatter than the $M_{\rm t}$ ones.

\subsection{The breaks we fit are likely to be the same as those observed in face-on galaxies}

\begin{figure}
\begin{center}
\begin{tabular}{c}
{\includegraphics[width=0.45\textwidth]{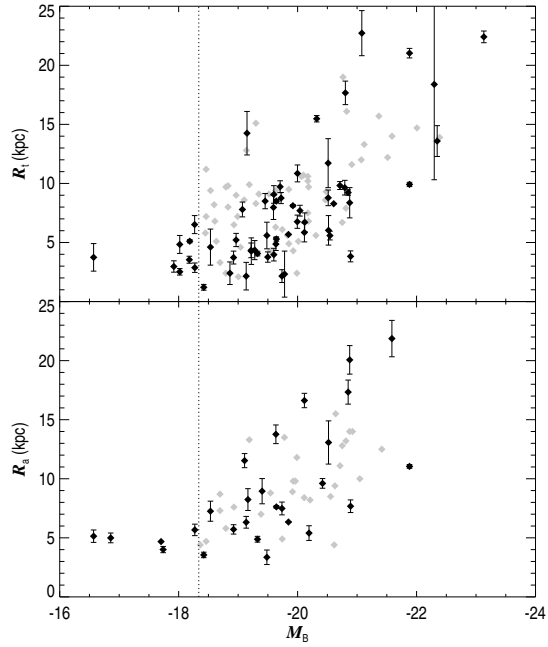}}\\
\end{tabular}
\caption{\label{truji} Top Panel: radii of the fitted truncations for the total disc as a function of the galaxy absolute blue magnitude. Bottom panel: radii of the fitted antitruncations for the total disc as a function of galaxy absolute blue magnitude. Black symbols correspond to the edge-on galaxies in this paper and gray symbols correspond to galaxies in Pohlen \& Trujillo (2006). The vertical pointed lines represent the lower brightness limit of Pohlen and Trujillo's (2006) sample. Error bars represent $2\sigma$ fitting errors.}
\end{center}
\end{figure}

Truncations and antitruncations in edge-on galaxies have long been assumed to correspond to those seen in face-on galaxies (see, e.g., Kregel et al.~2002 and Pohlen et al.~2004b review). However, before extracting any conclusions from our horizontal luminosity profile fits, it is important to assess whether our fitting approach is accurate enough when finding and describing breaks.

Truncation (Type~II breaks) and antitruncation (Type~III breaks) radii are known to have some correlation with galaxies properties such as brightness. We compared the distribution of our fitted break radius for the total luminosity profile as a function of galaxy absolute blue magnitude with those in the large face-on galaxy sample from Pohlen \& Trujillo (2006), which consists of almost a hundred nearby galaxies ranging from Sb to Sdm galaxies (thus, unlike us, excluding early-type disc galaxies). The reason for using the total luminosity profiles is that Pohlen \& Trujillo (2006) were not able to distinguish between thin and thick disc light due to the use of a sample of face-on galaxies.

Our galaxy absolute magnitudes were calculated by using HyperLEDA's internal dust extinction-corrected blue brightnesses and a distance modulus which makes use of the average redshift-independent distance measurements appearing in the NASA/IPAC Extragalactic Database (NED). For a few galaxies, no redshift-independent distance measurements were available and distances were determined from NED's Virgo Infall corrected radial velocities and a Hubble-Lema\^itre constant $H_{0}=73\,{\rm km\,s^{-1}\,Mpc^{-1}}$.

The results are presented in Figure~\ref{truji} and we see how the distribution of truncation radii (top panel) and that of antitruncation radii (bottom panel), are very similar for the face-on galaxy sample (gray dots) and our edge-on sample (black dots). It is thus reasonable to think that our fitting method has properly captured break radii.

\subsection{Classification of the horizontal luminosity profile fits}

\begin{table}
\begin{center}
\caption{\label{tablefit} Classification of the horizontal luminosity profile fits.}
\begin{tabular}{l | c c c c}
\hline
\hline
Profile type  & Total & Thin & Thick & PT06\\
\hline
Type~I        & 7  (10\%) & 9  (13\%) & 45 (64\%) & 9  (11\%)\\
Type~II       & 32 (46\%) & 32 (46\%) & 21 (30\%) & 46 (54\%)\\
Type~III      & 9  (13\%) & 7  (10\%) & 3  (4\%)  & 21 (25\%)\\
Type~II+II    & 3  (4\%)  & 3  (4\%)  & 0  (0\%)  & 2  (2\%) \\
Type~II+III   & 15 (21\%) & 15 (21\%) & 0  (0\%)  & 7  (8\%) \\
Type~III+II   & 1  (1\%)  & 1  (1\%)  & 1  (1\%)  & 0  (0\%) \\
Type~II+III+II& 3  (4\%)  & 3  (4\%)  & 0  (0\%)  & 0  (0\%) \\
\hline
\end{tabular}
\end{center}
Note.~-- Total, thin, and thick refer to total, thin, and thick horizontal luminosity profiles. PT06 refers to the galaxies classified in Pohlen \& Trujillo (2006).
\end{table}

\begin{figure}
\begin{center}
\begin{tabular}{c}
{\includegraphics[width=0.45\textwidth]{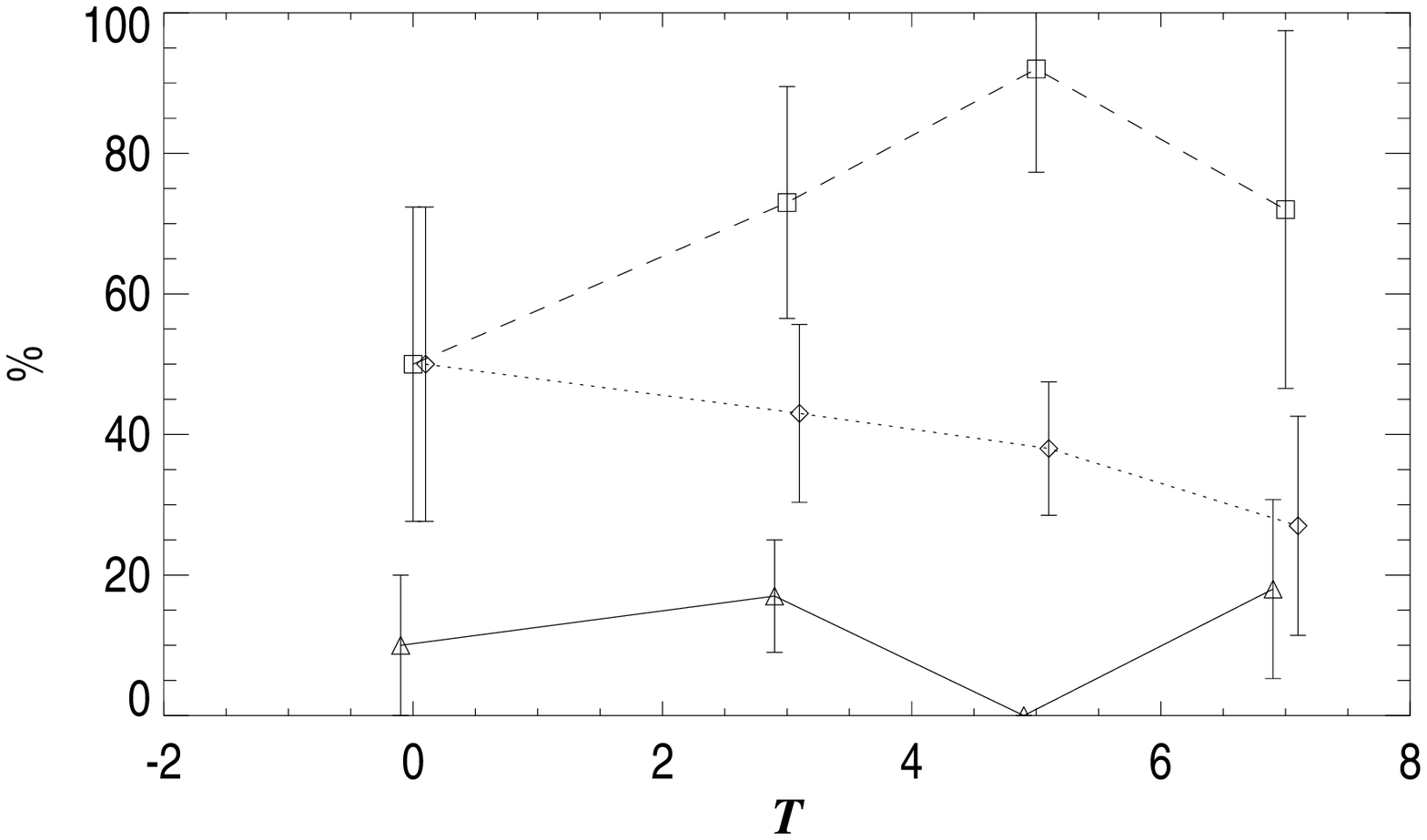}}\\
\end{tabular}
\caption{\label{truncdist} Fraction of galaxies having no breaks (triangle symbols and solid line), at least one truncation (square symbols and dashed line), and at least one antitruncation (diamond symbols and dotted line). The galaxies have been classified in four morphological type bins with $T\leq2$, $2<T\leq4$, $4<T\leq6$, and $6<T\leq8$. The error bars are obtained by using Poisson binomial statistics.}
\end{center}
\end{figure}

The fits to total, thin, and thick horizontal luminosity profiles were classified, using the Erwin et al.~(2008) criteria, into Type~I, Type~II, Type~III or combinations of Type~II and Type~III profiles (composite luminosity profiles). For example, a galaxy with a Type~II+III+II profile presents a truncation followed by an antitruncation and then a second truncation.

The individual classifications for each galaxy are presented in Tables~\ref{bigtable1}, \ref{bigtable2}, and \ref{bigtable3} for the total, the thin and the thick disc respectively. Pohlen \& Trujillo (2006) and Erwin et al.~(2008) use further subdivisions for Type~II and Type~III profiles, but those subdivisions require one to know properties, mostly related to the bar length, only easily measureable in face-on galaxies, so we have not been able to apply them to our sample. The result of the classification is summarized in Table~\ref{tablefit}.

Because the total luminosity profile is dominated by the light used to prepare the thin luminosity profile, the classifications for the total and the thin luminosity profiles are usually the same and the break radii are usually similar for both profiles.

Two of our total horizontal luminosity profiles -- those made for NGC~3600 and NGC~5084 -- have an antitruncation which is an artifact caused by the presence of an extended bulge like described in 15\% of the galaxies in Maltby et al.~(2012).

Because we can only detect breaks down to $\mu_{3.6\mu{\rm m}}(AB)=25-26\,{\rm mag\,arcsec^{-2}}$ (see Section~\ref{sectrunc}) our antitruncations are found at much brighter surface brightness than those which would be caused by a halo as described in Bakos \& Trujillo (2012).

When compared with the profile classification in Pohlen \& Trujillo (2006), we find that our total luminosity profiles have similar percentages of Type~I and Type~II profiles. However, there is a significant difference in the percentage of Type~III galaxies (25\% for them compared to 13\% for us) and the sum of Type~II+III and Type~II+III+II galaxies (8\% for them compared to 25\% for us). Since the sum of the fraction of Type~III, Type~II+III, and Type~II+III+II galaxies is nearly the same for both samples, we checked whether we could be classifying some Type~III galaxies as Type~II+III or Type~II+III+II. By looking at the luminosity profiles in Pohlen \& Trujillo (2006) we found that 6 out of their 21 Type~III galaxies have shoulders caused by star formation at the end of bars, inner rings, and/or prominent spiral arms which would have led us to classify those galaxies as Type~II+III if seen in an edge-on view. These galaxies are NGC~1084, NGC~1087, NGC~1299, NGC~4668, UGC~09741, and UGC~10721. If the fraction of shoulders caused by these reasons was the same in our sample $\sim3-6$ of our Type~II+III (Type~II+III+II) galaxies would be classified as Type~III (Type~III+II) by Pohlen \& Trujillo (2006). Thus, some of the truncations we only see in the thin disc may actually be an artifact of components (bars, rings) with a low $\Upsilon$. However, the number of misclassified galaxies could be smaller than the estimate we give, because these shoulders would get diluted due to line-of-sight integration and because the effect of star formation regions in rings and spiral arms should be less pronounced in the infrared (see, e.g., Buta et al.~2010).

As shown in Figure~\ref{truncdist}, we found that earlier-type galaxies tend to have less truncations than later-type galaxies in accordance with Figure~9 in Guti\'errez et al.~(2011). The fraction of untruncated and antitruncated discs seems to remain roughly constant with type changes (especially if we consider that two of the antitruncations in the earlier-type bin are an artifact caused by bulges), but large error bars do not allow us to be sure. Again, these results seem to agree with those by Guti\'errez et al.~(2011) which, for a larger sample of face-on galaxies, found a small decay in the frequency of untruncated discs when going to later types and a constant fraction of antitruncated discs.

\subsection{Truncations in thin and thick discs}

\label{sectrunc}

\begin{figure}
\begin{center}
\begin{tabular}{c}
{\includegraphics[width=0.45\textwidth]{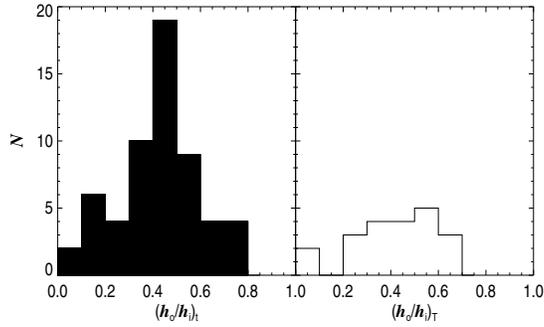}}\\
\end{tabular}
\caption{\label{slratio} Ratio of the outer disc and the inner disc scale lengths in truncated profiles (Type~II breaks) for the thin disc (left panel) and the thick disc (right panel). In the case of Type~II+II and Type~II+III+II profiles, both truncations have been considered.}
\end{center}
\end{figure}

\begin{figure}
\begin{center}
\begin{tabular}{c}
{\includegraphics[width=0.45\textwidth]{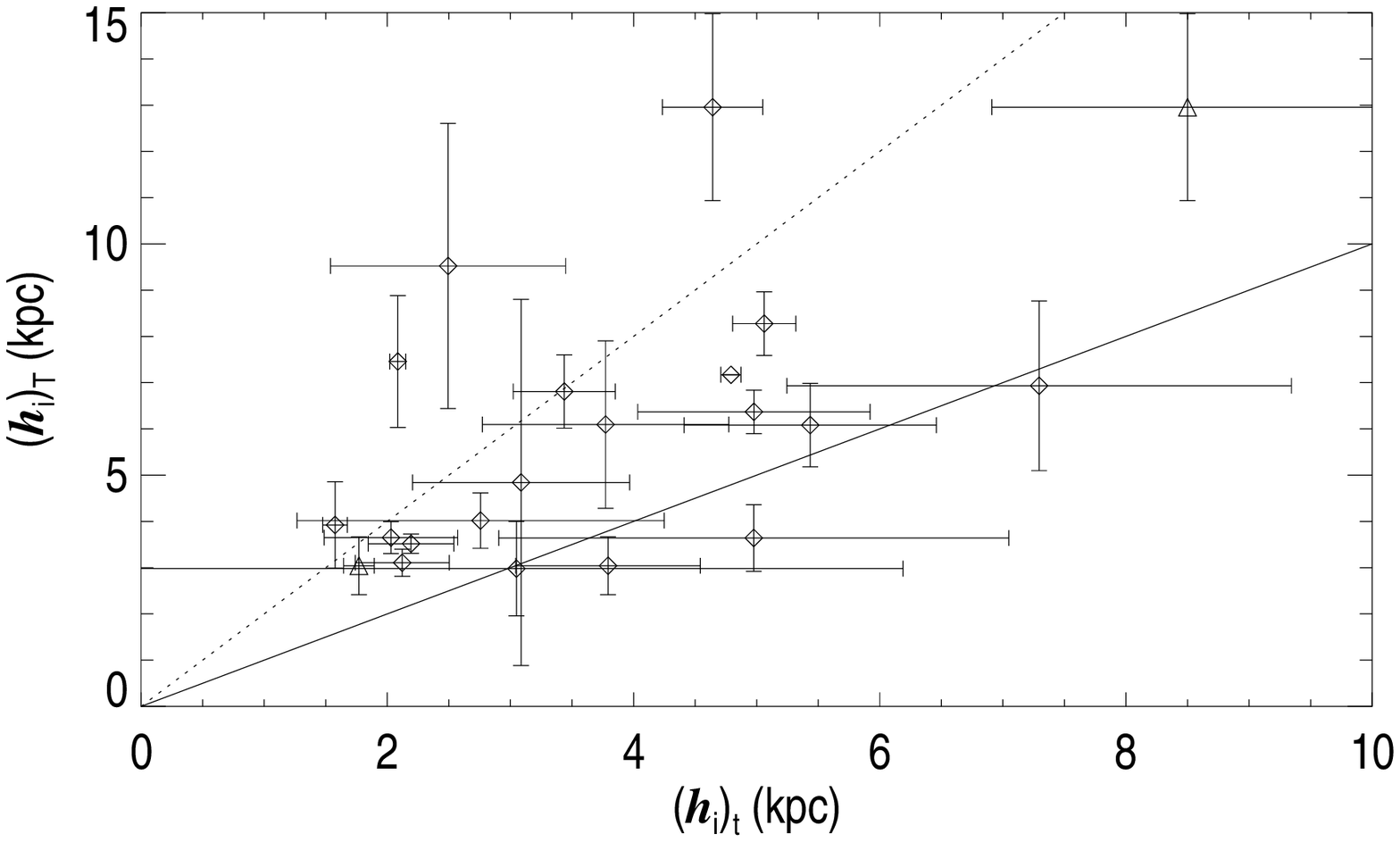}}\\
{\includegraphics[width=0.45\textwidth]{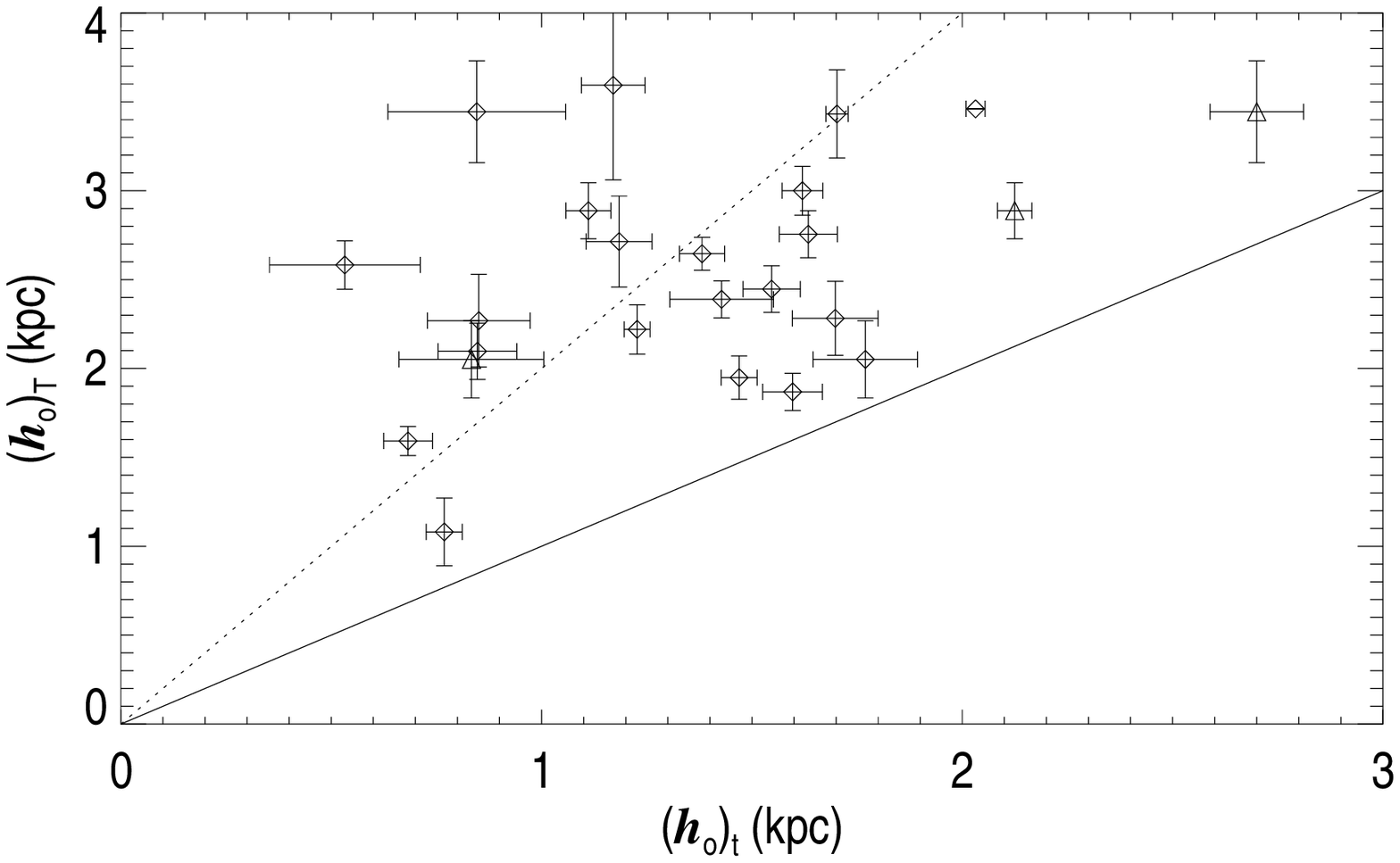}}\\
\end{tabular}
\caption{\label{hthinhthickltrunc} Scale lengths of inner thick discs as a function of their inner thin discs scale lengths (top panel) and scale lengths of outer thick discs as a function of their thin disc scale lengths (bottom panel) for galaxies with both thin and thick disc truncated. Solid lines trace a one-to-one relation between the thin and the thick disc scale lengths and the dotted lines indicates thick discs with a scale length two times larger than that of the thin disc. Triangle symbols stand for the second truncation in Type~II+II and Type~II+III+II profiles. Error bars represent $2\sigma$ fitting errors. }
\end{center}
\end{figure}

\begin{figure}
\begin{center}
\begin{tabular}{c}
{\includegraphics[width=0.45\textwidth]{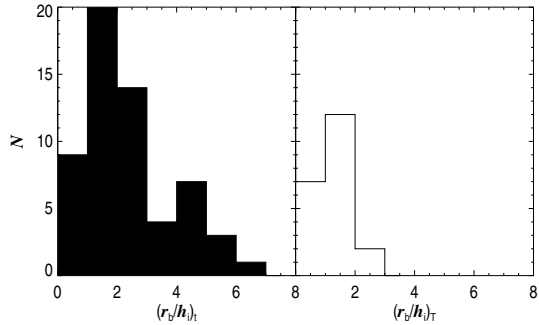}}\\
\end{tabular}
\caption{\label{rbsl} Ratio of the truncation break radius and the inner disc scale length for the thin disc (left panel) and the thick disc (right panel). In the case of Type~II+II and Type~II+III+II profiles, both truncations have been considered.}
\end{center}
\end{figure}

\begin{figure}
\begin{center}
\begin{tabular}{c}
{\includegraphics[width=0.45\textwidth]{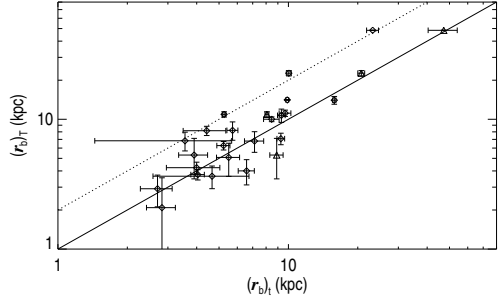}}\\
\end{tabular}
\caption{\label{rb} Truncation radii of thick discs as a function of truncation radii for thin discs. The solid line traces a one-to-one relation between the thin and the thick disc truncation radius and the dotted lines indicates thick discs with a truncation radius two times larger than that of the thin disc. Triangle symbols stand for the second truncation in Type~II+II and Type~II+III+II profiles. Error bars represent $2\sigma$ fitting errors.}
\end{center}
\end{figure}

\begin{figure}
\begin{center}
\begin{tabular}{c}
{\includegraphics[width=0.45\textwidth]{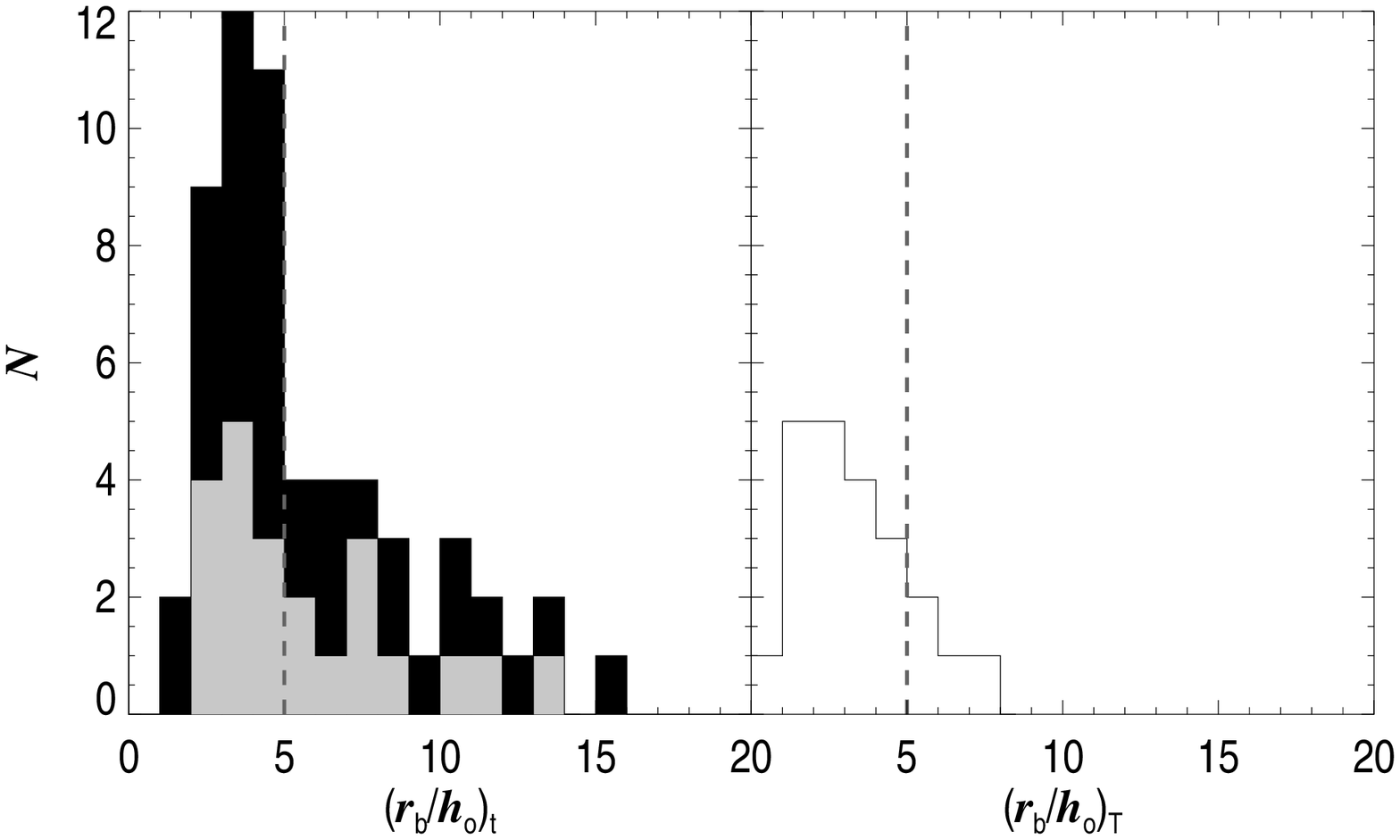}}\\
\end{tabular}
\caption{\label{rbsl2} Ratio of the truncation break radius and the outer disc scale length for the thin disc (left panel) and the thick disc (right panel). In the case of Type~II+II and Type~II+III+II profiles, both truncations have been considered. The gray histogram indicates thin disc truncations which are associated to a thick disc truncation. The vertical dashed line shows the limit between inner truncations and outer truncations according to the definition in Mart\'in et al.~(2012, submitted to MNRAS).}
\end{center}
\end{figure}

We searched whether the truncations (Type~II breaks) in thin and thick discs are qualitatively different. We first looked at how strong the truncations are by comparing the outer scale length ($h_{\rm o}$) and the inner scale length ($h_{\rm i}$) in both thin and thick discs (Figure~\ref{slratio}). We found that the $h_{\rm o}/h_{\rm i}$ distributions are quite similar, except for a narrow peak in the thin disc scale length ratio ($\left(h_{\rm o}/h_{\rm i}\right)_{\rm t}$). We also found that $\left(h_{\rm o}/h_{\rm i}\right)_{\rm t}$ and $\left(h_{\rm o}/h_{\rm i}\right)_{\rm T}$ are little correlated (linear Pearsons correlation coefficient $\rho\sim0.3$).

When both the thin and the thick disc are truncated, the inner scale length of the thick disc is generally between one and two times that of the thin disc (Figure~\ref{hthinhthickltrunc}). The outer scale length of the thick disc is even less correlated with that of the outer thin disc than the inner thin and thick disc scale lengths. Because there is some contribution of thick disc light in the range of heights where we have measured thin disc properties, it is expected that the thin disc scale lengths, especially $\left(h_{\rm o}\right)_{\rm t}$, have been overestimated.

When plotting a histogram of the distribution of break radii, $r_{\rm b}$, in units of the fitted inner disc scale lengths we find that thick discs do generally truncate at a lower relative radius than thin discs (Figure~\ref{rbsl}), which is generally due to thick discs having a longer inner scale length.

A difference between truncations in thin and thick discs is their frequency. Fifty-four of our thin discs are truncated (77\%), but only twenty-two of the thick discs are (31\%). Thus, apparently, thin discs have more than double the frequency of truncations in thick discs. This could be due to an intrinsically lower fraction of truncations in thick discs or because we may not be able to detect them due to their low surface brightness. To test that, we measured the surface brightness difference between the thin disc luminosity profile at $r=0$ and that at the radius of the truncation, $\Delta m_{\rm t}$. The thin disc surface brightness at the truncation radius was found directly from the luminosity profile, but that at $r=0$ was found by using the fits of the truncated disc function integrated over the line of sight. The reason to do so was to avoid the bulge influence. We then searched for the surface brightness difference between the luminosity profile at $r=0$ and the truncation detection threshold ($\Delta m_{\rm T}$) in the thick discs. In Section~\ref{secfithor} we explained that truncations were found visually before producing the actual fitting. Because of the subjective nature of this identification, the truncation detection threshold is unclear and we can only estimate it, making the $\Delta m_{\rm T}$ values rather uncertain. Based on our experience visually classifying the breaks, we are confident that we detect truncations down to $\mu_{3.6\mu{\rm m}}(AB)=25-26\,{\rm mag\,arcsec^{-2}}$.

We compared the $\Delta m_{\rm T}$ for each thick disc to a random set of $\Delta m_{\rm t}$ measured from real thin discs as explained before. If $\Delta m_{\rm t}>\Delta m_{\rm T}$, we considered that that particular thin disc truncation could not be detected on that particular thick disc. If our detection threshold was $\mu_{3.6\mu{\rm m}}(AB)=26\,{\rm mag\,arcsec^{-2}}$, we would detect $\sim60\%$ of thick disc truncations. If our detection threshold was $\mu_{3.6\mu{\rm m}}(AB)=25\,{\rm mag\,arcsec^{-2}}$, we would be detecting only $\sim30\%$ of the truncations. These numbers would be compatible with a similar frequency of truncations in thin and thick discs if thin and thick disc truncations were completely unrelated. 

However, truncations in thin and thick discs are not completely uncorrelated because we found the truncation radius for the thin disc to be similar to that of the thick disc in most cases (Figure~\ref{rb}). In cases of galaxies with two thin disc truncations, the thick disc truncation has a radius similar to that of one of the thin disc truncations. It thus seems that thick disc truncations tend to be associated with a thin disc truncation. This is further discussed in Section~\ref{secdisctrunc}.

Mart\'in et al.~(2012, submitted to MNRAS) divided truncations into inner truncations and outer truncations. Inner truncations are those for which the ratio of the truncation radius and the outer disc scale length is smaller than five, $r_{\rm b}/h_{\rm o}<5$, and outer truncations are those for which this ratio is larger than five, $r_{\rm b}/h_{\rm o}>5$. We have checked in which category our truncations would fall. We found that our thin discs have both inner and outer truncations and that thick discs only have inner truncations or borderline cases. We have also checked how thin disc truncations associated to a thick disc truncation are distributed. Again, we found that they may be inner or outer.

\subsection{Antitruncations in thin and thick discs}

\label{secantitrunc}

\begin{figure}
\begin{center}
\begin{tabular}{c}
{\includegraphics[width=0.45\textwidth]{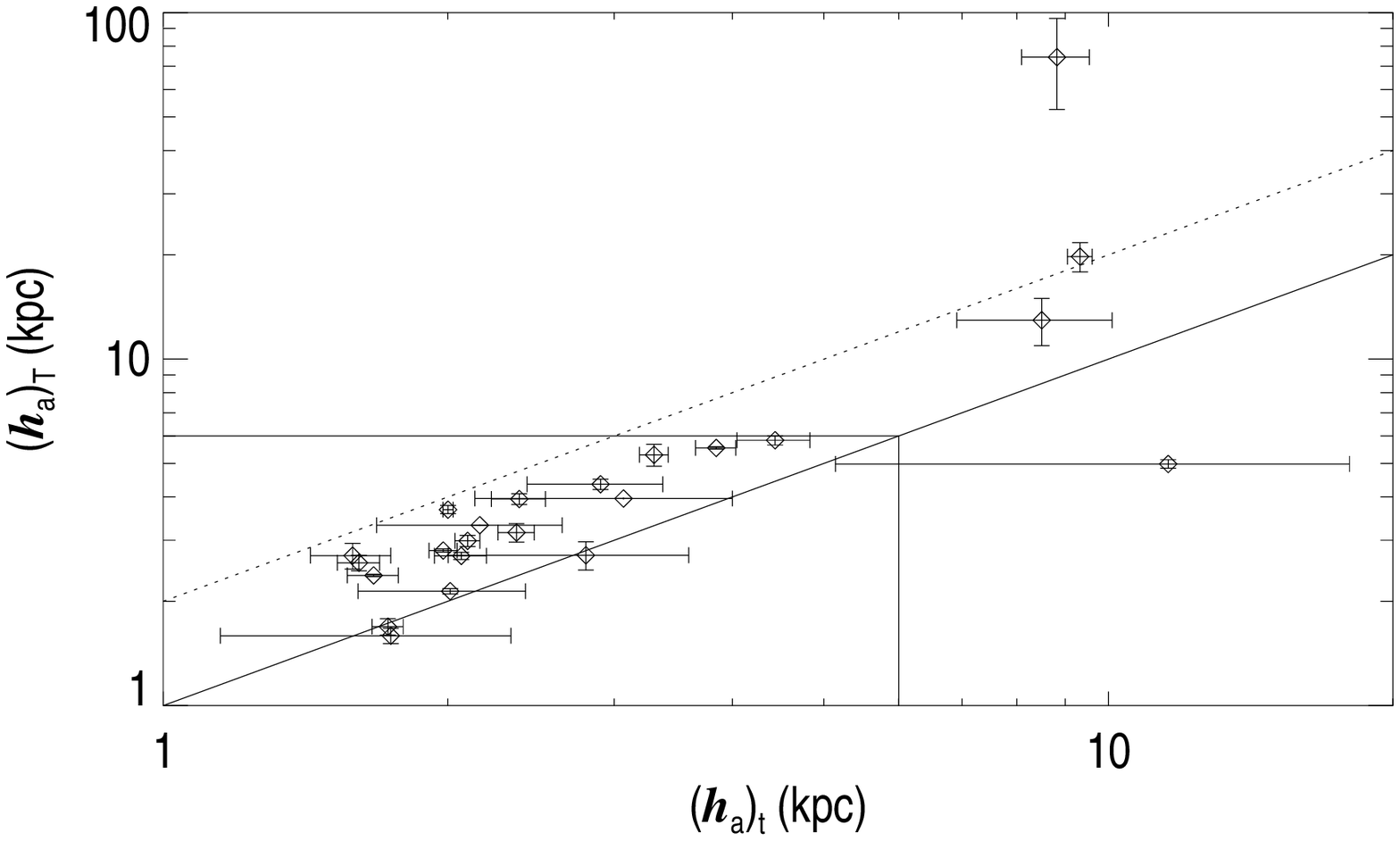}}\\
{\includegraphics[width=0.45\textwidth]{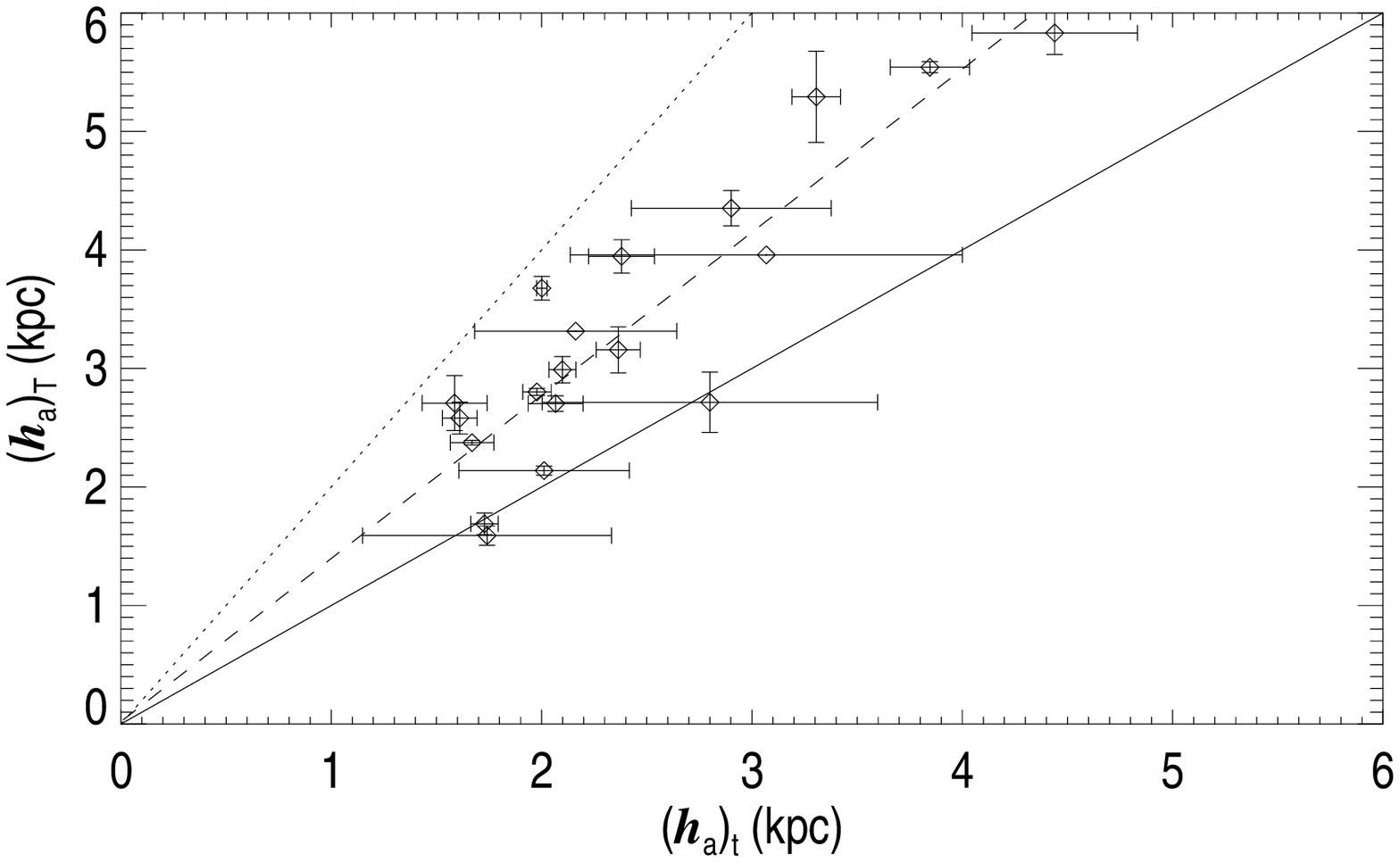}}\\
\end{tabular}
\caption{\label{anti} Thick disc scale length as a function of thin disc scale length for regions after a Typer~III break (antitruncation). In the bottom panel, the solid line represents a slope equal to one and the dotted line represents a slope equal to two; this factor of two ratio is indicated by the dotted line in the top panel too. The top panel shows the axis is a logarithmic scale and the bottom panel shows the range indicated by a box in the top panel with the axis displayed in a linear scale. The dashed line in the bottom panel represent the best linear fit for the galaxies displayed on it. Error bars represent $2\sigma$ fitting errors.}
\end{center}
\end{figure}

We define $\left(h_{\rm a}\right)_{\rm t}$ as the scale length of the thin disc in the section after an antitruncation (Typer~III break). We define  $\left(h_{\rm a}\right)_{\rm T}$ as the scale length of the thick disc section which has the largest overlap in galactocentric projected radius with the thin disc section with scale length $\left(h_{\rm a}\right)_{\rm t}$.

We found that the fitted thin disc scale length after an antitruncation ($\left(h_{\rm a}\right)_{\rm t}$) and the fitted thick disc scale length at that radius ($\left(h_{\rm a}\right)_{\rm T}$) correlate quite well (Figure~\ref{anti}). If we do consider the galaxies appearing in the lower panel, the slope is roughly 1.4 and the Pearsons correlation coefficient is  $\rho=0.89$. The galaxies with $\left(h_{\rm a}\right)_{\rm t}>6$\,kpc have not been considered because they are outliers which highly affect the fit.

According to the data in Table~\ref{tablefit} thick discs are very rarely found to antitruncate (6\%) within the surface brightness limits we are studying. We individually checked the origin of the antitruncations in thick discs. That in NGC~5529 seems to be a genuine antitruncation, with the host galaxy having a significant warp. Additionally, NGC~5529 has at least four companion galaxies (Irwin et al.~2007) within 10\,arcmin or 150\,kpc in projection. The brighter of these satellite galaxies, MCG~+06-31-085a, is around four magnitudes fainter than the main galaxy and is connected to it by an \hi\ bridge (Kregel et al.~2004), suggesting that the unusual thick disc antitruncation could be the consequence of an interaction. The antitruncations in NGC~3600 and NGC~5084 are related to the presence of extended spherical bulges, whose light makes the scale height at radii smaller than the truncation radius steeper than it would be in the absence of a bulge. The case of IC~5176 seems to be rather similar, but the bulge is much flatter and thus not a classical bulge.

\section{Discussion}
\label{secdiscussion}

\subsection{Truncations in thick discs could be linked to a thin disc truncation}

\label{secdisctrunc}

We found 60 truncations in thin discs (taking into account those truncated twice), for only 22 truncations in thick discs. In Section~\ref{sectrunc} we showed that if the difference in brightness between the disc center and the truncation radius was the same in thin and thick discs and we were considering truncations in both discs to be uncorrelated, we would be detecting only between $\sim30\%$ and $\sim60\%$ of truncations in thick discs.

However, most of the truncations we discovered in thick discs are found at a radius similar to that of the truncation of the thin disc. If thick disc truncations are systematically found at the same radius as their thin disc counterpart, we should be detecting most of them. As a consequence of that, we can say that some thick discs in galaxies with a truncated thin disc are untruncated.

Further evidence of a truncation happening at the same radius for the thin and the thick disc appears in Radburn-Smith et al.~(2012), who studied the truncation of the face-on galaxy NGC~7793 and found that if that galaxy had a significant thick disc -- which is likely to be the case because of its low mass -- it has to truncate as the same radius as the thin disc. 

This suggests two possible formation mechanisms for truncations: one creating truncations simultaneously both in the thin and the thick disc and one creating them only in the thin disc. A mechanism creating a truncation in both discs, should be a dynamical one, because it is hard to conceive a star formation threshold at the same radius in the two discs at different times of galaxy evolution. This would make sense if we consider that Pohlen \& Trujillo (2006) and Erwin et al.~(2008) divide face-on truncations into four groups based on their morphology and radius compared to the bar length. The most frequent truncation types in their papers, and thus studied with most detail, are OLR breaks or Type~II.o-OLR -- thought to be close to bar Outer Lindbald Resonances -- and classical truncations or Type~II.o-CT, which do not seem to be associated with the bar and the authors suggest they are related to star formation thresholds. Unfortunately, no galaxy property, except the ratio between the truncation radius and that of the bar, is clearly able to distinguish between those two kinds of truncations. To further complicate the picture, Mu\~noz-Mateos et al.~(2012; in preparation) suggest that some of the Type~II.o-CT truncations could also be related to the OLR in the case of a bar+spiral rotation pattern coupling. Another possibility is that the same mechanism creates a truncation in one disc or in both depending on some galactic property which would not be related those we have studied in this paper (kinematics, spiral arms/bar properties\ldots).

We checked whether thin disc truncations associated with untruncated thick discs are qualitatively different than those associated with truncated thick discs but we found no significant difference in galaxy morphological type, absolute magnitude, outer disc scale length compared to that of the inner disc ($\left(h_{\rm o}/h_{\rm i}\right)_{\rm t}$), break radius in inner disc scale length units ($\left(r_{\rm b}/h_{\rm i}\right)_{\rm t}$), or break radius in outer disc scale length units ($\left(r_{\rm b}/h_{\rm o}\right)_{\rm t}$). This would be consistent with Pohlen \& Trujillo (2006) findings of two kinds of truncations -- Type~II.o-OLR and Type~II.o-CT -- which can be only differentiated by the bar position with respect to the truncation radius.

If we consider that our classification of thin disc truncations into inner and outer types is valid, then an extra puzzle appears. Mart\'in et al.~(2012, submitted to MNRAS) suggested that inner and outer breaks may have been formed in two different kind of processes. However, having thick disc truncations associated to both inner and outer thin disc truncations shows that these two processes would both be able to build a thick disc truncation under the right circumstances. This variety of inner/outer truncations associated/non-associated to thick disc truncations indicates that the truncation mechanisms may be even more complex than expected.

\subsection{Genuine antitruncations are rare}

\begin{figure}
\begin{center}
\begin{tabular}{c}
{\includegraphics[width=0.45\textwidth]{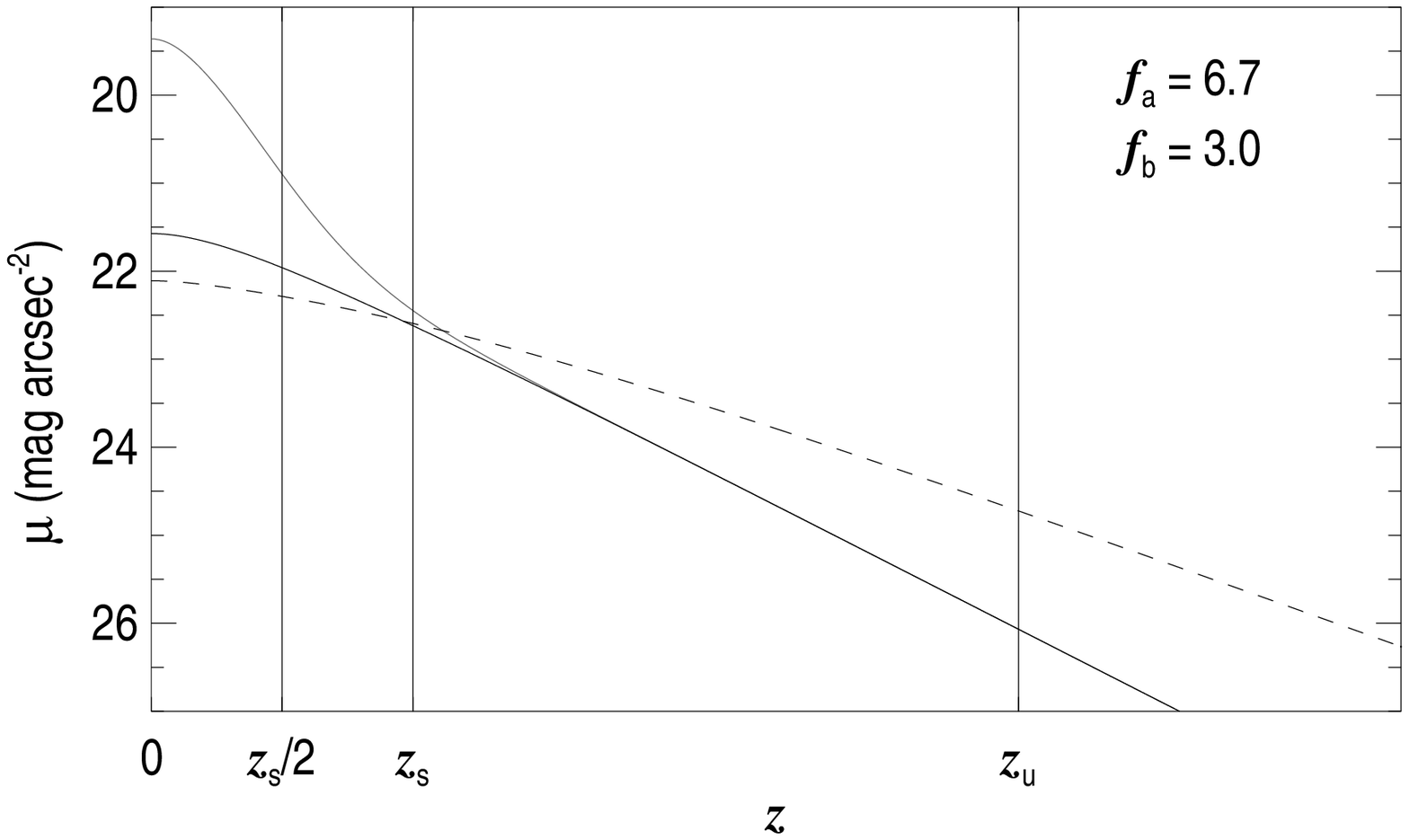}}\\
\end{tabular}
\caption{\label{thindiscremoved} Vertical luminosity profile for an arbitrary galaxy in our sample (gray). The thick disc is shown in black and a thick disc with the same velocity dispersion and face-on surface brightness but after the removal of the thin disc is shown as a dashed line. The $f_{\rm a}$ and $f_{\rm b}$ values are displayed in the top-right corner.}
\end{center}
\end{figure}

As seen in Section~\ref{secantitrunc} and Figure~\ref{anti}, the scale length of the thin disc profile after an antitruncation break seems to correlate with the thick disc scale length at that projected radius. A visual inspection of the fits in Figure~\ref{verticalfits} shows that, in many cases, the shape of the luminosity profiles of Type~III, Type~II+III, and Type~II+III+II thin discs after the antitruncation break radius, $r_{\rm a}$, are very similar to that of the thick disc in the same projected radius range. It is therefore possible that it is not the thin disc itself which antitruncates and that what we are actually seeing is that the thick disc dominates the light emission even in the range of heights $0<z<0.5z_{\rm s}$ for $R>r_{\rm a}$.

In order to test that possibility, we used the fits to vertical luminosity profiles and we calculated the mean flux per pixel emitted by the thick disc  in the range $0<z<0.5z_{\rm s}$ (height range for thin disc horizontal luminosity profiles) divided by the mean flux per pixel emitted at $z_{\rm s}<z<0.5z_{\rm u}$ (height range for thick disc horizontal luminosity profiles). This value, which we will call $f_{\rm a}$, was averaged over all the valid vertical bins. However, because the shape of the vertical thick disc luminosity profile changes in the absence of a thin disc, we calculated a second value, $f_{\rm b}$, which measures the same ratio if the thin disc was removed but the thick disc face-on column mass density and vertical velocity dispersion were kept constant. As seen in Figure~\ref{thindiscremoved}, $f_{\rm b}<f_{\rm a}$.

The $f_{\rm a}$ and $f_{\rm b}$ values can be used for estimating the brightness of the thick disc in the range of heights $0<z<0.5z_{\rm s}$, but we have to consider that they have been calculated at low $R$ and can only be considered a useful approximation to what happens in galaxy outskirts. In Figure~\ref{verticalfits}, the lower limit of the gray area is the estimate of the thick disc brightness at that range ($0<z<0.5z_{\rm s}$) in the absence of a thin disc calculated by moving the red symbols up by a factor $f_{\rm b}$. The upper limit of the gray area represents the thick disc surface brightness estimate for the range $0<z<0.5z_{\rm s}$ in the case of a thin disc whose face-on column mass density relative to the thick disc is similar to that found at low $R$ (moving the red symbols up by a factor $f_{\rm a}$). Thus, for the outskirts of a galaxy, the lower limit of the gray area symbols represents what we have defined to be the thin disc if it was only the low $z$ part of a thick disc and no significant genuine thin disc was found there.

Of the 26 thin discs with antitruncations 14 (ESO~157-049, ESO~440-027, IC~1553, NGC~0522, NGC~1163, NGC~1422, NGC~3454, NGC~4330, NGC~4359, NGC~4607, NGC~5529, UGC~06526, UGC~12518, and UGC~12692) are well fitted by the lower limit of the gray area, meaning that they are compatible with the range of projected radii $R>r_{\rm a}$ being dominated by the thick disc at all heights. In six galaxies (ESO~469-015, IC~1576, UGC~00903, UGC~01970, UGC~07086, and UGC~10288), the thick disc alone does not seem to be bright enough to account for all the light for heights $0<z<0.5z_{\rm s}$ at $R>r_{\rm a}$. In four galaxies (NGC~1596, NGC~3628, NGC~4081, and UGC~08737) our thick disc brightness estimate at $0<z<0.5z_{\rm s}$ overestimates the actual light emission. For NGC~5084, the antitruncation is an artifact caused by the presence of an extended bulge. Finally, for one galaxy (UGC~10297), the thick disc profile is too dim at $R>r_{\rm a}$ to be able to obtain any conclusion.

It thus seems that at least 14 out of 26 antitruncated galaxies, the antitruncation is actually an artifact caused by the presence of a thick disc. This number is probably a low estimate, because of the uncertainties of using $f_{\rm b}$ for inferring the behavior of the disc in its outskirts.

A legitimate question is that if for $R>r_{\rm a}$ we are mostly detecting the thick disc at all heights, why its horizontal luminosity profile is almost always flatter for $z_{\rm s}<z<z_{\rm u}$ (red horizontal luminosity profiles) than in the range $0<z<0.5z_{\rm s}$ (blue horizontal luminosity profiles)? This can be explained naturally if we consider that at $R\sim r_{\rm a}$ there is still a small fraction of stars belonging to the thin disc, meaning that the vertical thick disc luminosity profile becomes more peaked (see Figure~\ref{thindiscremoved}) than when the thin disc mass fraction becomes negligible for $R\gg r_{\rm a}$. This progressive broadening of the thick disc vertical luminosity profile with increasing projected radii causes the horizontal luminosity profile for $0<z<0.5z_{\rm s}$ to be steeper than that measured at $z_{\rm s}<z<0.5z_{\rm u}$. Additionally, unmasked faint stars and extended PSF wings are more likely to affect the much dimmer $z_{\rm s}<z<0.5z_{\rm u}$ range, potentially further contributing to make its horizontal luminosity profile shallower.

Here we have shown that genuine antitruncations are rare (twelve at maximum out of 70 or less than $\sim15\%$) and that most of the features which would be considered an antitrucation in a face-on view are actually an artifact caused by the superposition of a thin and a thick disc with different scale lengths. However, it seems that several of the found antitruncations are genuine and could be a signature of a past interaction -- as suggested by Laurikainen \& Salo (2001). Laurikainen \& Salo (2001) showed that many M~51-like interacting galaxies have antitruncations. They used numerical simulations to show that the redistribution of material caused by an interaction may last for several Gyr, which could explain the presence of genuine antitruncations in isolated galaxies. Also, some of the antitruncations which do not seem related to the thick disc could be caused by the presence of outer rings as suggested by Erwin et al.~(2005).

\subsection{Galaxies with circular velocities below $v_{\rm c}=120\,{\rm km\,s^{-1}}$ have more massive thick discs}
\label{secv120}

In Section~\ref{secmasses} we found that thick discs with $v_{\rm c}>120\,{\rm km\,s^{-1}}$ have roughly constant relative masses with respect the thin disc ($0.2<M_{\rm T}/M_{\rm t}<0.7$), and that the relative stellar mass of the thick disc increases for $v_{\rm c}<120\,{\rm km\,s^{-1}}$. This threshold was not noticed in CO11b due to smaller statistics and a larger scatter in data due to less restrictive quality selection criteria. However, it was detected by Yoachim \& Dalcanton (2006).

If, as indicated in CO11b, thick discs would have formed in situ at high redshift and have their thin discs formed from gas remaining from the formation process and posterior cold accretion, at least two factors could contribute to make $M_{\rm T}/M_{\rm t}$ larger in less massive galaxies:
\begin{itemize}
 \item Supernova feedback is more efficient at removing the gas from shallower potential wells. This is because in a high-mass galaxy, gas expelled by supernovae during the thick disc build-up could eventually come back to the galaxy and be used for building the thin disc. Gas expelled from low-mass galaxies would be less likely to come back.
 \item Lower-mass galaxies are dynamically younger than higher-mass ones. All galaxies have already formed their thick disc, and have later started to form the thin disc. Because of the slowest evolution of lower-mass galaxies, their thin discs are younger and less developed, implying a larger fraction of gas in them.
\end{itemize}

It is important to stress the point that less massive galaxies are likely to expel a larger gas fraction through supernova feedback than massive ones. Indeed, as pointed out by Elmegreen \& Elmegreen (2006), if a thick disc formed at high redshift had only $\sim20\%$ of the total mass of the disc (as it is the case for today's thick discs in some massive galaxies), its scale height would shrink considerably when the gas which formed the thin disc, the remaining 80\% in mass, was accreted. This could cause, if the thick disc scale height was not large enough (several kpc) that the thick discs would be indistinguishable from the thin discs in vertical luminosity profiles. The thick disc scale height shrinking would be smaller in a low-mass galaxy accreting a lower fraction of gas ($\sim50\%$), which would make the original thick disc formed at high redshift much easier to detect.

Thus, what we observe as a thick discs in today's low-mass galaxies, would be mostly the original in situ thick discs formed at high-redshifts. For massive galaxies, which have expelled less gas through supernova feedback, it is possible that this disc would have shrunk so much that it would be undetectable. In this case, there would be no genuine thick discs, but a continuous distribution of scale heights created by the secular heating of the thin disc by its own overdensities in accordance with the claim from Bovy et al.~(2012) for the Milky Way. Significant thin disc heating by satellites is likely to be discarded due to the absence of widespread flares as discussed in CO11b. The thick disc being mostly made of stars stripped from satellites is also discarded because that mechanism would lead thick discs significantly less massive than what is observed (see CO11b for a full discussion on these two points). Radial migration (Sch\"onrich \& Binney~2009) could also explain the presence of thick discs in massive galaxies, but fails at explaining the presence of very massive thick discs in low-mass galaxies which are dynamically younger than high-mass ones.

A way of testing whether $M_{\rm T}/M_{\rm t}$ is higher in low-mass galaxies due to a slower dynamical evolution is to recalculate $M_{\rm t}$ by considering that, in addition to the thin disc stars, it includes the gas disc. If low-mass galaxies were less efficient at forming stars, the new $M_{\rm T}/M_{\rm t}$ would show a flatter trend with the rotation velocity, $v_{\rm c}$. To calculate the mass of gas contained in a disc, we used the 21-cm corrected for self-absorption flux from HyperLEDA and we converted it to a gas mass using the expression
\begin{equation}
 M_{\rm HI}/M_{\sun}=236\,d^{2}\,f
\end{equation}
where $f$ is the area of the 21-cm line profile expressed in mJy\,km\,s$^{-1}$ and $d$ is the distance to the galaxy expressed in Mpc (Zwaan et al.~1997). As done in Yoachim \& Dalcanton (2006), we made the correction $M_{\rm gas}=1.4M_{\rm HI}$ to account for helium and metals and we did not apply a correction for molecular gas. The new $M_{\rm T}/M_{\rm t}$ values have been calculated considering that the absolute $3.6\mu{\rm m}$ magnitude of the Sun in AB system is $M_{\sun}=6.06\,{\rm mag}$ (Oh et al.~2008), that $\Upsilon_{\rm t}=1$, and that $\Upsilon_{\rm T}/\Upsilon_{\rm t}=1.2$. This has been done for all except three galaxies in the sample. Those three galaxies had no $f$ value available in HyperLEDA. The result of including the gas into the thin disc is shown in Figure~\ref{plotgas} in which the diamond symbols represent $M_{\rm T}/M_{\rm t}$ once the gas has been included in the thin disc. We note a slight tendency for low-mass galaxies to get their relative thick disc masses lowered by a larger factor by this operation than in the case of high-mass galaxies. When including the gas in the thin disc, the decrease in the thick disc relative mass fraction is $\Delta\left(M_{\rm T}/M_{\rm t}\right)=-0.27\pm0.04$ when $v_{\rm c}<120\,{\rm km\,s^{-1}}$ and $\Delta\left(M_{\rm T}/M_{\rm t}\right)=-0.11\pm0.03$ for galaxies with $v_{\rm c}>120\,{\rm km\,s^{-1}}$. This would support the idea of lower-mass discs having a slower evolution than higher-mass galaxies and having transformed less gas into thin disc stars. However, a definitive quantification of this effect cannot be obtained without knowing the mass-to-light ratio of the discs. We are working on this quantification and we will be publish it in a follow-up paper.

\section{Summary and conclusions}
\label{secconclusions}

Recent works suggest that thick discs are formed in situ at high redshift. So far, the easier way the study thin and thick discs as separate features is by looking at edge-on galaxies.

Breaks (truncations and antitruncations) in disc galaxies have been until now mostly studied for the disc as a whole, without distinguishing the light of the thin and the thick disc. Several theories compete for explaining the origin of such breaks. Truncations have been explained by dynamical arguments related with the conservation of angular momentum during galaxy formation, by star formation thresholds, and by the redistribution of angular momentum by a bar. Antitruncations have in some cases been linked to interactions and mergers.

In order to study a significant sample of edge-on galaxies (70), we produced luminosity profiles of thin and thick discs parallel to galaxy midplanes from the average of the 3.6 and 4.5$\mu m$ images from the Spitzer Survey of Stellar Structures in Galaxies (S$^4$G). The vertical height dominated by each disc was derived from luminosity profiles perpendicular to midplanes done in a similar way to that described in Comer\'on et al.~(2011b). In brief, we fitted the observed vertical luminosity profiles by comparing them with a grid of models made by solving the equations of three gravitationally coupled vertically isothermal discs (thin, thick, and gas discs). We defined the range of heights for which the thick disc dominates that height (named $z_{\rm s}$) above which it emits 90\% of the light. We defined a height, $z_{\rm u}$ at which the $26\,{\rm mag\,arcsec^{-2}}$ level was reached for the fitted vertical bins. The thick disc horizontal luminosity profile was made by averaging the light emission for $z_{\rm s}<z<z_{\rm u}$ and the thin disc one was made by averaging light in $0<z<0.5z_{\rm s}$.

Once the thin and thick disc luminosity profiles parallel to midplanes were created, we fitted them with a integrated over the line of sight generalization of the function used by Erwin et al.~(2008) to describe discs with breaks. 

The conclusions of this paper are:

\begin{itemize}

\item The position of our breaks are compatible with those fitted in face-on galaxies.

\item Horizontal luminosity profiles for thin discs are often truncated (77\% of cases) and also have a significant number of antitruncations (40\% of cases). According to our fits, thick discs truncate with a frequency of 31\%, although some extra truncations can not be discarded at a lower surface brightness than our detection threshold. Antitruncations are rare in our thick disc sample (6\%) and are in three out of four cases artifacts caused by an extended bulge.

\item When thick discs truncate, they do so at a radius usually compatible with that of the truncation of the thin disc. However, not all truncated thin discs are associated with a truncated thick disc. This dichotomy suggests two different mechanisms creating truncations: one -- necessarily of dynamical origin -- affecting both discs and one affecting only the thin disc. Another possibility is that the same mechanism creates a truncation in the thin disc or in both depending on some galactic property which has not been studied in this paper (such as bar or kinematical properties).

\item Antitruncations in thin discs seem to be related, in more than half of the cases, to the superposition of a thin and a thick disc with different scale lengths. Thus, for those thin discs, the part outside the antitruncation would actually be the thick disc. We estimate the fraction of genuine thin disc antitruncations to be at maximum 15\%. This is less than half of those that would be expected from a deep enough fit to a face-on galaxy sample.

\item Because our sample is much larger and better selected than that used in Comer\'on et al.~(2011b), we have been able to further constrain the shape of the relation between the stellar mass of the thick disc relative to that of the thin disc, $M_{\rm T}/M_{\rm t}$, and the galaxy circular velocity, $v_{\rm c}$ originally found by Yoachim \& Dalcanton (2006) (Figure~\ref{plotgas}). We found that galaxies with $v_{\rm c}>120\,{\rm km\,s^{-1}}$ have a roughly constant $M_{\rm T}/M_{\rm t}$. From $v_{\rm c}=120\,{\rm km\,s^{-1}}$ downwards, $M_{\rm T}/M_{\rm t}$ increases dramatically with, in some cases, $M_{\rm T}/M_{\rm t}>1$. For these galaxies, the high $M_{\rm T}/M_{\rm t}$ could be caused by a combination of a high efficiency of supernova feedback that removes gas from the weak potential well of the young galaxy and limits the eventual mass of the thin disc and a slower dynamical evolution causing the stellar thin discs to be younger and less massive than in higher-mass galaxies. In the latter case we showed that, if we assume the same $\Upsilon_{\rm T}/\Upsilon_{\rm t}$ for all galaxies, when gas is included in the mass of the thin disc, galaxies with $v_{\rm c}<120\,{\rm km\,s^{-1}}$ have $M_{\rm T}/M_{\rm t}$  values that are more similar to those in galaxies with $v_{\rm c}>120\,{\rm km\,s^{-1}}$. This similarity implies that part of the high value for $M_{\rm T}/M_{\rm t}$ low mass galaxies is the result of thin disc youth: the thin disc mass has not formed stars yet.

\end{itemize}

\section*{Acknowledgements}

The authors wish to thank the entire S$^4$G team for their efforts in this project. EA and AB thank the Centre National d'\'Etudes Spatiales for financial support. KS, J-CM-M, TK, and TM acknowledge support from the National Radio Astronomy Observatory, which is a facility of the National Science Foundation operated under cooperative agreement by Associated Universities, Inc.

We thank our anonymous referee for comments that helped to improve the paper.

This work is based on observations and archival data made with the Spitzer Space Telescope, which is operated by the Jet Propulsion Laboratory, California Institute of Technology under a contract with NASA. We are grateful to the dedicated staff at the Spitzer Science Center for their help and support in planning and execution of this Exploration Science program. We also gratefully acknowledge support from NASA JPL/Spitzer grant RSA 1374189 provided for the S$^4$G project.

Funding for SDSS-III has been provided by the Alfred P.~Sloan Foundation, the Participating Institutions, the National Science Foundation, and the U.S.~Department of Energy Office of Science. The SDSS-III web site is http://www.sdss3.org/.

SDSS-III is managed by the Astrophysical Research Consortium for the Participating Institutions of the SDSS-III Collaboration including the University of Arizona, the Brazilian Participation Group, Brookhaven National Laboratory, University of Cambridge, Carnegie Mellon University, University of Florida, the French Participation Group, the German Participation Group, Harvard University, the Instituto de Astrofisica de Canarias, the Michigan State/Notre Dame/JINA Participation Group, Johns Hopkins University, Lawrence Berkeley National Laboratory, Max Planck Institute for Astrophysics, New Mexico State University, New York University, Ohio State University, Pennsylvania State University, University of Portsmouth, Princeton University, the Spanish Participation Group, University of Tokyo, University of Utah, Vanderbilt University, University of Virginia, University of Washington, and Yale University. 

This research has made use of the NASA/IPAC Extragalactic Database (NED) which is operated by the Jet Propulsion Laboratory, California Institute of Technology, under contract with the National Aeronautics and Space Administration.

\clearpage

\begin{longtable}{l c c c c c c}
\caption{\label{sampledata} Properties of the 70 galaxies in our final sample.}\\
\hline
\hline
Galaxy ID  & $T$ & Distance & $B$  & $v_{\rm c}$ & $r_{25}$ & PA  \\
           &     & (Mpc)    & (mag) & km\,s$^{-1}$ & (\arcsec) & (\deg)  \\
\hline
\endfirsthead
\caption{(Continued)}\\

\hline
\hline
Galaxy ID  & $T$ & Distance & $B$  & $v_{\rm c}$ & $r_{25}$ & PA  \\
           &     & (Mpc)    & (mag) & km\,s$^{-1}$ & (\arcsec) & (\deg)  \\
\hline
\endhead

ESO~157-049&      3.0&      23.9&      13.55&      90.48&      52&     28.1 \\ 
ESO~240-011&      4.8&      42.4&      11.52&      267.46&      16&    127.2\\ 
ESO~292-014&      6.5&      24.9&      12.47&      107.73&      89&     83.9\\ 
ESO~346-001&      5.1&      34.9&      13.29&      47.68&      64&      64.3\\ 
ESO~440-027&      6.6&      21.6&      11.50&      117.53&      10&     79.7\\ 
ESO~443-021&      5.8&      48.1&      13.24&      161.71&      37&    159.3\\ 
ESO~466-014&      3.8&      47.7&      14.23&      121.00&      40&     50.4\\ 
ESO~469-015&      3.3&      27.0&      13.69&      93.01&      56&     149.6\\ 
ESO~533-004&      5.2&      36.2&      13.05&      147.54&      55&    150.5\\ 
ESO~544-027&      3.4&      40.3&      14.20&      92.00&      46&     153.3\\ 
IC~0217&      5.9&      24.2&      13.86&      100.15&      60&         35.7\\ 
IC~0610&      3.9&      20.7&      13.44&      133.84&      57&         29.0\\ 
IC~1197&      6.0&      25.7&      13.14&      87.90&      77&          56.4\\ 
IC~1553&      5.2&      39.9&      13.38&      67.93&      40&          15.0\\ 
IC~1711&      3.0&      52.3&      13.19&      173.41&      75&         43.9\\ 
IC~1913&      3.4&      21.4&      13.38&      77.88&      58&         147.7\\ 
IC~2058&      6.5&      19.4&      12.13&      82.78&      102&         17.4\\ 
IC~2135&      6.0&      29.2&      11.93&      107.08&      97&        108.6\\ 
IC~5176&      4.5&      26.4&      12.02&      164.38&      134&        27.9\\ 
NGC~0489&      2.7&      33.5&      12.42&      180.91&      46&       120.8\\ 
NGC~0522&      4.1&      35.0&      12.28&      169.10&      72&        32.9\\ 
NGC~0678&      3.0&      27.1&      12.15&      169.03&      93&        77.4\\ 
NGC~1032&     0.4&      35.6&      12.13&      283.94&      106&        67.5\\ 
NGC~1163&      4.1&      30.9&      13.40&      142.07&      69&       143.0\\ 
NGC~1422&      2.3&      16.9&      13.22&      65.87&      75&         65.5\\ 
NGC~1495&      5.0&      17.3&      12.19&      90.73&      67&        104.5\\ 
NGC~1596&     -2.0&      15.6&      11.94&      98.32&      117&        18.7\\ 
NGC~2732&     -2.0&      31.8&      12.73&      32.30&      55&         65.7\\ 
NGC~3098&     -1.5&      22.4&      12.68&      129.73&      70&        88.7\\ 
NGC~3279&      6.5&      32.5&      12.37&      155.85&      74&       152.0\\ 
NGC~3454&      5.5&      21.3&      12.44&      91.25&      72&        116.1\\ 
NGC~3501&      5.9&      23.3&      12.09&      133.61&      128&       28.0\\ 
NGC~3592&      5.3&      22.7&      13.44&      79.72&      64&        117.7\\ 
NGC~3600&      1.0&      13.6&      12.78&      86.90&      56&          4.4\\ 
NGC~3628&      3.1&      12.2&      9.150&      215.16&      329&      102.6\\ 
NGC~4081&      1.0&      26.1&      15.12&      116.53&      42 &130.0\\
NGC~4111&     -1.4&      16.0&      11.60&      71.63&      53  &152.2\\
NGC~4330&      6.3&      19.5&      11.99&      117.85&      69 & 59.2\\
NGC~4359&      5.0&      13.3&      13.43&      102.83&      41 &107.7\\
NGC~4437&      6.0&      9.8&      9.620&      140.13&      274 & 82.4\\
NGC~4565&      3.2&      13.3&      8.970&      244.94&      498&135.1\\
NGC~4607&      3.4&      20.0&      12.63&      98.93&      87  &  2.7\\
NGC~4747&      7.2&      12.3&      12.36&      69.60&      63  & 31.9\\
NGC~5084&     -2.0&      26.4&      11.07&      309.88&      300& 81.0\\
NGC~5470&      3.1&      18.0&      13.00&      109.45&      77 & 62.2\\
NGC~5529&      5.3&      50.0&      11.20&      284.14&      173&114.3\\
NGC~5981&      4.3&      29.2&      12.48&      251.12&      83 &139.8\\
NGC~6010&     0.4&      21.6&      12.33&      148.80&      55  &103.3\\
NGC~7347&      4.6&      38.7&      13.25&      115.11&      46 &130.8\\
PGC~013646&      5.1&      32.6&      12.32&      168.16&      102&34.4\\
PGC~028308&      6.8&      40.5&      12.87&      167.27&      60&125.8\\
PGC~030591&      6.8&      29.8&      13.36&      111.59&      45&168.4\\
PGC~032548&    -0.2&      37.3&      15.06&      93.33&      27  &149.7\\
PGC~052809&      5.9&      24.9&      12.46&      100.97&      95&169.2\\
UGC~00903&      3.9&      51.6&      13.09&      163.23&      50 & 52.6\\
UGC~01970&      5.9&      33.9&      13.32&      95.20&      55  & 22.5\\
UGC~05347&      6.5&      33.8&      13.53&      95.64&      41  & 17.0\\
UGC~05689&      6.4&      43.4&      14.07&      119.67&      40 &153.4\\
UGC~05958&      4.0&      29.4&      14.26&      78.15&      49  &178.9\\
UGC~06526&      7.0&      32.3&      12.64&      72.57&      49  & 87.2\\
UGC~07086&      3.1&      32.8&      14.02&      126.01&      69 & 70.5\\
UGC~08737&      4.0&      41.9&      13.27&      163.31&      72 &150.7\\
UGC~09448&      3.2&      37.8&      13.95&      111.50&      49 & 85.8\\
UGC~09665&      4.0&      42.5&      13.57&      126.05&      51 &141.5\\
UGC~10043&      4.1&      37.3&      13.80&      143.59&      66 &150.5\\
UGC~10288&      5.3&      32.4&      12.09&      166.59&      144& 90.4\\
UGC~10297&      5.1&      39.2&      13.43&      102.82&      64 &  2.9\\
UGC~12518&      3.0&      51.6&      14.11&      49.58&      42  & 25.2\\
UGC~12692&      3.9&      49.6&      14.21&      101.34&      38 & 53.2\\
UGC~12857&      4.0&      34.5&      13.24&      107.94&      61 & 33.5\\

\hline
\end{longtable}

Note.~-- $T$ values from HyperLEDA. Distance values from the average of redshift-independent measurements in NED when available and if not from Virgo Infall corrected radial velocities and a Hubble-Lema\^itre constant $H_{0}=73\,{\rm km\,s^{-1}\,Mpc^{-1}}$. $B$ values are HyperLEDA's internal absorption-corrected values. $v_{\rm c}$ values from HyperLEDA. $r_{25}$ values from HyperLEDA except for NGC~4111, for which we used the RC3 value. PA values from HyperLEDA, except for ESO~440-027, IC~5176, and NGC~3098, for which we measured our own PA values using ellipse fitting.

\clearpage

\begin{longtable}{lccccc}
\caption{\label{bigtable1} {Results of the fits to horizontal luminosity profiles for the whole disc}}\\
\hline
\hline
ID  & Type & $I_{0}$ & $r_{\rm i}$ & $r_{\rm o}$ & $h$\\
& &(mag/as$^3$)  &(\arcsec) &(\arcsec) &(\arcsec)\\
\hline
\endfirsthead
\caption{(Continued)}\\
\hline
\hline
ID  & Type & $I_{0}$ & $r_{\rm i}$ & $r_{\rm o}$ & $h$\\
& &(mag/as$^3$)  &(\arcsec) &(\arcsec) &(\arcsec)\\
\hline
\endhead
ESO~157-049&Type~II+III&$24.81\pm0.29$&$4.0\pm0.0$&$24.7\pm3.4$&$24.6\pm12.7$\\
&&&$24.7\pm3.4$&$48.9\pm4.2$&$6.7\pm0.7$\\
&&&$48.9\pm4.2$&$72.4\pm0.0$&$14.7\pm1.9$\\
\hline
ESO~240-011&Type~II&$25.11\pm0.08$&$11.7\pm0.0$&$89.4\pm39.3$&$29.2\pm1.3$\\
&&&$89.4\pm39.3$&$197.8\pm0.0$&$27.0\pm0.7$\\
\hline
ESO~292-014&Type~II&$25.60\pm0.09$&$13.9\pm0.0$&$80.5\pm4.2$&$20.2\pm0.8$\\
&&&$80.5\pm4.2$&$87.9\pm0.0$&$7.1\pm1.9$\\
\hline
ESO~346-001&Type~II+II&$25.43\pm0.12$&$7.6\pm0.0$&$23.5\pm3.2$&$21.3\pm3.2$\\
&&&$23.5\pm3.2$&$53.6\pm4.4$&$11.1\pm0.7$\\
&&&$53.6\pm4.4$&$59.0\pm0.0$&$5.9\pm1.4$\\
\hline
ESO~440-027&Type~II+III&$26.81\pm0.42$&$24.3\pm0.0$&$57.6\pm11.9$&$51.3\pm26.3$\\
&&&$57.6\pm11.9$&$124.8\pm17.5$&$23.8\pm1.9$\\
&&&$124.8\pm17.5$&$172.9\pm0.0$&$34.6\pm4.0$\\
\hline
ESO~443-021&Type~II&$24.73\pm0.22$&$11.7\pm0.0$&$37.7\pm2.8$&$18.0\pm3.0$\\
&&&$37.7\pm2.8$&$66.8\pm0.0$&$8.2\pm0.3$\\
\hline
ESO~466-014&Type~II&$24.65\pm0.17$&$1.5\pm0.0$&$18.8\pm3.5$&$11.4\pm2.0$\\
&&&$18.8\pm3.5$&$47.5\pm0.0$&$6.7\pm0.2$\\
\hline
ESO~469-015&Type~II+III&$24.64\pm0.26$&$1.5\pm0.0$&$9.2\pm1.9$&$114.4\pm382.6$\\
&&&$9.2\pm1.9$&$27.1\pm1.8$&$4.7\pm0.4$\\
&&&$27.1\pm1.8$&$59.0\pm0.0$&$14.9\pm0.9$\\
\hline
ESO~533-004&Type~II&$25.10\pm0.22$&$12.8\pm0.0$&$38.5\pm3.2$&$24.0\pm4.2$\\
&&&$38.5\pm3.2$&$87.9\pm0.0$&$11.1\pm0.2$\\
\hline
ESO~544-027&Type~II&$24.74\pm0.39$&$2.3\pm0.0$&$12.3\pm4.9$&$18.9\pm24.4$\\
&&&$12.3\pm4.9$&$49.6\pm0.0$&$7.4\pm0.2$\\
\hline
IC~0217&Type~II&$26.06\pm0.22$&$7.6\pm0.0$&$25.3\pm4.1$&$33.9\pm14.6$\\
&&&$25.3\pm4.1$&$91.3\pm0.0$&$15.1\pm0.2$\\
\hline
IC~0610&Type~II+II&$25.24\pm0.18$&$11.7\pm0.0$&$25.1\pm2.7$&$48.0\pm19.8$\\
&&&$25.1\pm2.7$&$48.2\pm7.6$&$10.8\pm1.2$\\
&&&$48.2\pm7.6$&$69.5\pm0.0$&$8.2\pm0.5$\\
\hline
IC~1197&Type~II&$26.72\pm0.22$&$9.6\pm0.0$&$41.8\pm4.5$&$38.5\pm13.4$\\
&&&$41.8\pm4.5$&$98.4\pm0.0$&$15.3\pm0.4$\\
\hline
IC~1553&Type~II+III&$24.26\pm0.07$&$7.6\pm0.0$&$29.4\pm0.2$&$11.7\pm0.3$\\
&&&$29.4\pm0.2$&$32.8\pm0.2$&$1.6\pm0.1$\\
&&&$32.8\pm0.2$&$49.6\pm0.0$&$10.8\pm0.8$\\
\hline
\\
IC~1711&Type~II&$26.76\pm0.13$&$28.9\pm0.0$&$69.7\pm3.9$&$25.9\pm1.6$\\
&&&$69.7\pm3.9$&$87.9\pm0.0$&$12.6\pm1.2$\\
\hline
IC~1913&Type~II&$27.36\pm0.20$&$17.6\pm0.0$&$34.2\pm2.9$&$43.6\pm13.6$\\
&&&$34.2\pm2.9$&$69.5\pm0.0$&$12.0\pm0.5$\\
\hline
IC~2058&Type~II&$26.29\pm0.06$&$1.5\pm0.0$&$90.5\pm6.9$&$26.1\pm0.9$\\
&&&$90.5\pm6.9$&$118.0\pm0.0$&$15.7\pm1.5$\\
\hline
IC~2135&Type~II&$26.10\pm0.18$&$13.9\pm0.0$&$68.2\pm4.4$&$41.0\pm7.1$\\
&&&$68.2\pm4.4$&$122.3\pm0.0$&$15.4\pm0.6$\\
\hline
IC~5176&Type~III&$24.22\pm0.06$&$4.9\pm0.0$&$75.0\pm3.2$&$16.2\pm0.4$\\
&&&$75.0\pm3.2$&$172.9\pm0.0$&$32.0\pm0.8$\\
\hline
NGC~0489&Type~I&$23.01\pm0.11$&$4.9\pm0.0$&$56.6\pm0.0$&$7.6\pm0.2$\\
\hline
NGC~0522&Type~II+III&$25.75\pm0.10$&$12.8\pm0.0$&$54.4\pm1.3$&$44.5\pm5.2$\\
&&&$54.4\pm1.3$&$102.2\pm6.0$&$8.9\pm0.4$\\
&&&$102.2\pm6.0$&$98.4\pm0.0$&$94.5\pm88.5$\\
\hline
NGC~0678&Type~II&$31.81\pm0.38$&$49.6\pm0.0$&$118.0\pm2.1$&$-43.6\pm8.4$\\
&&&$118.0\pm2.1$&$140.8\pm0.0$&$12.0\pm1.0$\\
\hline
NGC 1032&Type~II&$27.36\pm0.04$&$47.5\pm0.0$&$131.7\pm11.1$&$39.6\pm0.6$\\
&&&$131.7\pm11.1$&$185.0\pm0.0$&$31.8\pm1.4$\\
\hline
NGC~1163&Type~II+III&$24.50\pm0.33$&$5.7\pm0.0$&$14.4\pm7.8$&$14.4\pm7.8$\\
&&&$14.4\pm7.8$&$42.2\pm3.3$&$9.1\pm0.8$\\
&&&$42.2\pm3.3$&$98.4\pm0.0$&$18.9\pm0.7$\\
\hline
NGC~1422&Type~III&$25.48\pm0.09$&$4.0\pm0.0$&$49.1\pm3.2$&$13.1\pm0.6$\\
&&&$49.1\pm3.2$&$118.0\pm0.0$&$30.2\pm1.2$\\
\hline
NGC~1495&Type~II&$25.35\pm0.13$&$27.3\pm0.0$&$93.1\pm7.5$&$21.2\pm0.9$\\
&&&$93.1\pm7.5$&$113.8\pm0.0$&$12.9\pm1.8$\\
\hline
NGC 1596&Type~III&$26.21\pm0.10$&$27.3\pm0.0$&$152.5\pm7.9$&$27.6\pm0.9$\\
&&&$152.5\pm7.9$&$241.0\pm0.0$&$107.1\pm17.6$\\
\hline
NGC 2732&Type~II&$24.33\pm0.03$&$10.6\pm0.0$&$49.9\pm3.0$&$13.1\pm0.2$\\
&&&$49.9\pm3.0$&$78.3\pm0.0$&$10.5\pm0.2$\\
\hline
NGC 3098&Type~III&$23.94\pm0.13$&$27.3\pm0.0$&$75.9\pm8.5$&$11.8\pm0.4$\\
&&&$75.9\pm8.5$&$91.3\pm0.0$&$18.1\pm3.6$\\
\hline
NGC~3279&Type~II&$25.00\pm0.11$&$0.0\pm0.0$&$35.4\pm2.3$&$35.2\pm6.5$\\
&&&$35.4\pm2.3$&$94.8\pm0.0$&$12.6\pm0.2$\\
\hline
NGC~3454&Type~II+III&$26.42\pm0.13$&$4.9\pm0.0$&$39.4\pm1.9$&$58.6\pm18.9$\\
&&&$39.4\pm1.9$&$47.3\pm2.4$&$5.0\pm1.7$\\
&&&$47.3\pm2.4$&$94.8\pm0.0$&$18.5\pm0.7$\\
\hline
NGC~3501&Type~II&$25.15\pm0.12$&$27.3\pm0.0$&$96.0\pm6.3$&$24.6\pm1.2$\\
&&&$96.0\pm6.3$&$135.9\pm0.0$&$15.9\pm0.8$\\
\hline
NGC~3592&Type~II&$25.57\pm0.12$&$11.7\pm0.0$&$59.1\pm6.9$&$16.3\pm0.9$\\
&&&$59.1\pm6.9$&$75.3\pm0.0$&$10.5\pm1.3$\\
\hline
NGC~3600&Type~III&$26.33\pm0.00$&$24.3\pm0.0$&$71.1\pm0.0$&$17.6\pm0.0$\\
&&&$71.1\pm0.0$&$118.0\pm0.0$&$43.9\pm0.0$\\
\hline
NGC~3628&Type~II+III+II&$26.70\pm0.08$&$59.0\pm0.0$&$167.9\pm3.1$&$94.6\pm5.8$\\
&&&$167.9\pm3.1$&$187.0\pm2.7$&$18.5\pm2.6$\\
&&&$187.0\pm2.7$&$356.0\pm6.8$&$475.9\pm184.4$\\
&&&$356.0\pm6.8$&$532.0\pm0.0$&$55.2\pm1.6$\\
\hline
NGC~4081&Type~II+III&$23.87\pm0.05$&$0.0\pm0.0$&$29.6\pm9.2$&$9.4\pm0.3$\\
&&&$29.6\pm9.2$&$40.6\pm4.2$&$7.6\pm2.0$\\
&&&$40.6\pm4.2$&$64.1\pm0.0$&$13.8\pm0.6$\\
\hline
NGC 4111&Type~II&$25.84\pm0.12$&$32.2\pm0.0$&$102.7\pm13.1$&$31.4\pm1.7$\\
&&&$102.7\pm13.1$&$145.8\pm0.0$&$24.0\pm1.4$\\
\hline
NGC~4330&Type~II+III+II&$26.77\pm0.05$&$15.1\pm0.0$&$56.0\pm2.0$&$53.3\pm3.6$\\
&&&$56.0\pm2.0$&$80.6\pm1.2$&$18.3\pm1.3$\\
&&&$80.6\pm1.2$&$90.0\pm1.4$&$-29.4\pm10.7$\\
&&&$90.0\pm1.4$&$172.9\pm0.0$&$26.7\pm0.4$\\
\hline
NGC~4359&Type~III&$26.47\pm0.06$&$0.0\pm0.0$&$77.2\pm6.4$&$20.4\pm0.7$\\
&&&$77.2\pm6.4$&$140.8\pm0.0$&$35.5\pm2.1$\\
\hline
NGC~4437&Type~II&$27.26\pm0.07$&$39.4\pm0.0$&$207.5\pm7.1$&$104.5\pm5.8$\\
&&&$207.5\pm7.1$&$389.5\pm0.0$&$50.2\pm0.8$\\
\hline
NGC~4565&Type~II+II&$27.13\pm0.19$&$98.4\pm0.0$&$210.0\pm20.2$&$133.1\pm20.6$\\
&&&$210.0\pm20.2$&$431.4\pm12.3$&$80.4\pm2.4$\\
&&&$431.4\pm12.3$&$484.7\pm0.0$&$43.8\pm3.7$\\
\hline
NGC~4607&Type~II+III&$25.26\pm0.33$&$17.6\pm0.0$&$38.4\pm5.6$&$26.2\pm8.3$\\
&&&$38.4\pm5.6$&$58.9\pm4.1$&$11.1\pm2.1$\\
&&&$58.9\pm4.1$&$135.9\pm0.0$&$24.3\pm0.5$\\
\hline
NGC~4747&Type~I&$26.59\pm0.05$&$7.6\pm0.0$&$126.7\pm0.0$&$27.5\pm0.4$\\
\hline
NGC 5084&Type~III&$26.65\pm0.15$&$37.5\pm0.0$&$170.9\pm12.0$&$42.0\pm2.8$\\
&&&$170.9\pm12.0$&$427.9\pm0.0$&$112.0\pm5.8$\\
\hline
NGC~5470&Type~II&$25.62\pm0.07$&$15.1\pm0.0$&$58.6\pm1.6$&$26.8\pm1.3$\\
&&&$58.6\pm1.6$&$91.3\pm0.0$&$11.0\pm0.3$\\
\hline
\\
\\
NGC~5529&Type~II+III+II&$25.77\pm0.03$&$13.9\pm0.0$&$92.5\pm2.0$&$34.0\pm0.5$\\
&&&$92.5\pm2.0$&$139.2\pm1.2$&$18.7\pm0.4$\\
&&&$139.2\pm1.2$&$181.7\pm3.4$&$\infty\pm\infty$\\
&&&$181.7\pm3.4$&$204.5\pm0.0$&$33.1\pm3.6$\\
\hline
NGC~5981&Type~II&$24.78\pm0.22$&$20.2\pm0.0$&$47.4\pm5.6$&$21.8\pm3.2$\\
&&&$47.4\pm5.6$&$102.1\pm0.0$&$14.4\pm0.3$\\
\hline
NGC 6010&Type~II&$25.19\pm0.14$&$20.2\pm0.0$&$36.0\pm4.2$&$20.9\pm2.1$\\
&&&$36.0\pm4.2$&$94.8\pm0.0$&$13.4\pm0.3$\\
\hline
NGC~7347&Type~II&$24.02\pm0.02$&$13.9\pm0.0$&$43.3\pm0.5$&$11.0\pm0.1$\\
&&&$43.3\pm0.5$&$56.6\pm0.0$&$5.8\pm0.1$\\
\hline
PGC~013646&Type~II&$26.24\pm0.01$&$13.9\pm0.0$&$52.5\pm0.2$&$116.5\pm2.2$\\
&&&$52.5\pm0.2$&$122.3\pm0.0$&$14.1\pm0.0$\\
\hline
PGC~028308&Type~II&$23.92\pm0.09$&$4.9\pm0.0$&$59.7\pm10.6$&$10.2\pm0.3$\\
&&&$59.7\pm10.6$&$64.1\pm0.0$&$7.2\pm3.0$\\
\hline
PGC~030591&Type~I&$24.01\pm0.15$&$13.9\pm0.0$&$54.2\pm0.0$&$8.8\pm0.3$\\
\hline
PGC 032548&Type~I&$25.77\pm0.00$&$5.7\pm0.0$&$54.2\pm0.0$&$10.5\pm0.0$\\
\hline
PGC~052809&Type~II&$26.07\pm0.12$&$11.7\pm0.0$&$72.5\pm3.9$&$34.7\pm3.3$\\
&&&$72.5\pm3.9$&$126.7\pm0.0$&$15.2\pm0.5$\\
\hline
UGC~00903&Type~II+III&$23.19\pm0.14$&$4.9\pm0.0$&$15.3\pm1.9$&$12.8\pm2.0$\\
&&&$15.3\pm1.9$&$30.7\pm2.1$&$5.8\pm0.5$\\
&&&$30.7\pm2.1$&$64.1\pm0.0$&$10.8\pm0.4$\\
\hline
UGC~01970&Type~III+II&$24.70\pm0.22$&$5.7\pm0.0$&$20.4\pm3.7$&$7.6\pm1.1$\\
&&&$20.4\pm3.7$&$34.0\pm6.9$&$30.8\pm26.6$\\
&&&$34.0\pm6.9$&$78.3\pm0.0$&$13.6\pm0.4$\\
\hline
UGC~05347&Type~II&$25.88\pm0.32$&$6.7\pm0.0$&$26.5\pm3.8$&$18.0\pm6.7$\\
&&&$26.5\pm3.8$&$47.5\pm0.0$&$7.0\pm0.5$\\
\hline
UGC~05689&Type~II&$24.81\pm0.08$&$4.0\pm0.0$&$20.3\pm5.3$&$9.7\pm0.5$\\
&&&$20.3\pm5.3$&$43.3\pm0.0$&$8.0\pm0.3$\\
\hline
UGC~05958&Type~I&$25.12\pm0.15$&$7.6\pm0.0$&$64.1\pm0.0$&$10.8\pm0.4$\\
\hline
UGC~06526&Type~III&$24.42\pm0.20$&$6.7\pm0.0$&$34.5\pm4.0$&$8.8\pm0.8$\\
&&&$34.5\pm4.0$&$87.9\pm0.0$&$18.3\pm0.9$\\
\hline
UGC~07086&Type~II+III&$24.58\pm0.10$&$0.0\pm0.0$&$29.0\pm9.6$&$13.1\pm1.1$\\
&&&$29.0\pm9.6$&$45.6\pm5.3$&$9.4\pm2.2$\\
&&&$45.6\pm5.3$&$87.9\pm0.0$&$17.9\pm0.7$\\
\hline
\\
\\
UGC~08737&Type~II+III&$24.75\pm0.28$&$15.1\pm0.0$&$28.8\pm3.9$&$23.5\pm7.3$\\
&&&$28.8\pm3.9$&$81.8\pm3.0$&$11.9\pm0.4$\\
&&&$81.8\pm3.0$&$122.3\pm0.0$&$31.2\pm2.7$\\
\hline
UGC~09448&Type~I&$24.75\pm0.11$&$8.6\pm0.0$&$59.0\pm0.0$&$9.6\pm0.3$\\
\hline
UGC~09665&Type~II&$24.28\pm0.34$&$4.9\pm0.0$&$11.3\pm9.5$&$15.8\pm9.7$\\
&&&$11.3\pm9.5$&$59.0\pm0.0$&$9.6\pm0.4$\\
\hline
UGC~10043&Type~II&$25.49\pm0.16$&$18.8\pm0.0$&$78.8\pm10.2$&$17.2\pm1.0$\\
&&&$78.8\pm10.2$&$81.4\pm0.0$&$4.6\pm6.2$\\
\hline
UGC~10288&Type~II+III&$26.05\pm0.18$&$13.9\pm0.0$&$53.3\pm8.2$&$38.7\pm7.7$\\
&&&$53.3\pm8.2$&$127.8\pm7.7$&$22.0\pm1.1$\\
&&&$127.8\pm7.7$&$191.3\pm0.0$&$41.8\pm3.2$\\
\hline
UGC~10297&Type~II+III&$25.34\pm0.17$&$0.0\pm0.0$&$11.3\pm2.9$&$24.3\pm14.6$\\
&&&$11.3\pm2.9$&$39.4\pm2.9$&$8.3\pm0.4$\\
&&&$39.4\pm2.9$&$64.1\pm0.0$&$15.0\pm0.9$\\
\hline
UGC~12518&Type~II+III&$25.29\pm0.26$&$7.6\pm0.0$&$19.5\pm2.2$&$29.9\pm17.0$\\
&&&$19.5\pm2.2$&$55.0\pm3.2$&$8.2\pm0.3$\\
&&&$55.0\pm3.2$&$72.4\pm0.0$&$18.3\pm2.7$\\
\hline
UGC~12692&Type~III&$23.58\pm0.21$&$7.6\pm0.0$&$37.3\pm4.4$&$6.1\pm0.4$\\
&&&$37.3\pm4.4$&$54.2\pm0.0$&$14.9\pm3.7$\\
\hline
UGC~12857&Type~I&$24.62\pm0.11$&$9.6\pm0.0$&$66.8\pm0.0$&$10.9\pm0.3$\\
\hline
\end{longtable}
Note.~-- $r_{\rm i}$ and $r_{\rm o}$ stand for the inner and the outer limit of each fitted section and $h$ corresponds to the scale length of that section. The lower limit of the innermost section and the upper limit of the outermost section have no error bar because they have been set manually.

\begin{longtable}{lccccc}
\caption{\label{bigtable2} {Results of the fits to horizontal luminosity profiles for the thin disc}}\\
\hline
\hline
ID  & Type & $I_{0}$ & $r_{\rm i}$ & $r_{\rm o}$ & $h$\\
& &(mag/as$^3$)  &(\arcsec) &(\arcsec) &(\arcsec)\\
\hline
\endfirsthead
\caption {(Continued)}\\
\hline
\hline
ID  & Type & $I_{0}$ & $r_{\rm i}$ & $r_{\rm o}$ & $h$\\
& &(mag/as$^3$)  &(\arcsec) &(\arcsec) &(\arcsec)\\
\hline
\endhead
ESO~157-049&Type~II+III&$23.14\pm0.42$&$4.0\pm0.0$&$23.3\pm3.7$&$26.3\pm27.1$\\
&&&$23.3\pm3.7$&$52.9\pm5.2$&$5.9\pm0.5$\\
&&&$52.9\pm5.2$&$69.5\pm0.0$&$15.0\pm5.1$\\
\hline
ESO~240-011&Type~II&$23.86\pm0.11$&$11.7\pm0.0$&$100.5\pm14.8$&$29.6\pm1.7$\\
&&&$100.5\pm14.8$&$197.8\pm0.0$&$23.2\pm0.7$\\
\hline
\\
ESO~292-014&Type~II&$23.84\pm0.09$&$13.9\pm0.0$&$76.8\pm3.0$&$17.3\pm0.5$\\
&&&$76.8\pm3.0$&$87.9\pm0.0$&$7.0\pm1.0$\\
\hline
ESO~346-001&Type~II+II&$24.20\pm0.15$&$7.6\pm0.0$&$23.0\pm3.3$&$22.4\pm4.4$\\
&&&$23.0\pm3.3$&$52.5\pm3.5$&$10.5\pm0.7$\\
&&&$52.5\pm3.5$&$59.0\pm0.0$&$4.9\pm1.0$\\
\hline
ESO~440-027&Type~II+III&$25.20\pm0.43$&$24.3\pm0.0$&$64.5\pm9.4$&$52.9\pm29.8$\\
&&&$64.5\pm9.4$&$142.3\pm24.9$&$19.0\pm1.5$\\
&&&$142.3\pm24.9$&$172.9\pm0.0$&$29.3\pm8.9$\\
\hline
ESO~443-021&Type~II&$23.41\pm0.25$&$11.7\pm0.0$&$36.4\pm2.6$&$18.5\pm3.5$\\
&&&$36.4\pm2.6$&$66.8\pm0.0$&$7.6\pm0.3$\\
\hline
ESO~466-014&Type~II&$23.23\pm0.23$&$1.5\pm0.0$&$19.5\pm3.5$&$11.3\pm2.7$\\
&&&$19.5\pm3.5$&$47.5\pm0.0$&$5.8\pm0.2$\\
\hline
ESO~469-015&Type~II+III&$22.90\pm0.07$&$1.5\pm0.0$&$10.2\pm0.3$&$100.2\pm70.1$\\
&&&$10.2\pm0.3$&$25.6\pm0.5$&$3.8\pm0.1$\\
&&&$25.6\pm0.5$&$59.0\pm0.0$&$13.2\pm0.5$\\
\hline
ESO~533-004&Type~II&$23.29\pm0.32$&$12.8\pm0.0$&$40.6\pm3.9$&$21.5\pm5.7$\\
&&&$40.6\pm3.9$&$87.9\pm0.0$&$9.2\pm0.3$\\
\hline
ESO~544-027&Type~II&$22.94\pm0.17$&$2.3\pm0.0$&$14.5\pm2.1$&$15.8\pm4.5$\\
&&&$14.5\pm2.1$&$49.6\pm0.0$&$6.3\pm0.2$\\
\hline
IC~0217&Type~II&$24.12\pm0.37$&$7.6\pm0.0$&$34.1\pm8.9$&$21.3\pm8.1$\\
&&&$34.1\pm8.9$&$91.3\pm0.0$&$11.8\pm0.5$\\
\hline
IC~0610&Type~II+II&$23.54\pm0.16$&$11.7\pm0.0$&$24.9\pm1.8$&$67.8\pm34.7$\\
&&&$24.9\pm1.8$&$50.7\pm2.6$&$9.7\pm0.7$\\
&&&$50.7\pm2.6$&$69.5\pm0.0$&$5.7\pm0.3$\\
\hline
IC~1197&Type~II&$25.22\pm0.31$&$9.6\pm0.0$&$44.2\pm5.1$&$39.9\pm16.6$\\
&&&$44.2\pm5.1$&$98.4\pm0.0$&$12.4\pm0.5$\\
\hline
IC~1553&Type~II+III&$22.69\pm0.09$&$7.6\pm0.0$&$29.5\pm0.2$&$11.5\pm0.5$\\
&&&$29.5\pm0.2$&$32.2\pm0.2$&$1.1\pm0.1$\\
&&&$32.2\pm0.2$&$47.5\pm0.0$&$8.2\pm0.8$\\
\hline
IC~1711&Type~II&$25.18\pm0.26$&$28.9\pm0.0$&$79.8\pm3.1$&$23.4\pm2.7$\\
&&&$79.8\pm3.1$&$91.3\pm0.0$&$5.7\pm1.0$\\
\hline
IC~1913&Type~II&$26.46\pm0.29$&$17.6\pm0.0$&$35.5\pm2.9$&$49.7\pm25.8$\\
&&&$35.5\pm2.9$&$69.5\pm0.0$&$9.2\pm0.5$\\
\hline
IC~2058&Type~II&$24.91\pm0.39$&$1.5\pm0.0$&$37.9\pm22.5$&$29.4\pm15.9$\\
&&&$37.9\pm22.5$&$118.0\pm0.0$&$18.1\pm1.1$\\
\hline
\\
IC~2135&Type~II&$24.17\pm0.23$&$13.9\pm0.0$&$68.5\pm4.0$&$35.2\pm6.7$\\
&&&$68.5\pm4.0$&$122.3\pm0.0$&$11.6\pm0.5$\\
\hline
IC~5176&Type~III&$22.18\pm0.10$&$4.9\pm0.0$&$71.7\pm5.6$&$15.0\pm0.6$\\
&&&$71.7\pm5.6$&$172.9\pm0.0$&$25.8\pm0.9$\\
\hline
NGC~0489&Type~I&$21.42\pm0.11$&$4.9\pm0.0$&$56.6\pm0.0$&$6.9\pm0.1$\\
\hline
NGC~0522&Type~II+III&$24.05\pm0.23$&$12.8\pm0.0$&$54.9\pm2.1$&$43.0\pm12.1$\\
&&&$54.9\pm2.1$&$88.2\pm5.0$&$7.0\pm0.5$\\
&&&$88.2\pm5.0$&$105.9\pm0.0$&$16.5\pm4.7$\\
\hline
NGC~0678&Type~II&$31.12\pm0.55$&$49.6\pm0.0$&$120.6\pm2.0$&$-36.3\pm8.7$\\
&&&$120.6\pm2.0$&$150.9\pm0.0$&$8.9\pm0.6$\\
\hline
NGC 1032&Type~II&$25.50\pm0.20$&$47.5\pm0.0$&$135.1\pm21.8$&$32.3\pm2.3$\\
&&&$135.1\pm21.8$&$191.3\pm0.0$&$24.9\pm1.9$\\
\hline
NGC~1163&Type~II+III&$22.88\pm0.22$&$5.7\pm0.0$&$16.8\pm5.2$&$13.8\pm3.3$\\
&&&$16.8\pm5.2$&$40.3\pm3.9$&$8.1\pm0.9$\\
&&&$40.3\pm3.9$&$105.9\pm0.0$&$15.8\pm0.7$\\
\hline
NGC~1422&Type~III&$23.43\pm0.16$&$4.0\pm0.0$&$51.2\pm4.5$&$10.7\pm0.7$\\
&&&$51.2\pm4.5$&$126.7\pm0.0$&$25.3\pm1.6$\\
\hline
NGC~1495&Type~II&$23.72\pm0.23$&$27.3\pm0.0$&$95.2\pm5.6$&$19.1\pm1.4$\\
&&&$95.2\pm5.6$&$113.8\pm0.0$&$8.0\pm1.3$\\
\hline
NGC 1596&Type~III&$23.78\pm0.11$&$27.3\pm0.0$&$143.8\pm6.3$&$21.2\pm0.6$\\
&&&$143.8\pm6.3$&$241.0\pm0.0$&$57.4\pm5.3$\\
\hline
NGC 2732&Type~II&$22.77\pm0.23$&$10.6\pm0.0$&$28.7\pm6.0$&$14.2\pm2.3$\\
&&&$28.7\pm6.0$&$84.6\pm0.0$&$9.5\pm0.3$\\
\hline
NGC 3098&Type~I&$21.66\pm0.18$&$27.3\pm0.0$&$91.3\pm0.0$&$10.2\pm0.3$\\
\hline
NGC~3279&Type~II&$23.38\pm0.12$&$0.0\pm0.0$&$36.3\pm2.0$&$34.5\pm6.5$\\
&&&$36.3\pm2.0$&$94.8\pm0.0$&$10.8\pm0.2$\\
\hline
NGC~3454&Type~II+III&$24.68\pm0.20$&$4.9\pm0.0$&$39.2\pm2.6$&$48.7\pm18.4$\\
&&&$39.2\pm2.6$&$50.6\pm3.4$&$5.2\pm1.7$\\
&&&$50.6\pm3.4$&$98.4\pm0.0$&$15.6\pm0.8$\\
\hline
NGC~3501&Type~II&$23.64\pm0.22$&$27.3\pm0.0$&$91.5\pm8.2$&$22.6\pm1.9$\\
&&&$91.5\pm8.2$&$145.8\pm0.0$&$13.9\pm0.7$\\
\hline
NGC~3592&Type~II&$23.93\pm0.16$&$11.7\pm0.0$&$59.7\pm4.8$&$14.3\pm0.9$\\
&&&$59.7\pm4.8$&$78.3\pm0.0$&$7.7\pm0.8$\\
\hline
NGC~3600&Type~I&$25.51\pm0.09$&$24.3\pm0.0$&$118.0\pm0.0$&$21.0\pm0.4$\\
\hline
\\
\\
NGC~3628&Type~II+III+II&$24.36\pm0.14$&$59.0\pm0.0$&$170.0\pm4.2$&$78.6\pm6.9$\\
&&&$170.0\pm4.2$&$185.5\pm4.0$&$14.3\pm3.6$\\
&&&$185.5\pm4.0$&$350.8\pm12.1$&$143.9\pm26.9$\\
&&&$350.8\pm12.1$&$532.0\pm0.0$&$45.7\pm1.9$\\
\hline
NGC~4081&Type~II+III&$21.76\pm0.09$&$0.0\pm0.0$&$23.1\pm5.3$&$8.3\pm0.5$\\
&&&$23.1\pm5.3$&$51.4\pm3.2$&$6.6\pm0.3$\\
&&&$51.4\pm3.2$&$64.1\pm0.0$&$15.9\pm3.2$\\
\hline
NGC 4111&Type~II&$24.32\pm0.13$&$32.2\pm0.0$&$100.6\pm7.2$&$28.7\pm1.6$\\
&&&$100.6\pm7.2$&$167.2\pm0.0$&$19.5\pm0.6$\\
\hline
NGC~4330&Type~II+III+II&$25.04\pm0.04$&$15.1\pm0.0$&$55.6\pm1.0$&$53.5\pm2.7$\\
&&&$55.6\pm1.0$&$72.9\pm0.9$&$11.8\pm0.6$\\
&&&$72.9\pm0.9$&$85.4\pm1.6$&$\infty\pm\infty$\\
&&&$85.4\pm1.6$&$172.9\pm0.0$&$22.5\pm0.4$\\
\hline
NGC~4359&Type~III&$24.55\pm0.10$&$0.0\pm0.0$&$70.7\pm9.6$&$17.0\pm0.8$\\
&&&$70.7\pm9.6$&$140.8\pm0.0$&$25.8\pm1.6$\\
\hline
NGC~4437&Type~II&$25.88\pm0.03$&$39.4\pm0.0$&$208.8\pm2.9$&$101.2\pm1.7$\\
&&&$208.8\pm2.9$&$401.9\pm0.0$&$42.9\pm0.5$\\
\hline
NGC~4565&Type~II+II&$25.53\pm0.24$&$98.4\pm0.0$&$211.9\pm21.6$&$133.7\pm28.8$\\
&&&$211.9\pm21.6$&$436.3\pm5.7$&$75.5\pm2.5$\\
&&&$436.3\pm5.7$&$500.0\pm0.0$&$27.7\pm1.5$\\
\hline
NGC~4607&Type~II+III&$23.16\pm0.18$&$17.6\pm0.0$&$39.0\pm2.7$&$22.1\pm2.8$\\
&&&$39.0\pm2.7$&$60.7\pm2.9$&$8.5\pm0.9$\\
&&&$60.7\pm2.9$&$135.9\pm0.0$&$20.4\pm0.7$\\
\hline
NGC~4747&Type~I&$24.22\pm0.07$&$7.6\pm0.0$&$126.7\pm0.0$&$19.4\pm0.4$\\
\hline
NGC 5084&Type~III&$23.82\pm0.17$&$37.5\pm0.0$&$146.4\pm8.6$&$30.4\pm1.8$\\
&&&$146.4\pm8.6$&$414.7\pm0.0$&$72.4\pm2.3$\\
\hline
NGC~5470&Type~II&$23.77\pm0.27$&$15.1\pm0.0$&$60.0\pm3.9$&$24.3\pm4.4$\\
&&&$60.0\pm3.9$&$98.4\pm0.0$&$8.8\pm0.5$\\
\hline
NGC~5529&Type~II+III+II&$24.39\pm0.12$&$13.9\pm0.0$&$95.8\pm5.7$&$35.3\pm2.5$\\
&&&$95.8\pm5.7$&$119.3\pm7.0$&$10.8\pm3.3$\\
&&&$119.3\pm7.0$&$194.8\pm28.1$&$40.2\pm6.3$\\
&&&$194.8\pm28.1$&$204.5\pm0.0$&$19.5\pm16.7$\\
\hline
NGC~5981&Type~II&$23.17\pm0.13$&$20.2\pm0.0$&$52.0\pm2.1$&$20.7\pm1.4$\\
&&&$52.0\pm2.1$&$102.1\pm0.0$&$11.5\pm0.2$\\
\hline
NGC 6010&Type~II&$23.78\pm0.40$&$20.2\pm0.0$&$40.2\pm3.9$&$22.9\pm8.1$\\
&&&$40.2\pm3.9$&$87.9\pm0.0$&$10.7\pm0.2$\\
\hline
NGC~7347&Type~II&$22.79\pm0.00$&$13.9\pm0.0$&$41.3\pm0.0$&$11.0\pm0.0$\\
&&&$41.3\pm0.0$&$59.0\pm0.0$&$4.9\pm0.0$\\
\hline
PGC~013646&Type~II&$24.73\pm0.00$&$13.9\pm0.0$&$57.4\pm0.0$&$84.4\pm0.1$\\
&&&$57.4\pm0.0$&$122.3\pm0.0$&$11.5\pm0.0$\\
\hline
PGC~028308&Type~II&$22.04\pm0.09$&$4.9\pm0.0$&$59.9\pm8.5$&$8.9\pm0.2$\\
&&&$59.9\pm8.5$&$64.1\pm0.0$&$4.9\pm3.0$\\
\hline
PGC~030591&Type~I&$22.25\pm0.21$&$16.3\pm0.0$&$61.5\pm0.0$&$7.6\pm0.3$\\
\hline
PGC 032548&Type~I&$24.10\pm0.08$&$5.7\pm0.0$&$59.0\pm0.0$&$9.2\pm0.2$\\
\hline
PGC~052809&Type~II&$24.16\pm0.19$&$11.7\pm0.0$&$70.0\pm5.4$&$28.4\pm3.4$\\
&&&$70.0\pm5.4$&$126.7\pm0.0$&$13.2\pm0.6$\\
\hline
UGC~00903&Type~II+III&$21.14\pm0.19$&$4.9\pm0.0$&$18.4\pm1.7$&$12.7\pm2.8$\\
&&&$18.4\pm1.7$&$22.7\pm1.8$&$3.1\pm1.2$\\
&&&$22.7\pm1.8$&$64.1\pm0.0$&$8.0\pm0.1$\\
\hline
UGC~01970&Type~III+II&$22.91\pm0.10$&$5.7\pm0.0$&$16.5\pm1.1$&$6.1\pm0.2$\\
&&&$16.5\pm1.1$&$47.7\pm3.4$&$17.8\pm1.4$\\
&&&$47.7\pm3.4$&$78.3\pm0.0$&$9.9\pm0.5$\\
\hline
UGC~05347&Type~II&$24.70\pm0.33$&$6.7\pm0.0$&$30.1\pm4.2$&$14.7\pm4.2$\\
&&&$30.1\pm4.2$&$49.6\pm0.0$&$6.0\pm0.5$\\
\hline
UGC~05689&Type~II&$23.15\pm0.33$&$4.0\pm0.0$&$22.2\pm9.9$&$9.6\pm2.6$\\
&&&$22.2\pm9.9$&$43.3\pm0.0$&$6.8\pm0.6$\\
\hline
UGC~05958&Type~I&$23.17\pm0.12$&$7.6\pm0.0$&$59.0\pm0.0$&$8.5\pm0.2$\\
\hline
UGC~06526&Type~III&$22.81\pm0.11$&$6.7\pm0.0$&$40.4\pm3.2$&$8.3\pm0.3$\\
&&&$40.4\pm3.2$&$87.9\pm0.0$&$15.2\pm1.0$\\
\hline
UGC~07086&Type~II+III&$22.58\pm0.09$&$0.0\pm0.0$&$31.7\pm2.2$&$12.8\pm0.9$\\
&&&$31.7\pm2.2$&$36.7\pm2.2$&$5.2\pm1.6$\\
&&&$36.7\pm2.2$&$87.9\pm0.0$&$13.2\pm0.4$\\
\hline
UGC~08737&Type~II+III&$22.61\pm0.21$&$15.1\pm0.0$&$32.4\pm3.2$&$19.3\pm3.1$\\
&&&$32.4\pm3.2$&$95.8\pm5.0$&$9.3\pm0.3$\\
&&&$95.8\pm5.0$&$118.0\pm0.0$&$56.9\pm31.6$\\
\hline
UGC~09448&Type~I&$22.94\pm0.07$&$8.6\pm0.0$&$47.5\pm0.0$&$8.4\pm0.2$\\
\hline
UGC~09665&Type~II&$22.95\pm0.34$&$4.9\pm0.0$&$13.7\pm5.3$&$19.8\pm13.1$\\
&&&$13.7\pm5.3$&$64.1\pm0.0$&$8.5\pm0.3$\\
\hline
UGC~10043&Type~II&$23.78\pm0.19$&$18.8\pm0.0$&$69.7\pm4.4$&$15.5\pm1.1$\\
&&&$69.7\pm4.4$&$81.4\pm0.0$&$6.6\pm1.2$\\
\hline
\\
\\
UGC~10288&Type~II+III&$24.27\pm0.31$&$13.9\pm0.0$&$55.7\pm13.2$&$38.9\pm12.7$\\
&&&$55.7\pm13.2$&$105.5\pm25.0$&$19.2\pm3.4$\\
&&&$105.5\pm25.0$&$197.8\pm0.0$&$24.5\pm1.2$\\
\hline
UGC~10297&Type~II+III&$24.37\pm0.12$&$0.0\pm0.0$&$12.8\pm2.3$&$18.6\pm4.9$\\
&&&$12.8\pm2.3$&$37.4\pm3.4$&$7.8\pm0.5$\\
&&&$37.4\pm3.4$&$64.1\pm0.0$&$13.2\pm0.9$\\
\hline
UGC~12518&Type~II+III&$23.09\pm0.42$&$7.6\pm0.0$&$27.2\pm4.9$&$14.2\pm5.7$\\
&&&$27.2\pm4.9$&$51.3\pm5.5$&$5.9\pm0.7$\\
&&&$51.3\pm5.5$&$72.4\pm0.0$&$11.6\pm1.9$\\
\hline
UGC~12692&Type~III&$21.63\pm0.21$&$7.6\pm0.0$&$32.5\pm5.2$&$5.2\pm0.3$\\
&&&$32.5\pm5.2$&$54.2\pm0.0$&$9.0\pm2.0$\\
\hline
UGC~12857&Type~I&$22.99\pm0.12$&$9.6\pm0.0$&$69.5\pm0.0$&$9.7\pm0.3$\\
\hline
\end{longtable}
Note.~-- $r_{\rm i}$ and $r_{\rm o}$ stand for the inner and the outer limit of each fitted section and $h$ corresponds to the scale length of that section. The lower limit of the innermost section and the upper limit of the outermost section have no error bar because they have been set manually.

\begin{longtable}{lccccc}
\caption{\label{bigtable3} {Results of the fits to horizontal luminosity profiles for the thick disc}}\\
\hline
\hline
ID& Type & $I_{0}$ & $r_{\rm i}$ & $r_{\rm o}$ & $h$\\
&& (mag/as$^3$)  &(\arcsec) &(\arcsec) &(\arcsec)\\
\hline
\endfirsthead
\caption {(Continued)}\\
\hline
\hline
ID& Type & $I_{0}$ & $r_{\rm i}$ & $r_{\rm o}$ & $h$\\
&& (mag/as$^3$)  &(\arcsec) &(\arcsec) &(\arcsec)\\
\hline
\endhead
ESO~157-049&Type~II&$27.08\pm0.19$&$1.5\pm0.0$&$25.1\pm6.9$&$25.7\pm8.8$\\
&&&$25.1\pm6.9$&$72.4\pm0.0$&$13.7\pm0.7$\\
\hline
ESO~240-011&Type~I&$28.37\pm0.08$&$0.0\pm0.0$&$131.3\pm0.0$&$33.1\pm1.1$\\
\hline
ESO~292-014&Type~II&$28.99\pm0.15$&$6.7\pm0.0$&$58.7\pm6.0$&$61.8\pm11.8$\\
&&&$58.7\pm6.0$&$84.6\pm0.0$&$18.8\pm2.2$\\
\hline
ESO~346-001&Type~II&$27.33\pm0.21$&$5.7\pm0.0$&$31.3\pm10.8$&$18.0\pm3.7$\\
&&&$31.3\pm10.8$&$56.6\pm0.0$&$12.1\pm1.3$\\
\hline
ESO~440-027&Type~I&$27.95\pm0.00$&$9.6\pm0.0$&$172.9\pm0.0$&$37.8\pm0.0$\\
\hline
ESO~443-021&Type~I&$28.00\pm0.12$&$11.7\pm0.0$&$64.1\pm0.0$&$19.4\pm0.9$\\
\hline
ESO~466-014&Type~I&$26.93\pm0.15$&$7.6\pm0.0$&$41.3\pm0.0$&$11.4\pm0.6$\\
\hline
ESO~469-015&Type~I&$27.21\pm0.16$&$4.9\pm0.0$&$54.2\pm0.0$&$12.9\pm0.7$\\
\hline
ESO~533-004&Type~II&$27.73\pm0.22$&$7.6\pm0.0$&$38.7\pm7.0$&$34.7\pm10.3$\\
&&&$38.7\pm7.0$&$84.6\pm0.0$&$17.1\pm0.8$\\
\hline
\\
ESO~544-027&Type~II&$27.07\pm0.28$&$2.3\pm0.0$&$10.7\pm7.5$&$24.8\pm20.3$\\
&&&$10.7\pm7.5$&$45.3\pm0.0$&$11.4\pm0.7$\\
\hline
IC~0217&Type~II&$28.73\pm0.11$&$4.9\pm0.0$&$36.1\pm3.7$&$81.2\pm26.3$\\
&&&$36.1\pm3.7$&$91.3\pm0.0$&$22.5\pm0.8$\\
\hline
IC~0610&Type~I&$26.66\pm0.00$&$4.9\pm0.0$&$69.5\pm0.0$&$16.2\pm0.0$\\
\hline
IC~1197&Type~II&$27.49\pm0.21$&$17.6\pm0.0$&$40.8\pm11.7$&$29.2\pm5.8$\\
&&&$40.8\pm11.7$&$94.8\pm0.0$&$19.6\pm1.1$\\
\hline
IC~1553&Type~I&$27.55\pm0.24$&$16.3\pm0.0$&$49.6\pm0.0$&$14.0\pm1.2$\\
\hline
IC~1711&Type~I&$28.37\pm0.17$&$22.9\pm0.0$&$87.9\pm0.0$&$24.8\pm1.5$\\
\hline
IC~1913&Type~I&$29.60\pm0.13$&$0.0\pm0.0$&$69.5\pm0.0$&$34.1\pm2.7$\\
\hline
IC~2058&Type~II&$28.31\pm0.15$&$9.6\pm0.0$&$72.7\pm12.0$&$42.8\pm6.4$\\
&&&$72.7\pm12.0$&$113.8\pm0.0$&$24.3\pm2.2$\\
\hline
IC~2135&Type~II&$27.49\pm0.07$&$0.8\pm0.0$&$78.8\pm4.4$&$45.0\pm3.3$\\
&&&$78.8\pm4.4$&$122.3\pm0.0$&$19.5\pm0.9$\\
\hline
IC~5176&Type~III&$27.12\pm0.26$&$24.3\pm0.0$&$75.9\pm12.1$&$23.2\pm3.0$\\
&&&$75.9\pm12.1$&$150.9\pm0.0$&$41.3\pm3.0$\\
\hline
NGC~0489&Type~I&$27.02\pm0.09$&$0.0\pm0.0$&$49.6\pm0.0$&$12.4\pm0.4$\\
\hline
NGC~0522&Type~II&$27.94\pm0.24$&$12.8\pm0.0$&$63.2\pm7.4$&$40.9\pm10.8$\\
&&&$63.2\pm7.4$&$98.4\pm0.0$&$16.0\pm1.5$\\
\hline
NGC~0678&Type~II&$32.14\pm0.51$&$39.4\pm0.0$&$107.0\pm7.4$&$-80.4\pm44.8$\\
&&&$107.0\pm7.4$&$145.8\pm0.0$&$27.4\pm4.1$\\
\hline
NGC 1032&Type~I&$29.50\pm0.06$&$0.0\pm0.0$&$172.9\pm0.0$&$53.7\pm1.5$\\
\hline
NGC~1163&Type~I&$27.73\pm0.18$&$9.6\pm0.0$&$87.9\pm0.0$&$21.1\pm1.3$\\
\hline
NGC~1422&Type~I&$28.46\pm0.06$&$4.0\pm0.0$&$109.8\pm0.0$&$33.1\pm0.8$\\
\hline
NGC~1495&Type~I&$28.10\pm0.08$&$9.6\pm0.0$&$98.4\pm0.0$&$30.7\pm1.0$\\
\hline
NGC 1596&Type~I&$29.62\pm0.07$&$0.0\pm0.0$&$241.0\pm0.0$&$72.9\pm2.3$\\
\hline
NGC 2732&Type~II&$27.04\pm0.08$&$0.0\pm0.0$&$52.9\pm4.3$&$22.8\pm1.4$\\
&&&$52.9\pm4.3$&$78.3\pm0.0$&$12.6\pm0.8$\\
\hline
NGC 3098&Type~I&$26.99\pm0.08$&$4.9\pm0.0$&$87.9\pm0.0$&$19.0\pm0.5$\\
\hline
NGC~3279&Type~II&$28.04\pm0.12$&$4.9\pm0.0$&$52.1\pm8.4$&$38.6\pm5.7$\\
&&&$52.1\pm8.4$&$84.6\pm0.0$&$21.8\pm1.6$\\
\hline
NGC~3454&Type~II&$29.61\pm0.07$&$0.0\pm0.0$&$36.3\pm3.3$&$\infty\pm\infty$\\
&&&$36.3\pm3.3$&$91.3\pm0.0$&$25.0\pm1.3$\\
\hline
NGC~3501&Type~I&$28.26\pm0.07$&$18.8\pm0.0$&$126.7\pm0.0$&$34.3\pm0.8$\\
\hline
\\
\\
NGC~3592&Type~II&$28.24\pm0.18$&$11.7\pm0.0$&$36.3\pm8.0$&$35.6\pm8.5$\\
&&&$36.3\pm8.0$&$72.4\pm0.0$&$19.0\pm1.4$\\
\hline
NGC~3600&Type~III&$27.58\pm0.11$&$4.9\pm0.0$&$74.6\pm9.1$&$17.7\pm1.0$\\
&&&$74.6\pm9.1$&$118.0\pm0.0$&$120.6\pm51.8$\\
\hline
NGC~3628&Type~II&$28.95\pm0.16$&$69.5\pm0.0$&$383.5\pm18.1$&$219.3\pm34.2$\\
&&&$383.5\pm18.1$&$532.0\pm0.0$&$58.3\pm4.9$\\
\hline
NGC~4081&Type~I&$26.77\pm0.06$&$9.6\pm0.0$&$64.1\pm0.0$&$16.9\pm0.3$\\
\hline
NGC 4111&Type~I&$29.00\pm0.10$&$13.9\pm0.0$&$145.8\pm0.0$&$45.2\pm2.0$\\
\hline
NGC~4330&Type~II&$29.09\pm0.06$&$0.0\pm0.0$&$115.4\pm5.4$&$87.5\pm7.3$\\
&&&$115.4\pm5.4$&$167.2\pm0.0$&$30.5\pm1.7$\\
\hline
NGC~4359&Type~I&$28.70\pm0.03$&$13.9\pm0.0$&$135.9\pm0.0$&$36.7\pm0.3$\\
\hline
NGC~4437&Type~II&$30.66\pm0.00$&$39.4\pm0.0$&$297.9\pm0.0$&$151.5\pm0.0$\\
&&&$297.9\pm0.0$&$332.5\pm0.0$&$73.1\pm0.0$\\
\hline
NGC~4565&Type~I&$30.13\pm0.04$&$78.3\pm0.0$&$469.9\pm0.0$&$130.4\pm1.8$\\
\hline
NGC~4607&Type~I&$27.29\pm0.03$&$1.5\pm0.0$&$135.9\pm0.0$&$28.9\pm0.3$\\
\hline
NGC~4747&Type~I&$27.93\pm0.06$&$7.6\pm0.0$&$126.7\pm0.0$&$37.4\pm0.9$\\
\hline
NGC 5084&Type~III&$28.10\pm0.13$&$37.5\pm0.0$&$217.6\pm19.3$&$58.8\pm3.8$\\
&&&$217.6\pm19.3$&$427.9\pm0.0$&$150.0\pm13.9$\\
\hline
NGC~5470&Type~II&$28.07\pm0.11$&$4.0\pm0.0$&$71.8\pm5.4$&$35.6\pm3.4$\\
&&&$71.8\pm5.4$&$87.9\pm0.0$&$12.4\pm2.2$\\
\hline
NGC~5529&Type~III+II&$28.20\pm0.01$&$27.3\pm0.0$&$132.1\pm0.5$&$36.6\pm0.1$\\
&&&$132.1\pm0.5$&$200.0\pm0.9$&$\infty\pm\infty$\\
&&&$200.0\pm0.9$&$204.5\pm0.0$&$25.5\pm1.1$\\
\hline
NGC~5981&Type~I&$28.15\pm0.06$&$6.7\pm0.0$&$102.1\pm0.0$&$29.6\pm0.7$\\
\hline
NGC 6010&Type~I&$28.20\pm0.04$&$0.0\pm0.0$&$91.3\pm0.0$&$25.2\pm0.5$\\
\hline
NGC~7347&Type~I&$27.93\pm0.15$&$6.7\pm0.0$&$54.2\pm0.0$&$17.6\pm1.1$\\
\hline
PGC~013646&Type~I&$28.25\pm0.06$&$7.6\pm0.0$&$113.8\pm0.0$&$31.1\pm0.7$\\
\hline
PGC~028308&Type~I&$28.00\pm0.21$&$13.9\pm0.0$&$56.6\pm0.0$&$22.3\pm2.1$\\
\hline
PGC~030591&Type~I&$27.59\pm0.20$&$9.6\pm0.0$&$45.3\pm0.0$&$13.6\pm1.0$\\
\hline
PGC 032548&Type~I&$28.82\pm0.29$&$0.0\pm0.0$&$43.3\pm0.0$&$17.5\pm2.8$\\
\hline
PGC~052809&Type~II&$28.01\pm0.10$&$2.3\pm0.0$&$82.8\pm3.5$&$56.3\pm6.6$\\
&&&$82.8\pm3.5$&$122.3\pm0.0$&$15.5\pm0.9$\\
\hline
UGC~00903&Type~I&$26.53\pm0.07$&$1.5\pm0.0$&$59.0\pm0.0$&$14.7\pm0.4$\\
\hline
UGC~01970&Type~I&$29.09\pm0.10$&$13.9\pm0.0$&$75.3\pm0.0$&$28.5\pm1.3$\\
\hline
UGC~05347&Type~I&$28.67\pm0.23$&$9.6\pm0.0$&$35.7\pm0.0$&$15.8\pm1.8$\\
\hline
\\
UGC~05689&Type~II&$27.66\pm0.08$&$2.3\pm0.0$&$17.3\pm3.4$&$17.3\pm1.6$\\
&&&$17.3\pm3.4$&$43.3\pm0.0$&$11.4\pm0.5$\\
\hline
UGC~05958&Type~I&$27.70\pm0.18$&$7.6\pm0.0$&$64.1\pm0.0$&$17.4\pm1.2$\\
\hline
UGC~06526&Type~I&$28.23\pm0.08$&$0.0\pm0.0$&$87.9\pm0.0$&$25.2\pm0.9$\\
\hline
UGC~07086&Type~I&$26.98\pm0.11$&$4.9\pm0.0$&$87.9\pm0.0$&$18.8\pm0.7$\\
\hline
UGC~08737&Type~I&$27.05\pm0.09$&$15.1\pm0.0$&$122.3\pm0.0$&$24.5\pm0.7$\\
\hline
UGC~09448&Type~I&$26.89\pm0.25$&$8.6\pm0.0$&$37.5\pm0.0$&$10.8\pm1.0$\\
\hline
UGC~09665&Type~I&$27.66\pm0.09$&$4.0\pm0.0$&$59.0\pm0.0$&$17.7\pm0.7$\\
\hline
UGC~10043&Type~I&$28.42\pm0.25$&$22.9\pm0.0$&$78.3\pm0.0$&$25.8\pm2.6$\\
\hline
UGC~10288&Type~I&$27.88\pm0.03$&$18.8\pm0.0$&$156.1\pm0.0$&$35.3\pm0.3$\\
\hline
UGC~10297&Type~I&$28.37\pm0.24$&$6.7\pm0.0$&$39.4\pm0.0$&$14.6\pm1.7$\\
\hline
UGC~12518&Type~I&$27.32\pm0.10$&$2.3\pm0.0$&$72.4\pm0.0$&$17.4\pm0.6$\\
\hline
UGC~12692&Type~I&$27.57\pm0.01$&$5.7\pm0.0$&$49.6\pm0.0$&$13.8\pm0.0$\\
\hline
UGC~12857&Type~I&$28.35\pm0.20$&$9.6\pm0.0$&$49.6\pm0.0$&$18.0\pm1.6$\\
\hline
\end{longtable}
Note.~-- $r_{\rm i}$ and $r_{\rm o}$ stand for the inner and the outer limit of each fitted section and $h$ corresponds to the scale length of that section. The lower limit of the innermost section and the upper limit of the outermost section have no error bar because they have been set manually.

\clearpage

\appendix

\section{Luminosity profile fits}
\label{appendixver}

The following pages include the vertical and horizontal luminosity profile fits for all the galaxies in our final sample. They also include the image of each galaxy. The information is organized as follows:

\begin{itemize}
 \item The three images in the top row show the average of the 3.6 and 4.5$\mu{\rm m}$-band background-subtracted S$^4$G frames (left), the same image after masking (center), and the used mask (right). The vertical red lines indicate the limits of the fitted vertical bins, the central one being ignored due to the possible presence of a bulge.
 \item  The next two rows show the fits to the luminosity profiles in these bins. The data points have $2\sigma$ statistical error bars, the dashed curve represents the best fit, the dotted curves indicate the contributions of the thin and the thin discs. The dash-dotted vertical lines indicate the limits of the range in vertical distance above the mid-plane used for the fit. The vertical solid line indicates $z_{\rm s}$ for each bin. The profiles in bins which have not been selected for the data processing in this paper (Section~\ref{secdefrs}) are indicated with a ``Not used'' label.
 \item The bottom panel shows the horizontal luminosity profiles integrated from $z=0$ to $z=z_{\rm u}$ (total; yellow), from $z=0$ to $z=0.5z_{\rm s}$ (thin disc; blue), and from $z=z_{\rm s}$ to $z=z_{\rm u}$ (thick disc; red). The black lines indicate fits to those profiles, with vertical intersecting lines indicating the fitting range and the truncation radii. The vertical solid gray lines represent $0.2r_{25}$, $0.5r_{25}$, and $0.8r_{25}$. The vertical dashed line represents $r_{25}$. The gray area represents the inferred average surface brightness of the thick disc in the range $z=0$ to $z=0.5z_{\rm s}$ in the presence of a thin disc with a relative face-on surface brightness similar to which is found where the vertical luminosity profiles have been successfully fitted (upper limit) and in the absence of it (lower limit) extrapolated from the vertical luminosity profile fits. The bins for which valid vertical luminosity fits have been obtained can be seen in the bottom left of each panel. The errors 
in the fitted parameters are $2\sigma$ fitting errors.
\end{itemize}

Because of file size issues, only the data for ESO~157-049 are presented for the arXiv version of the paper. All the remaining data are available online via the Data Conservancy Project (DC).

\clearpage

\begin{figure*}[h]
\begin{center}
\begin{tabular}{c c}
\multicolumn{2}{c}{\includegraphics[width=0.9\textwidth]{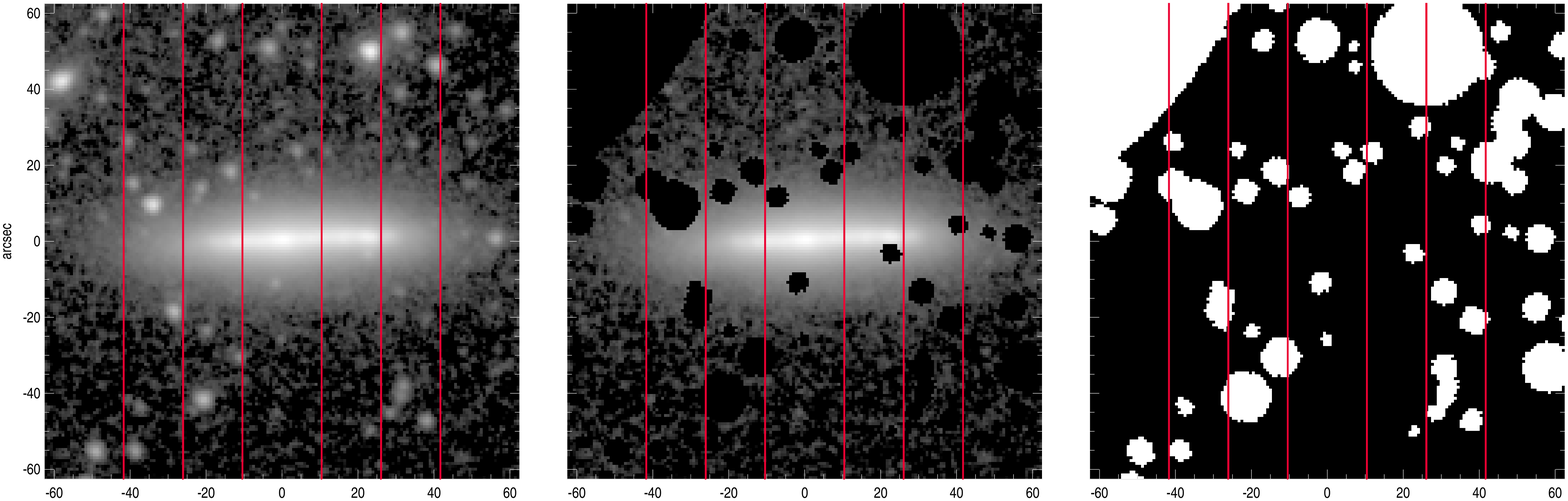}}\\
\includegraphics[width=0.45\textwidth]{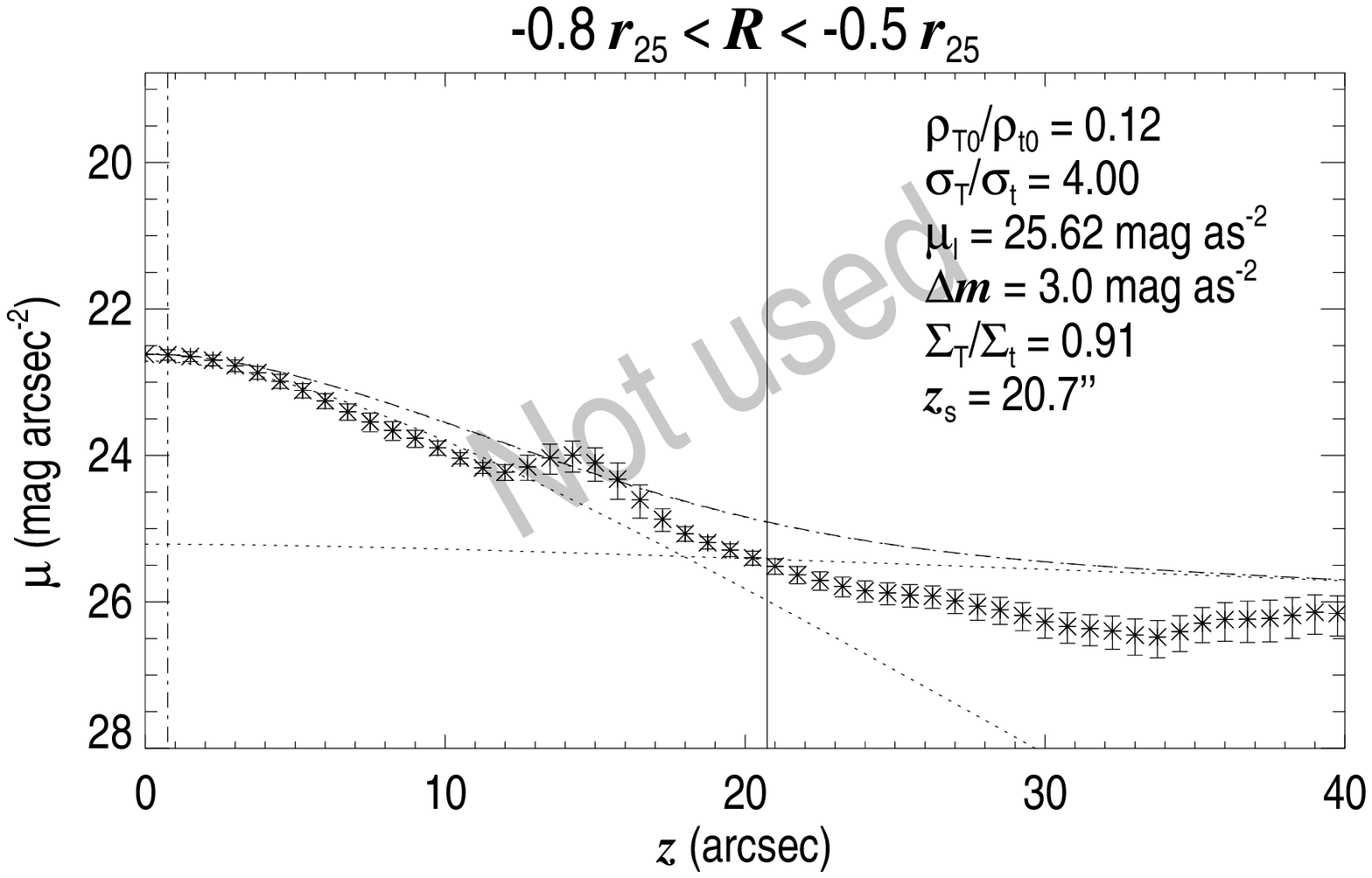}&
\includegraphics[width=0.45\textwidth]{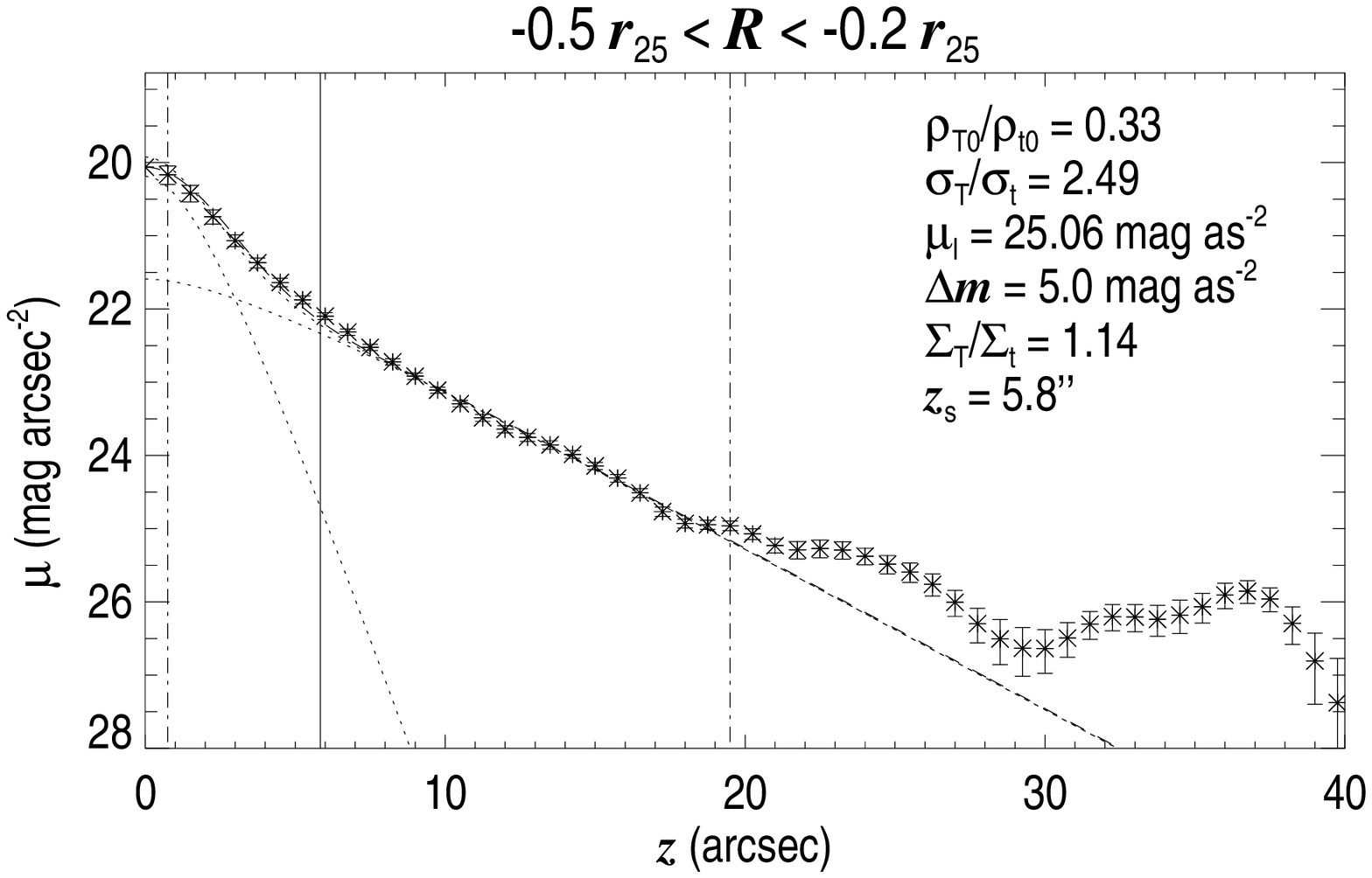}\\
\includegraphics[width=0.45\textwidth]{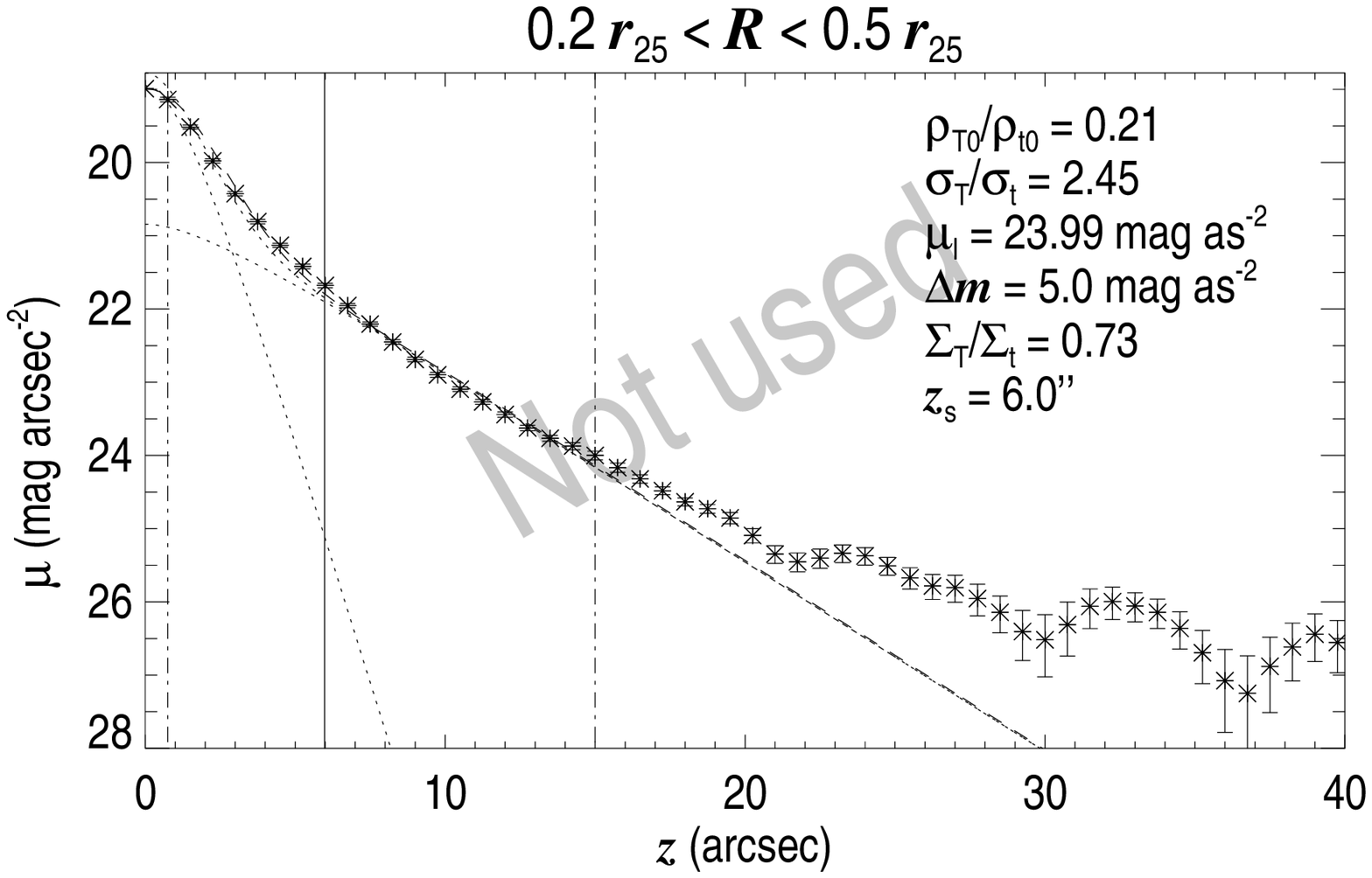}&
\includegraphics[width=0.45\textwidth]{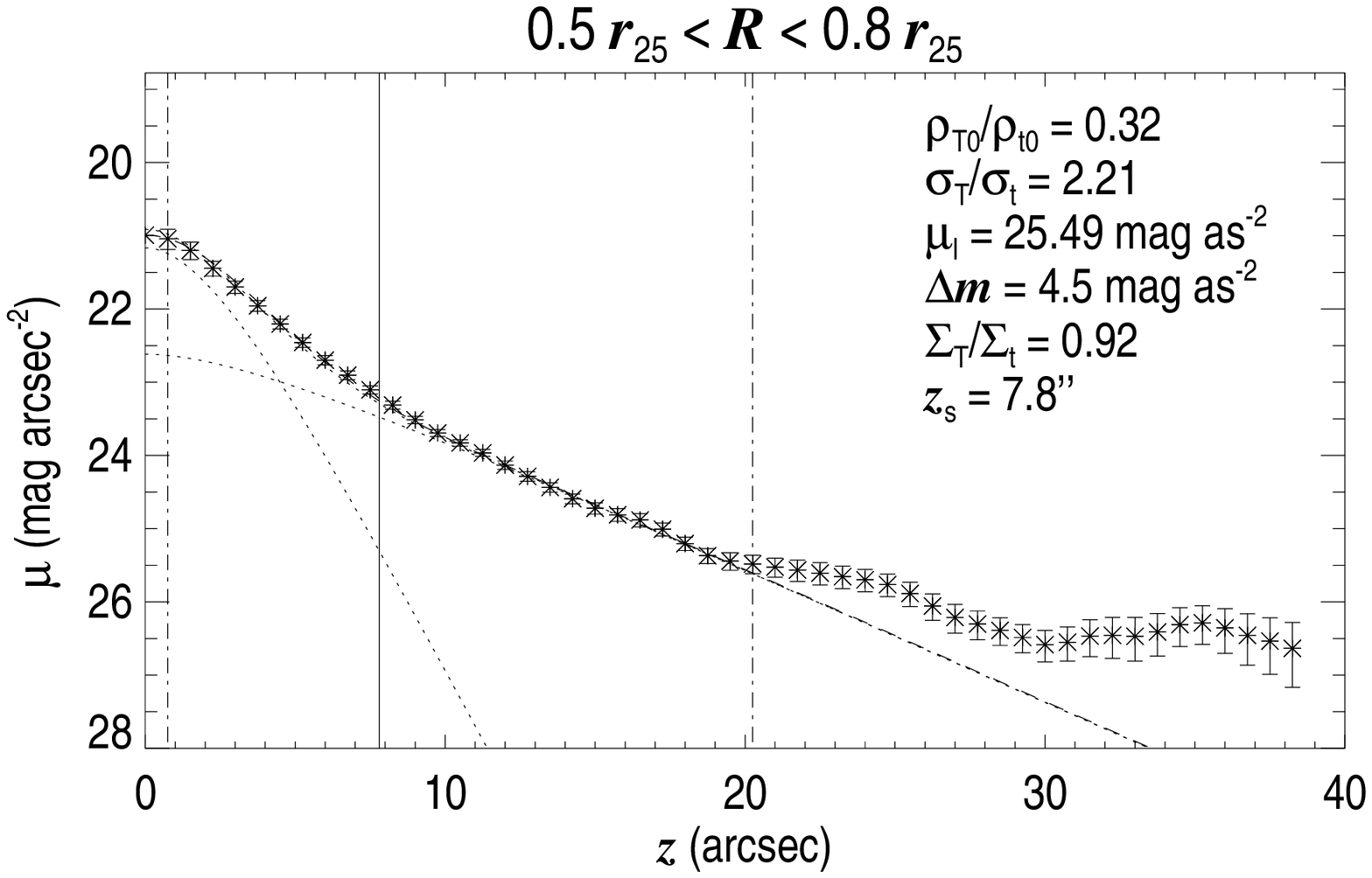}\\
\multicolumn{2}{c}{\includegraphics[width=0.7\textwidth]{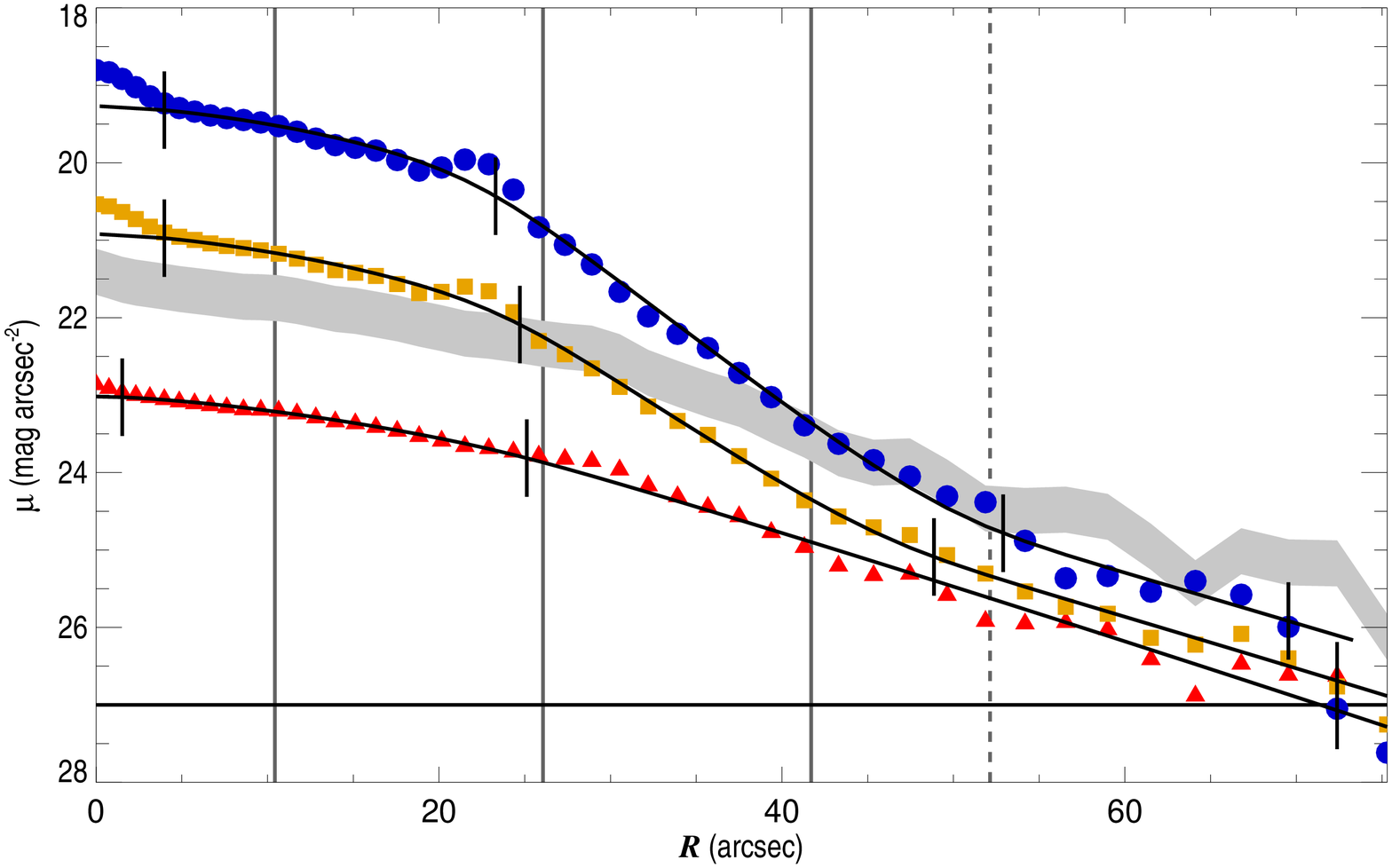}}
\end{tabular}
\caption{\label{verticalfits} ESO157-049.}
\end{center}
\end{figure*}
\clearpage

\end{document}